\documentclass[12pt, notitlepage,aps,preprintnumbers,superscriptaddress,nofootinbib,aps,prd,floatfix]{revtex4-2}
\pdfoutput=1
\usepackage{graphicx}% Include figure files
\usepackage{subfigure}
\usepackage{dcolumn}% Align table columns on decimal point
\usepackage{bm}% bold math
\usepackage{slashed}
\usepackage{extarrows}
\usepackage{multirow,array}
\usepackage{amsmath,amssymb,physics}
\usepackage{slashed}
\usepackage{url}
\usepackage{hyperref}
\usepackage{xcolor}
\definecolor{nicered}{rgb}{0.5,0.,0.}
\definecolor{nicegreen}{rgb}{0.,0.5,0.}
\definecolor{niceblue}{rgb}{0.,0.,0.5}
\definecolor{darkmagenta}{rgb}{0.4,0,0.4}
\hypersetup{colorlinks,citecolor=nicegreen,linkcolor=nicered,urlcolor=niceblue}
\usepackage{braket}
\usepackage{colortbl}
\usepackage{gensymb}%degree
\usepackage{comment}
\usepackage{enumitem}
\setlist{nolistsep}
\usepackage{setspace}

\newcommand{\GeV}{\textrm{GeV}}
\newcommand{\PDF}{\textrm{PDF}}
\newcommand{\QED}{\textrm{QED}}
\newcommand{\bea}{\begin{equation}\begin{aligned}}
\newcommand{\eea}{\end{aligned}\end{equation}}

\newcommand{\calO}{\mathcal{O}}
\newcommand{\msbar}{\overline{\textrm{MS}}}
\newcommand{\el}{\textrm{el}}
\newcommand{\inel}{\textrm{inel}}
\newcommand{\HT}{\textrm{HT}}

\begin{document}
\title{The photon PDF within the CT18 global analysis}
\author{Keping Xie}
\email{xiekeping@pitt.edu}
\affiliation{Pittsburgh Particle Physics, Astrophysics, and Cosmology Center, 
Department of Physics and Astronomy, University of Pittsburgh, Pittsburgh, PA 15260, USA\looseness=-1}
\author{T.~J.~Hobbs}
\email{thobbs@fnal.gov}
\affiliation{Fermi National Accelerator Laboratory, Batavia, IL 60510, USA}
\affiliation{Department of Physics, Southern Methodist University, Dallas, Texas 75275, USA}
\affiliation{Jefferson Lab, Newport News, VA 23606, USA}
\affiliation{Department of Physics, Illinois Institute of Technology, Chicago, IL 60616, USA}
\author{Tie-Jiun Hou}
\email{houtiejiun@mail.neu.edu.cn}
\affiliation{Department of Physics, College of Sciences, Northeastern University, Shenyang 110819, China\looseness=-1}
\author{Carl Schmidt}
\email{schmi113@msu.edu}
\affiliation{Department of Physics and Astronomy, Michigan State University, East Lansing, MI 48824, USA\looseness=-1}
\author{Mengshi Yan}
\email{msyan@pku.edu.cn}
\affiliation{School of Physics and State Key Laboratory of Nuclear Physics and Technology,Peking University, Beijing 100871, China\looseness=-1}
\author{C.-P. Yuan}
\email{yuanch@msu.edu}
\collaboration{CTEQ-TEA Collaboration}
\date{\today}
\singlespacing

\preprint{FERMILAB-PUB-21-370-QIS-SCD-T, MSUHEP-21-013, PITT-PACC-2112, SMU-HEP-21-06}

\begin{abstract}
Building upon the most recent CT18 global fit, we present a new calculation of the photon content of the proton based on an application of the LUX formalism. In this work, we explore two principal variations of the LUX ansatz. In one approach, which we designate ``CT18lux,'' the photon PDF is calculated directly using the LUX formula for all scales, $\mu$. In an alternative realization, ``CT18qed,'' we instead initialize the photon PDF in terms of the LUX formulation at a lower scale, $\mu\! \sim\! \mu_0$, and evolve to higher scales with a combined QED+QCD kernel at $\mathcal{O}(\alpha),~\mathcal{O}(\alpha\alpha_s)$ and $\mathcal{O}(\alpha^2)$.
While we find these two approaches generally agree, especially at intermediate $x$ ($10^{-3}\lesssim x\lesssim0.3$), we discuss some moderate discrepancies that can occur toward the end-point regions at very high or low $x$.
We also study effects that follow from variations of the inputs to the LUX calculation originating outside the pure deeply-inelastic scattering (DIS) region, including from elastic form factors and other contributions to the photon PDF. Finally, we investigate the phenomenological implications of these photon PDFs for the LHC, including high-mass Drell-Yan, vector-boson pair, top-quark pair, and Higgs associated with vector-boson production.     
\end{abstract}

\maketitle

\newpage
\tableofcontents{}

\section{Introduction}
\label{sec:intro}

With the steady accumulation of copious experimental data at the LHC, we have entered into a high-precision era for hadron-collider physics. In parallel, theoretical computations of higher-order QCD corrections to standard LHC processes have reached to next-to-next-to-leading order (NNLO) in general, and, in some instances, even next-to-NNLO (N$^3$LO) accuracy has now become available (see Ref.~\cite{Heinrich:2017una} for an overview). At this level of precision, electroweak (EW) corrections begin to have an observable impact, as $\alpha_{\QED}\sim \alpha_S^2$. To that end, next-to-leading order (NLO) EW corrections to hard-scattering matrix elements have been computed for many LHC processes of interest, and the automation of the NLO EW corrections has also been achieved in recent years \cite{Biedermann:2017yoi,Frederix:2018nkq}.
To perform consistent higher-order calculations with EW corrections included in the initial state of parton-scattering processes at the LHC, it is necessary to employ a set of parton distribution functions (PDFs) in which the photon appears as an active, partonic constituent of the proton. 

The first such PDF set to include the photon as a parton of the proton was the 2004
release of the MRST group, the MRST2004QED PDFs \cite{Martin:2004dh}.
The MRST group used a parametrization for the photon PDF based on radiation off of  ``primordial'' up and down quarks, with the photon radiation cut off at low scales governed by constituent- or current-quark masses.
Another approach to include the photon PDF is to constrain it in an analogous way to other partons by fitting available Deep Inelastic Scattering (DIS) and Drell-Yan data \cite{Ball:2013hta}, an approach first developed by the NNPDF Collaboration and released in the NNPDF2.3QED~\cite{Ball:2013hta} and NNPDF3.0QED~\cite{Ball:2014uwa} PDFs. A similar determination is also performed by the \textsc{xFitter} group~\cite{xFitterDevelopersTeam:2017fxf}.
The constraints on the photon PDFs from the data were rather weak, due to the small size of the photon-initiated contributions, so that the NNPDF  photon PDF was consistent with zero at the initial scale of $\sqrt{2}$ GeV,  with large photon PDF uncertainty at high $x$.

Contemporaneously, the CT14QED PDFs were constructed by implementing Quantum Electrodynamics (QED)-informed
evolution at leading order (LO) along with QCD
evolution at NLO within the CTEQ-TEA (CT) global analysis
package~\cite{Schmidt:2015zda}.  The inelastic contribution to the photon PDF was described by a two-parameter ansatz, coming from radiation off the valence quarks, and based on the CT14 NLO PDFs. 
The inelastic photon PDFs were specified in
terms of the inelastic momentum fraction carried by the photon, at the initial scale $\mu_0$, and they were constrained by comparing with ZEUS data~\cite{Chekanov:2009dq}  on
the production of isolated photons in deeply-inelastic scattering (DIS), $ep\rightarrow e\gamma+X$.
The advantage of using this process is that the initial-state photon contributions are at leading
order in the perturbation expansion.  In contrast, the initial-state photon contribution to Drell-Yan or $W$ and $Z$ production
is suppressed by factors of $(\alpha/\alpha_s)$ relative to the leading quark-antiquark production. 
As discussed by Martin and Ryskin~\cite{Martin:2014nqa}, the photon PDF also receives a large {\em elastic} contribution in which the proton remains intact, in addition to
the {\em inelastic} contribution in which the proton breaks into a multihadron final state.\footnote{In Ref.~\cite{Martin:2014nqa} these two contributions are referred to as ``coherent'' and ``incoherent'', respectively.}  
In addition to the default CT14QED set, CT therefore also released the CT14QEDinc PDFs, in which
the {\em inclusive} photon PDF at the scale $\mu_0$ is
defined by a sum of an inelastic photon PDF and a corresponding elastic photon distribution obtained from the Equivalent Photon Approximation (EPA) \cite{Budnev:1974de}; in this latter formulation, the EPA allows the elastic
photon contribution to be determined through the nucleon's elastic form factors, which can themselves be directly measured
via elastic scattering experiments.
Neither MRST nor NNPDF directly addressed these separated contributions to the photon PDF, although
the NNPDF photon can be assumed to be {\em inclusive}, containing both inelastic and elastic components, since it was constrained using inclusive Drell-Yan and vector boson data.
In all cases, the available data were unable to constrain the photon PDF to a high degree of accuracy.

In order to overcome these deficiencies, more accurate determinations of the photon distribution are necessary. The
key to higher accuracy for photon PDFs is again the abovementioned EPA~\cite{Budnev:1974de}, in this case, extended
to include inelastic contributions evaluated in terms of inelastic structure functions~\cite{Anlauf:1991wr,Blumlein:1993ef,Mukherjee:2003yh}.
This structure function approach can then be employed to predict photon-initiated production processes of high-mass particles at hadron colliders~\cite{Luszczak:2015aoa,Harland-Lang:2016apc,Harland-Lang:2016kog}. This idea was revived by the LUXqed group~\cite{Manohar:2016nzj,Manohar:2017eqh}, who demonstrated it in a rigorous framework based upon collinear factorization extending beyond leading order via the inclusion of the proper $\msbar$ matching term; this
work additionally released the first public photon PDF based upon this theoretical approach. 
As the elastic and inelastic proton structure functions have been determined experimentally to high precision, the photon PDFs can be constrained at the level of $1{-}2\%$ over a wide range of the momentum fraction $x$.
Furthermore, based on the anomalous dimensions in light-cone gauge,\footnote{See Ref.~\cite{Blumlein:2012bf} and the references therein for a complete review.} the QED splitting kernels in the  DGLAP~\cite{Dokshitzer:1977sg,Gribov:1972ri,Lipatov:1974qm,Altarelli:1977zs} evolution equations have now been calculated up to $\mathcal{O}(\alpha\alpha_S)$ \cite{deFlorian:2015ujt} and $\mathcal{O}(\alpha^2)$ \cite{deFlorian:2016gvk}, whose effects are important to the determination of precise photon PDFs at a large energy scale $\mu$ as a function of $x$. 
Subsequently, the NNPDF group adopted the LUX formalism and introduced a photon PDF in a global PDF fit, named NNPDF3.1luxQED PDFs~\cite{Bertone:2017bme}.
Likewise, the MMHT2015qed PDF set was released a number of years ago. These PDFs were generated in a global fit by adopting  
the LUX formalism at a low starting scale, $\mu_0=1$ GeV, for the photon PDF, and evolving to higher scales using QED-corrected DGLAP evolution equations~\cite{Harland-Lang:2019pla}. These features were inherited in the more recent MSHT20qed~\cite{Cridge:2021pxm} analysis, which was based on the latest release of the MSHT20 global fit~\cite{Cridge:2021qfd}.

In this paper, we follow the original CT14QED strategy of separating the photon PDF into its respective {\it elastic} and {\it inelastic} components. 
The photon PDFs are generated based on 
two different approaches of applying the LUX formalism within the framework of the CT18 NNLO global analysis~\cite{Hou:2019efy}, where the NNLO QCD kernels have been used in the evolution of partons. 
In the first approach, CT18lux, the photon PDF is calculated directly using the LUX formula at any scale $\mu$. In the second approach, CT18qed, we instead initialize the photon PDF in terms of the LUX formulation at a lower scale, $\mu\! \sim\! \mu_0$, and evolve to higher scales with a combined QED and QCD kernels at $\mathcal{O}(\alpha),~\mathcal{O}(\alpha\alpha_s)$ and $\mathcal{O}(\alpha^2)$.  
For convenience, we shall refer to the former as the 
LUX formalism approach, and the latter as the DGLAP evolution approach.
While the PDFs generated by both approaches generally agree, particularly at the intermediate $x$ region ($10^{-3}\lesssim x\lesssim0.3$), they differ in the low $x$ region where the inelastic photon dominates and in the large $x$ region where the contributions from the elastic component of the photon PDF becomes important. Hence, in this work, we have also explored the impact on the photon PDF from various recent updates of the elastic component, which represents the photon contribution from elastic photon-proton scattering processes.
As we shall discuss in greater detail below, implementation of the LUX formalism, which involves integrations of the proton's unpolarized electromagnetic structure functions,
$F_{2,L}$, over broad $Q^2$, can be sensitive to higher-twist ({\it i.e.}, twist-4) and other nonperturbative QCD contributions. These effects are unsuppressed
at low $Q^2$ and must be explicitly modeled; this is necessary both for theoretical accuracy as well as uncertainty quantification, for which an estimate of the
possible model and parametric dependence is important. We point out that these considerations contrast with the situation in a typical NNLO PDF global analysis like
CT18, in which only leading-twist ({\it i.e.}, twist-2) dynamics are admitted into the relevant calculations. This is achieved by modeling only the twist-2 PDFs
at the boundary of QCD evolution, $\mu\! =\! \mu_0$, and constraining the resulting parametrization through an admixture of perturbative QCD parton-level
cross sections and hadronic data at sufficiently high $Q^2$ and $W^2$ (for DIS) to ensure that contributions from sub-leading twist are safely, kinematically
suppressed.

%
%\cpy{Is it true that the DGLAP equations only involve the inelastic component? In MMHT15qed, the elastic component is also evolved.}

Finally, owing to its importance to the LHC phenomenology discussed at the end of this article, in this work we concentrate on the photon PDF of the
{\it proton}. We note that it is also technically feasible to carry out analogous studies for the photon content of the neutron,  as considered in the MMHT2015qed~\cite{Harland-Lang:2019pla} and MSHT20qed~\cite{Cridge:2021pxm} analyses. To treat such scenarios in full
generality, it is necessary to consider the possibility of explicit charge-symmetry breaking at parton-level, and we defer
such issues to later work.

The organization of this paper is as follows.  In Sec.~\ref{sec:LUX2DGLAP},
we present both the LUX formalism and  DGLAP evolution approaches to generate photon PDFs. 
In Sec.~\ref{sec:CT18lux}, we discuss the CT18lux photon PDF based on the LUX formalism. Various sources of the photon PDF uncertainty are also discussed. 
In Sec.~\ref{sec:ct18qed}, we present the result of the DGLAP-driven CT18qed, and compare various photon PDF sets with different choices of the input scale, $\mu_0$, where the photon PDF is provided by the LUX master formula. 
The main difference between the CT18qed and CT18lux photon PDFs will also be explored. 
In Sec.~\ref{sec:pheno}, we investigate the phenomenological implications of these photon PDFs at the LHC, including high-mass Drell-Yan, vector-boson pair, top-quark pair, and Higgs associated with vector-boson production. 
In Sec.~\ref{sec:conc}, we discuss our findings and give conclusions.
Following the main body of the paper, we defer a number of technical details to a set of appendices. 
%In App.~\ref{sec:chi2}, we present $\chi^2$ values for the main PDFs released in this study. 
The separation of the photon PDF into elastic
and inelastic components is discussed in App.~\ref{app:sep_inel}. In App.~\ref{app:sep}, we detail the physical factorization and
$\msbar$ conversion terms that appear in the LUX formalism.

\section{The LUX formalism versus DGLAP evolution}
\label{sec:LUX2DGLAP}

As reviewed in Sec.~\ref{sec:intro}, the CT14QEDinc photon PDF \cite{Schmidt:2015zda} was comprised of two distinct sub-components. At the initial scale, $\mu_0$, CT14QEDinc is given by a sum of inelastic ($\gamma^{\inel}$, \emph{i.e.}, CT14QED) and elastic ($\gamma^{\rm el}$) pieces. The CT14QEDinc photon PDF at any higher energy scale, $\mu\!>\!\mu_0$, is obtained by 
solving the QED-corrected DGLAP evolution equation,
\begin{equation}\label{eq:DGLAP}
\frac{\dd\gamma}{\dd\log \mu^2}=\frac{\alpha}{2\pi}
\left[p_{\gamma\gamma}\otimes\gamma+\sum_i e_i^2p_{\gamma q}\otimes(q_{i}+\bar{q}_i)\right].
\end{equation}
The CT14QED photon PDFs were parametrized (by a two-parameter ansatz) and specified in
terms of the inelastic momentum fraction carried by the photon at the initial scale $\mu_0$, which was constrained by comparing with ZEUS data~\cite{Chekanov:2009dq} on
the production of isolated photons in DIS, $ep\rightarrow e\gamma+X$.
The ZEUS experiment imposed a cut requiring at least one reconstructed track, well-separated from the lepton, which removed the elastic component of the photon. Although it is possible that some inelastic photon contribution would also fail this cut, this is expected to be small relative to the experimental uncertainties of the experiment~\cite{Schmidt:2015zda}.
The elastic component was parametrized by the Equivalent Photon Approximation~\cite{Budnev:1974de}, which involves an integration over the proton electromagnetic form factors. 

\begin{figure}
	\centering
	\includegraphics[width=0.65\textwidth]{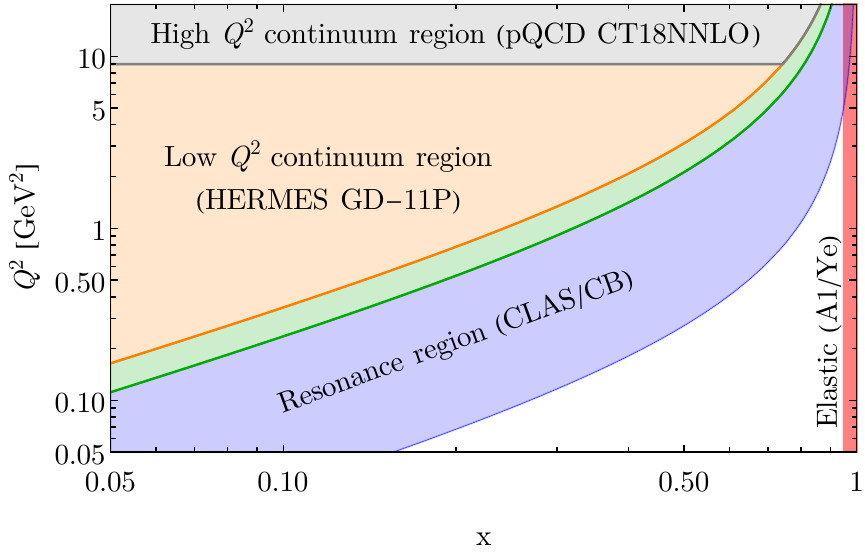}
	\caption{The breakup of $(x,Q^2)$ plane to determine the $F_{2}(x,Q^2)$ and $F_L(x,Q^2)$.}
	\label{fig:xQ2plane}
\end{figure}

As implied by the Equivalent Photon Approximation, an alternative access to the photon PDF is through its relationship to electromagnetic scattering off the proton~\cite{Budnev:1975poe}.   Early investigations on the connection between the photon PDF and the proton structure functions, $F_2(x,Q^2)$ and $F_L(x,Q^2)$,  were presented in Refs.~\cite{Anlauf:1991wr,Blumlein:1993ef,Mukherjee:2003yh,Luszczak:2015aoa}.  This idea  of obtaining the photon PDF by viewing the $e p\to e +X$ scattering process as an electron scattering off the photon parton in the proton was formalized systematically within perturbative QCD/QED by the LUX group in Refs.~\cite{Manohar:2016nzj,Manohar:2017eqh}.
In such a way, the photon PDF is fully determined by the structure functions, without the need of introducing a non-perturbative parameterization at an input scale, $\mu_0$. The master formula to determine the LUX photon PDF is\footnote{The $\msbar$ running coupling $\alpha(\mu^2)$ is related with the physical coupling $\alpha_{\rm ph}(q^2)$ as
	\begin{equation*}
	\alpha_{\rm ph}(q^2)=\frac{\alpha(\mu^2)}{1-\Pi(q^2,\mu^2)},
	\end{equation*}
	where $q^2=-Q^2$ corresponds to the spacelike region, and $\Pi(q^2,\mu^2)$ is the vacuum polarization. In the large momentum limit, $|q^2|\gg m_q^2$ or $m^2_{\ell}$, where $m_q(m_\ell)$ is the masses of light quarks (leptons), $\Pi(q^2,\mu^2)=\frac{2}{3}\frac{\alpha(\mu^2)}{2\pi}\left(\sum_{q}N_ce_q^2+\sum_\ell e_\ell^2\right)\log(|q^2|/\mu^2)$.
	We have the freedom to choose the renormalization scale as $\mu^2=|q^2|$ to make $\Pi(q^2,\mu^2)=0$ and, therefore, $\alpha(\mu^2)=\alpha_{\rm ph}(-Q^2)$.
}
\bea
x \gamma(x,\mu^2)
&=\frac{1}{2\pi\alpha(\mu^2)}\int_{x}^{1}\frac{\dd z}{z}
\Bigg\{\int_{\frac{x^{2}m_{p}^{2}}{1-z}}^{\frac{\mu^{2}}{1-z}}
\frac{\dd Q^{2}}{Q^{2}}\alpha_{\rm ph}^{2}(-Q^2)
\Bigg[\left(zp_{\gamma q}(z)+
\frac{2x^{2}m_{p}^{2}}{Q^{2}}\right)F_{2}(x/z,Q^{2})\\
&-z^{2}F_{L}(x/z,Q^{2})\Bigg]
-\alpha^{2}(\mu^2)z^{2}F_{2}(x/z,\mu^2) 
\Bigg\} + \mathcal{O}(\alpha^2, \alpha\alpha_s), 
\label{eq:LUX}
\eea
which includes all the 
$\alpha L(\alpha_sL)^n$, 
$\alpha (\alpha_sL)^n$,
and 
$\alpha^2 L(\alpha_sL)^n$ terms, with $L=\ln(Q^2/m_p^2)$. In this equation, $p_{\gamma q}(z)\equiv [1+(1-z)^2]/z$ is the leading order DGLAP splitting kernel, and the
$\mathcal{O}(\alpha^2, \alpha\alpha_s)$ term does not contain any  logarithmic enhancement for large $L$.
As explained in App.~\ref{app:sep}, the LUX photon consists of two components. 
The first term inside square brackets corresponds to the {\bf physical factorization} contribution, $\gamma^{\rm PF}(x,\mu^2)$, while the second term involving only $F_2$ is the {\bf $\msbar$ conversion} piece, $\gamma^{\rm con}(x,\mu^2)$, after a divergence is cancelled with the corresponding counterterms \cite{Manohar:2017eqh}.
To perform the integration in Eq.~(\ref{eq:LUX}), it is necessary to know the structure functions over the full $(x,Q^2)$ plane. Inside the high-$Q^2$ and high-$W^2$ region, $Q^2\!>\!Q^2_{\rm PDF}=9~\GeV^2$ and $W^2\!>\!W^2_{\rm high}=4~\GeV^2$, over which perturbative QCD is reliably applicable, the structure functions can be calculated directly from quark and gluon PDFs. In the low-$W^2$ region, $W^2\!<\!W^2_{\rm low}=3~\GeV^2$, the original LUX methodology adopts the structure functions directly from phenomenological fits of CLAS\footnote{The CLAS fit is bounded by a threshold $W^2>(m_p+m_{\pi})^2$ for nucleon resonance production, illustrated as the lower edge in Fig. \ref{fig:xQ2plane}.} \cite{Osipenko:2003bu} or Christy and Bosted (CB) \cite{Christy:2007ve}. In the continuum region, $Q^2<Q^2_{\rm PDF}$ and $W^2\!>\!W^2_{\rm high}$, the GD11-P fit of HERMES Collaboration \cite{Airapetian:2011nu} based on the ALLM funcion form \cite{Abramowicz:1991xz} was adopted. A smooth and continuous transition from the $W^2_{\rm low}$ to $W^2_{\rm high}$ (the green band in Fig. \ref{fig:xQ2plane}) was performed based on the a quadratic functional form, with details in Sec. \ref{sec:CT18lux}. The subdivision of the $(x,Q^2)$ plane
is illustrated in Fig.~\ref{fig:xQ2plane}.

In subsequent analyses, both the NNPDF \cite{Bertone:2017bme} and MMHT \cite{Harland-Lang:2019pla} groups released photon PDFs based on the incorporation of the LUX formalism into their respective frameworks, making a number of
different technical choices regarding the implementation of the LUX approach. Similar to LUXqed(17)\footnote{The LUXqed17 \cite{Manohar:2017eqh} slightly differs from the original LUXqed \cite{Manohar:2016nzj} in the photon PDF calculation and the error estimation.} \cite{Manohar:2016nzj,Manohar:2017eqh}, NNPDF3.1luxQED \cite{Bertone:2017bme} initializes the photon PDF with the LUX master formula, Eq.~(\ref{eq:LUX}), at a high scale, $\mu_0=100\,\GeV$, which falls within the high-$Q^2$ continuum region shown in Fig.~\ref{fig:xQ2plane}. In this approach, the photon PDF is mainly determined by the fitted quark and gluon PDFs with a scale dependence specified by DGLAP evolution. As a consequence, the PDFs evolve bidirectionally in $\mu^2$.

Representing an alternative general scheme, which we collectively designate the ``DGLAP approach'' for the purposes of this article, the MMHT2015qed study~\cite{Harland-Lang:2019pla} instead initialized the photon PDF at a low scale, $\mu_0=1\,\GeV$, with an important modification from the default LUX setup. In the LUX formalism embodied by Eq.~(\ref{eq:LUX}), to determine the photon PDF $x{\gamma}(x,\mu_0^2)$, the upper integration limit $\mu_0^2/(1-z)$ can become significantly larger than $\mu_0^2$ when $z$ is large. Therefore, MMHT breaks the integration over $Q^2$ into two parts, as
\begin{equation}
\left(\int_{\frac{x^2m_p^2}{1-z}}^{\mu_0^2}+\int_{\mu_0^2}^{\frac{\mu_0^2}{1-z}}\right)
\frac{\dd Q^2}{Q^2}\Big[\cdots\Big]\ ,
\end{equation}
where the dots above represent the expression inside the square brackets appearing in Eq.~(\ref{eq:LUX}).
In the second integration range of this quantity, the $F_L$ term is neglected because it is relatively suppressed by one additional order higher in $\alpha_S$ compared to $F_2$, {\it i.e.}, $F_L\sim \calO(\alpha_S)$. Also, given their
slow scale dependence, $F_2(Q^2)$ and $\alpha(Q^2)$ are approximately stationary, {\it i.e.},
\begin{equation}\label{eq:stationary}
\pdv{F_2}{Q^2}\sim0\ ,~\pdv{\alpha}{Q^2}\sim0\ ,
\end{equation}
an observation which permits the sum over the second integration region to be performed analytically:
\begin{equation}
\int_{\mu_0^2}^{\frac{\mu_0^2}{1-z}}\frac{\dd Q^2}{Q^2}\left[\cdots\right]
=-\alpha^2(\mu_0^2)\left(z^2+\ln(1-z)zp_{\gamma,q}-\frac{2x^2m_p^2z}{\mu_0^2}\right)F_{2}(x/z,\mu_0^2)\ ,
\end{equation}
very much like the modified conversion term in Eq. (\ref{eq:con}).

\begin{table}\centering
	\begin{tabular}{c|c|c|c|c}
		\hline
		PDF  & Reference & DGLAP evolution  & $\mu_0$ [GeV] & Momentum sum rule \\
		\hline
		LUXqed(17)  &\cite{Manohar:2016nzj,Manohar:2017eqh}   &  Yes   & 10 & Yes  \\
		NNPDF3.1luxQED &\cite{Bertone:2017bme} & Yes &100  & Yes \\
		MMHT2015qed &\cite{Harland-Lang:2019pla} & Yes  &1 & Yes \\
		CT18lux & This work  & No  & -- & No \\
		CT18qed(1.3GeV) & This work & Yes  & 3(1.3) & Yes \\
		\hline
	\end{tabular}
	\caption{The comparison of the photon PDFs of LUXqed(17), NNPDF3.1luxQED, MMHT2015qed, CT18lux and CT18qed(1.3GeV). 
	}
	\label{tab:comparePDF}
\end{table}
A concise summary of these different methods is listed in Tab.~\ref{tab:comparePDF}. In the pure LUX approach\footnote{We want to remind the reader that
	both LUXqed \cite{Manohar:2016nzj} and LUXqed17 \cite{Manohar:2017eqh} initialized the photon PDF in terms of the LUX formula at $\mu_0=10$ GeV and evolved the QCD and QED DGLAP equation to obtain the PDFs at other scales. In this sense, they are based on the high-$\mu_0$ DGLAP approach in our language.}, the photon PDF is fully determined by the structure functions which were either extracted from low-energy data, or calculated from the quark or gluon PDFs. In this way, the photon PDF is viewed essentially as an addition to the quark and gluon PDFs. The momentum sum rule
\begin{equation}
\label{eq:momSR}
\int_0^1 x\left[\Sigma(x,\mu^2)+g(x,\mu^2)+\gamma(x,\mu^2)\right]\dd x=1\ ,
\end{equation}
will be violated by a small amount\footnote{Typically, this effect is of the order of a few per mille, depending on the quark and gluon PDFs used in calculating the DIS structure functions in the master formula.}, where the singlet PDF appearing above is defined as
\begin{equation}
\Sigma(x,\mu^2)=\sum_{i}\left[q_i(x,\mu^2)+\bar{q}_i(x,\mu^2)\right]\ .
\end{equation}
This violation is remedied in the LUXqed(17), NNPDF and MMHT approaches at the starting scale and maintained at other scales as well through DGLAP evolution. Strictly speaking, however, the momentum sum rule is actually very slightly violated even after being imposed at the starting scale, because the elastic photon satisfies a different evolution \cite{Manohar:2017eqh,Harland-Lang:2019pla}. We expect this small violation to be only at the 0.01\% level, which is fully negligible.

Although the photon PDF in the LUX formalism is not obtained through DGLAP evolution, it still runs with the energy scale. It has been demonstrated in Ref.~\cite{Manohar:2017eqh} that the LUX expression gives the correct DGLAP evolution kernel up to one order higher than the input coefficient functions. In another words, the one-loop order $\alpha_s$ and $\alpha$ of Wilson coefficient functions\footnote{Pay attention that the leading order of longitudinal structure function $F_L$ starts at one loop $\calO(\alpha_s)$ and $\calO(\alpha)$.} give the two-loop order $\alpha\alpha_s$ and $\alpha^2$ $p_{\gamma i}$ splitting kernels. As feedback, the splitting diagram $(q\to q\gamma)$ will affect the quark distributions through the QED real correction to the $P_{qq}$ function. Similar diagrams occur for the gluon at higher orders. As a consequence, all parton distributions should run simultaneously, a feature that is not captured by the pure LUX formalism.  

\begin{figure}
	\centering
	\includegraphics[width=0.8\textwidth]{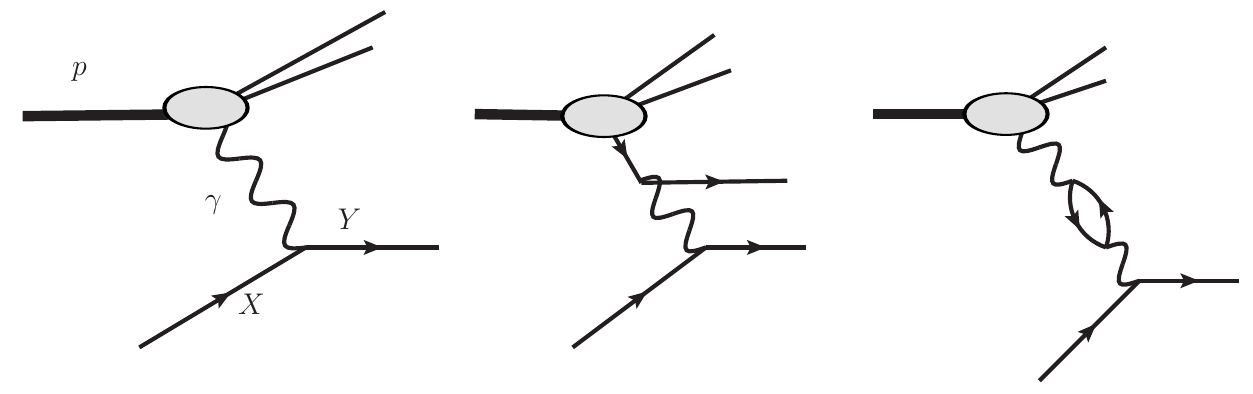}
	\caption{The representative Feynman diagrams for $(p\to\gamma)X\to Y$ scattering at the leading and next-to-leading orders.}
	\label{fig:gammaX}
\end{figure}

In a realistic scattering of the form $(p\to\gamma) X\to Y$ as shown in Fig.~\ref{fig:gammaX}, we can factorize the hadronic cross section as
\begin{equation}
\sigma=\gamma(x,\mu^2)\otimes\hat{\sigma}_{\gamma X\to Y}(x,\mu^2)+\cdots.
\end{equation}
Starting at next-to-leading order in the QED coupling, quark-initiated processes begin to participate in the hard scattering as depicted in the middle diagram of Fig.~\ref{fig:gammaX}. However, part of the inelastic photon originates within the $(q\to q\gamma)$ splitting, which contributes to the first diagram. The overlapping contribution should be subtracted in order to avoid double counting. After this subtraction as well as the cancellation of collinear singularities, the photon and quark (and gluon as well) distributions evolve according to the standard DGLAP equations.  A similar situation will also occur when a hard interaction involves the partonic $(q\to q\gamma)$ subprocess in the initial state. In this scenario, inclusion of the QED corrections to the DGLAP evolution ({\it i.e.}, to $P_{qq}$ in this case) ensures an explicit $\mu_F$ cancellation in order to get the complete NLO QED precision.
In the LUX formalism, this cancellation can only be achieved with proper $\msbar$ conversion terms, which must be determined order-by-order, even though the difference may not be numerically sizable. In the end, the QED-evolved PDF set is recommended for realizing the corresponding precision in terms of higher-order QED corrections.

In order to accommodate this subtlety, we release two PDF sets in this analysis. In one set, the photon PDF is directly calculated with the LUX master formula, {\it i.e.}, Eq.~(\ref{eq:LUX}), at all scales. In this scenario, the quark and gluon PDFs are taken from the CT18 NNLO PDF fit \cite{Hou:2019efy} without modification. We designate this PDF set ``CT18lux''.
In this sense, the momentum sum rule of CT18lux PDF will be weakly violated as the photon enters as an additional, small component. In the other set, the photon PDF is obtained through DGLAP evolution, and we call the result ``CT18qed''. In this set, we initialize the photon at a low starting scale, $\mu_0$, based on the LUX formalism, similarly to the approach taken by MMHT2015qed \cite{Harland-Lang:2019pla}.
 We emphasize that our ``CT18lux'' differs from the LUXqed(17) PDFs, which were obtained through a high-$\mu$ application
of Eq.~(\ref{eq:LUX}), and determined at other scales through DGLAP evolution.  As pointed out above, the difference between CT18lux and CT18qed mainly comes from higher-order matchings, which we explore in detail in the next two sections.

% - - - - - - - - - - - - - - - - - - - - - - - - - - - - - - - - - - - - - - - - - - - - - - - - - - -
\section{The CT18lux photon PDF}
\label{sec:CT18lux}
In this section, we first present a photon PDF determined through a more direct implementation of the LUX formalism \cite{Manohar:2016nzj,Manohar:2017eqh}
into the framework of CT18 NNLO global analysis \cite{Hou:2019efy} and obtain the CT18lux PDF. In the process, we will also illustrate the general use of the LUX master formula and the variety of
physics inputs necessary to compute the photon PDF and its uncertainty. These developments will be instructive in Sec.~\ref{sec:ct18qed}, wherein we construct our main recommended
photon PDF, CT18qed, which we base on a marriage of the LUX formalism and DGLAP evolution \cite{Dokshitzer:1977sg,Gribov:1972ri,Lipatov:1974qm,Altarelli:1977zs}.

\subsection{Numerical procedure}

The photon PDF in the CT18lux is generated with the LUX master formula, Eq. (\ref{eq:LUX}), at all scales above the CT18 starting one $\mu_0$, according to the numerical prescription described below.

\begin{itemize}
	
	\item Instead of taking PDF4LHC15 as input, we use the CT18 NNLO PDFs for the (anti)quark and gluon PDFs needed to calculate the structure functions $F_{2,L}$ appearing in the LUX master formula, Eq.~(\ref{eq:LUX}), in the high-$Q^2$ and high-$W^2$ region. The quark and gluon PDFs remain unchanged relevant to the default ones fitted in CT18 NNLO.
	
	\item Similar to the CT18 NNLO PDFs, the CT18lux contains one central set and 58 Hessian error sets, with all generated by applying the LUX master formula to the 1+58 CT18 NNLO PDFs.
	The error sets quantify a part of the photon PDF uncertainty induced by the quark and gluon partons through perturbative DIS structure functions, $F_{2,L}$, in Eq. (\ref{eq:LUX}).
	
	\item Because of this procedure, the proton momentum sum rule is slightly violated. The amount of violation is given by the additional fractional momentum of the proton carried by the photon. In
	Fig.~\ref{fig:mom}, we show this momentum fraction as a function of the energy scale, $\mu$. It is about 0.22\% at $\mu=1.3$ GeV and grows to about 0.65\% at $\mu=1$ TeV.

	\item In addition to the 58 photon error PDFs encapsulating the quark and gluon PDF uncertainty noted above, a number of other dynamical effects at smaller $Q$ contribute to the photon PDF calculation. We assess uncertainties associated with these in Sec.~\ref{sec:nonDIS} below, and ultimately include them into the final CT18lux photon PDF uncertainty. 
	
\end{itemize}

\subsection{The CT18lux photon PDF and its uncertainty}
\label{sec:nonDIS}
Based on the formalism developed in Sec.~\ref{sec:LUX2DGLAP}, it is clear that $x\gamma(x,\mu^2)$
in the master formula of Eq.~(\ref{eq:LUX}) involves a series of contributions away from the kinematical region dominated by inelastic processes as shown in Fig.~\ref{fig:xQ2plane} ---
the ``DIS'' or ``high-$Q^2$ continuum region'' in which the most appropriate description is in
terms of PDFs. These contributions enter via the direct evaluation of $\gamma(x,\mu^2)$
from the phenomenological $F_{2,L}$ structure functions, and arise from several distinct dynamical
sources and scattering processes in the electromagnetic interaction of the photon with the
proton. In turn, these contributions generally have a number of potential variations
in terms of their calculation and implementation in Eq.~(\ref{eq:LUX}) which
represent a significant source of uncertainty on $x\gamma(x,\mu^2)$.
We note that, while many of the nonperturbative error sources explored here may have been considered or analyzed
in some form in previous work(s), here we systematically revisit them, having considered an expanded set
of possible variations and, where relevant, updating inputs into Eq.~(\ref{eq:LUX}) with
more recent parametrizations of structure functions and form factors that have emerged since,
{\it e.g.}, Ref.~\cite{Manohar:2016nzj}.

\subsubsection{Construction of the photon PDF uncertainty}
%}
The most straightforward way to determine the full photon PDF uncertainty is to build it sequentially, by
first calculating a central PDF through Eq.~(\ref{eq:LUX}) as detailed above, and then computing eigensets
associated with variations among the PDF parameters used to compute $F_{2,L}$. This encapsulates
the photon PDF uncertainty from the ``continuum QCD'' or higher-$Q^2$ region.
A realistic assessment of the uncertainties of the extracted photon
PDF also depends, however, on the treatment of low-energy (or, ``low-$Q^2$'') contributions
to $F_{2,L}$ (through and including $Q\!\sim$ few GeV), especially at high $x$; we must therefore
account for variations of these low-energy effects within the larger Hessian uncertainty on $\gamma(x,\mu^2)$ as well.
In particular, we specify the uncertainty on $\gamma(x,\mu^2)$ through a number of additional error sets,
leading to a collection of
\begin{equation}
N_\mathrm{sets} = 1\ (\mathrm{central}) + 2 N_\mathrm{PDF} + 2n_{\textrm{low-}Q^2}\ ;
\end{equation}
PDF sets in a fit with $N_\mathrm{PDF}$ shape parameters for the parton distributions and $n_{\textrm{low-}Q^2}$ separate
low-$Q^2$ inputs to $F_{2,L}$. For the latter, we symmetrize the uncertainty for each of the low-$Q^2$ inputs, and rescale the error band to 90\% CL,
with the convention adopted in the CT PDF family. 
In such a way, the combined PDF uncertainty can be obtained through a Hessian approach
\bea\label{eq:unc}
\delta X&=\sqrt{\sum_{i=1}^{N_{\rm PDF}+n_{\textrm{low-}Q^2}}\left(\frac{X_{i}^{+}-X_{i}^-}{2}\right)^2} ,\\
\delta^+ X&=\sqrt{\sum_{i=1}^{N_{\rm PDF}+n_{\textrm{low-}Q^2}}\left[\max(X_{i}^{+}-X_{0},X_i^{-}-X_0,0)\right]^2}\ ,\\
\delta^- X&=\sqrt{\sum_{i=1}^{N_{\rm PDF}+n_{\textrm{low-}Q^2}}\left[\max(X_{0}-X_{i}^{+},X_0-X_{i}^{-},0)\right]^2}.
\eea
An overall factor 1.645 is applied to get the 68\% CL uncertainty, for the comparison with other PDFs throughout this work.

%We remind the reader that the $2N_{\rm PDF}$ Hessian eigenvector sets are defined at the 90\% confidence level (CL) in the CT18 family, while the low-$Q^2$ variations are understood as being at the $1\sigma$ level, corresponding to 68\% CL. Therefore, the $1\sigma$ PDF uncertainties of an observable $X$ are combined as
%\bea\label{eq:unc}
%\delta X&=\sqrt{\sum_{i=1}^{N_{\rm PDF}}\left(\frac{X_{i}^{+}-X_{i}^-}{2\times 1.645}\right)^2+\sum_{i=2N_{\rm PDF}+1}^{2N_\mathrm{PDF} +n_{\textrm{low-}Q^2}}\left(X_i-X_0\right)^2}\ ,\\
%\delta^+ X&=\sqrt{\frac{1}{1.645^2}\sum_{i=1}^{N_{\rm PDF}}\left[\max(X_{i}^{+}-X_{0},X_i^{-}-X_0,0)\right]^2+\sum_{i=2N_{\rm PDF}+1}^{2N_\mathrm{PDF} +n_{\textrm{low-}Q^2}}\left(X_i-X_0\right)^2}\ ,\\
%\delta^- X&=\sqrt{\frac{1}{1.645^2}\sum_{i=1}^{N_{\rm PDF}}\left[\max(X_{0}-X_{i}^{+},X_0-X_{i}^{-},0)\right]^2+\sum_{i=2N_{\rm PDF}+1}^{2N_\mathrm{PDF} +n_{\textrm{low-}Q^2}}\left(X_i-X_0\right)^2},
%\eea
%where the 1.645 is the converting factor from 90\% CL to 68\% CL. We adopt these conventions throughout this work.

\subsubsection{Contributions to the photon PDF and its uncertainty}
\label{sec:uncSource}
%}
Here, we describe a number of unique contributions or effects that enter
the photon PDF master formula of Eq.~(\ref{eq:LUX}) and are responsible for
the ultimate uncertainty of the photon PDF.\\

\begin{figure}
	\centering
	\includegraphics[width=0.65\textwidth]{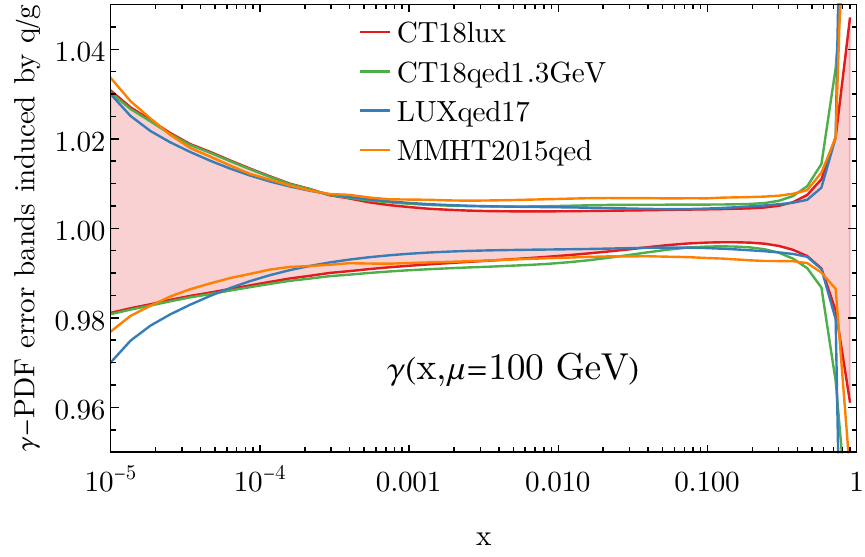}
	\caption{The photon PDF uncertainty bands purely induced by the DGLAP-evolved quark and gluon PDFs.}
	\label{fig:phUncqg}
\end{figure}

{\bf The quark and gluon PDF uncertainty}. The structure functions, $F_{2,L}$, in the high-$Q^2$ continuum region are determined through perturbative calculations. The uncertainty of high-$Q^2$ $F_{2,L}$ due to the quark and gluon uncertainty will propagate to the inelastic photon component. We show the photon PDF uncertainty purely induced by the high-$Q^2$ quark and gluon PDFs in Fig.~\ref{fig:phUncqg}. We have presented the same error bands from LUXqed17 and MMHT2015qed as well for comparison.\footnote{We do not include the NNPDF3.1luxQED here, as NNPDF does not provide separated sets purely from the quark and gluon PDF variations.} We see that the overall size of the error band agrees very well among these three groups, while MMHT2015qed gives a slightly larger band in the intermediate $x$ region.
%which is mainly due to its different method. 
Instead of directly calculating the photon PDF with the LUX master formula, MMHT2015qed evolves the photon together with the quark and gluon PDFs with the DGLAP equations, which includes the interplay between the photon and other partons, as well. This treatment is similar to that used in CT18qed (1.3GeV), which will be discussed in the next section.\\

{\bf Elastic form factors}.
Another necessary consideration is the interplay of parametric uncertainties in phenomenological
fits or models of Sach's electromagnetic form factors of the proton, $G_{E,M}$, and the ultimate photon PDF uncertainty.
Historically, there have been diverse attempts to simulate these form factors in QCD-inspired models \cite{Theberge:1981mq,Chung:1991st,Lu:1997sd,Hobbs:2014lea,Hobbs:2016xfz,Zhang:2019iyx}, compute them using lattice QCD or other theoretical methods
(see, {\it e.g.}, \cite{Perdrisat:2006hj}), or to extract them phenomenologically based on empirical data, analogously to the quark and gluon PDFs themselves.
Moreover, a number of simplified prescriptions have also been used for the $Q^2$ dependence of the electromagnetic form factors; most commonly, this
has been a dipole expression based on the observation
that the quark model and high-energy QCD prescribe an asymptotic 
$\sim\!1/Q^4$ dependence at large enough $Q^2$ values. At lower $Q^2$,
the observed form factor is known to deviate significantly from this
behavior, and, for the sake of comparisons, we take a dipole ansatz 
baseline of
\begin{equation}
    G^\mathrm{dip}_E(Q^2) = \left( {1 \over \mu_p} \right) G^\mathrm{dip}_M(Q^2) = F^\mathrm{dip}(Q^2) = {1 \over \left( 1 + Q^2 / \Lambda^2_\mathrm{dip} \right)^2}\ ,
\label{eq:dipole}
\end{equation}
where the value of the dipole mass parameter is $\Lambda^2_\mathrm{dip} = 0.71$ GeV$^2$ and $\mu_p=2.793$~\cite{Mohr:2015ccw} is the magnetic moment of the proton.
In this work, we have primarily used the empirical data-driven
approach, taking a series of phenomenological fits of the Sach's form factors in order to more fully determine the dependence of our photon PDF on choices
for $G_{E,M}(Q^2)$.

\begin{figure}[ht]\centering
	\includegraphics[width=0.49\textwidth]{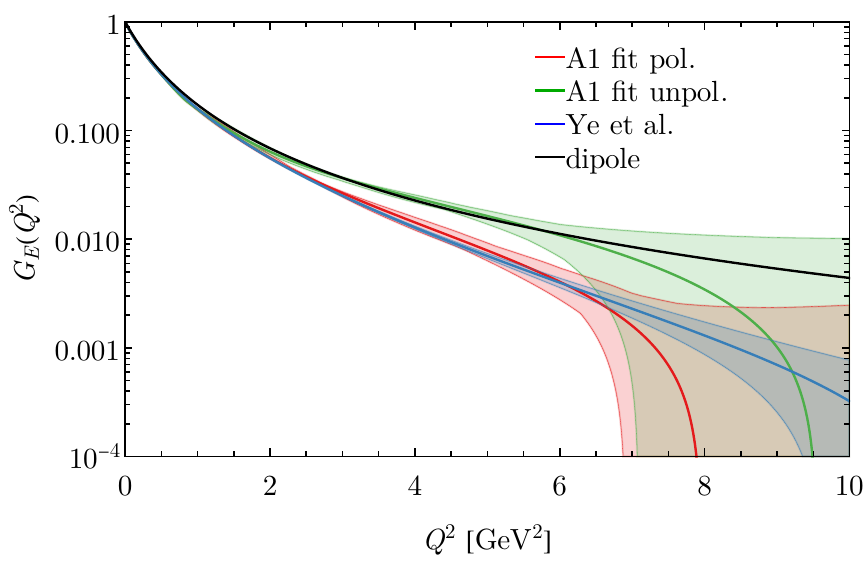}
	\includegraphics[width=0.49\textwidth]{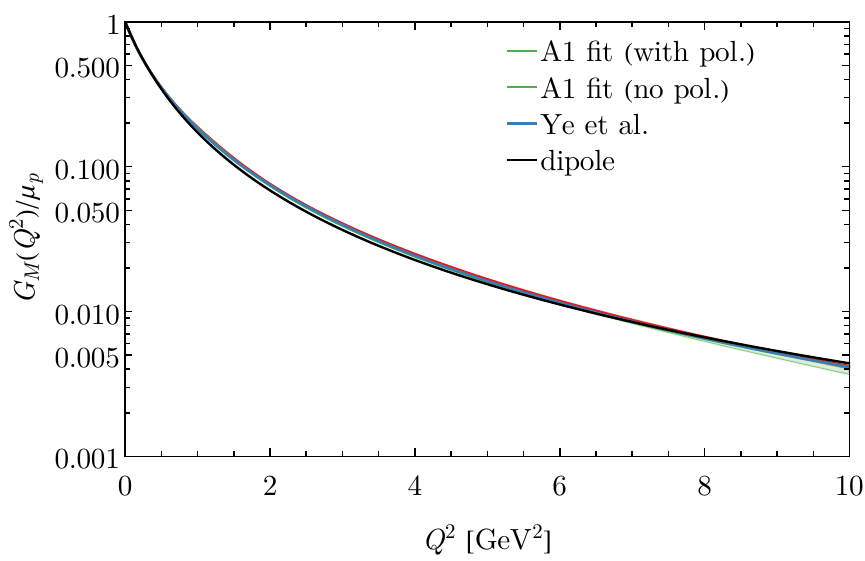}	
	\includegraphics[width=0.97\textwidth]{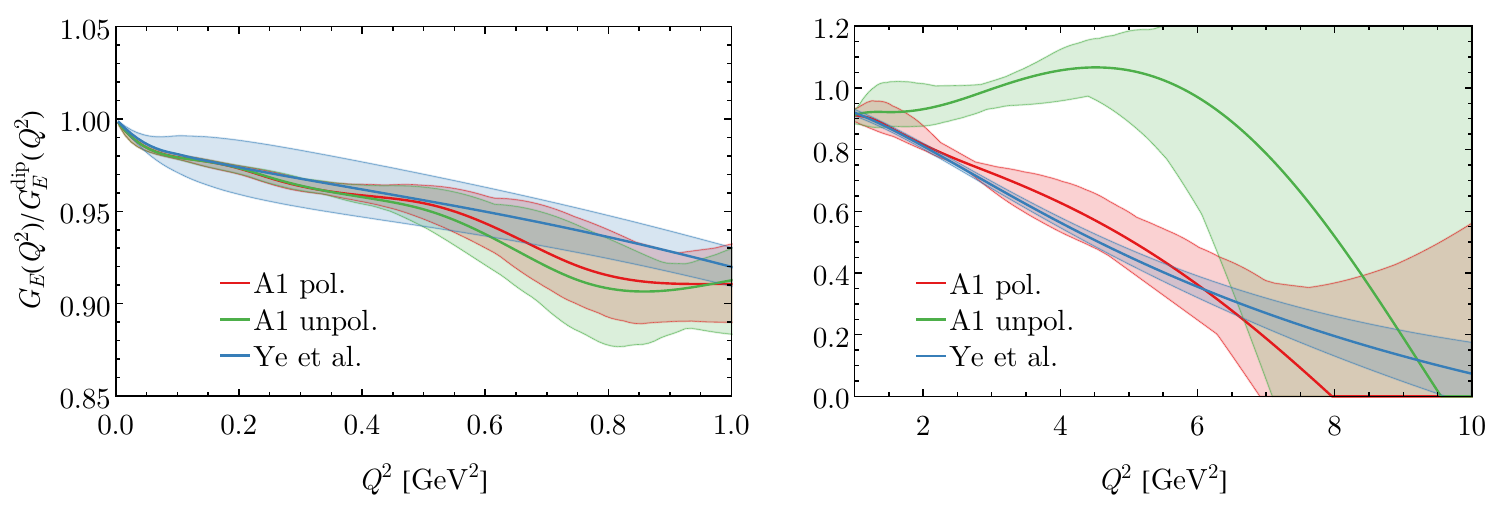}
	\includegraphics[width=0.97\textwidth]{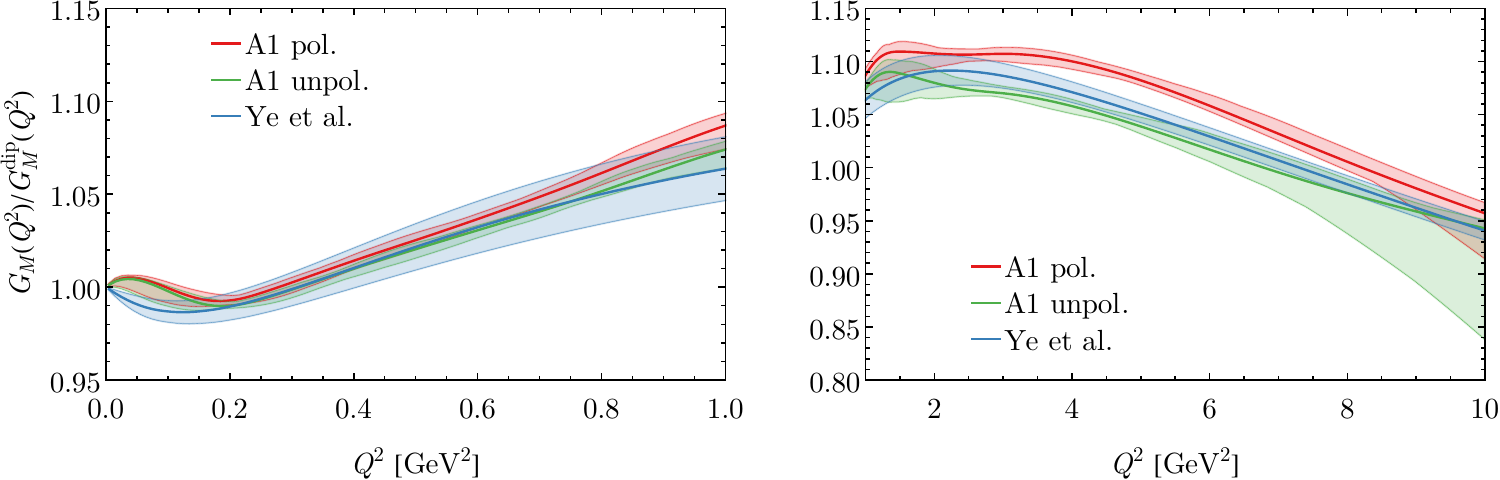}
	\caption{The elastic form factors normalized to the standard dipole form, obtained by the fits of the A1 Collaboration \cite{Bernauer:2013tpr} and Ye \emph{et al.} \cite{Ye:2017gyb}. We stress that the
	dipole calculation is provided {\it only for the sake of reference} and is not actively used in
	subsequent calculations of the photon PDF or related phenomenology, with the exception of the
	results shown in Fig.~\ref{fig:elphoton}.
	}
	\label{fig:GEGM}
\end{figure}

\begin{figure}\centering
\includegraphics[width=0.49\textwidth]{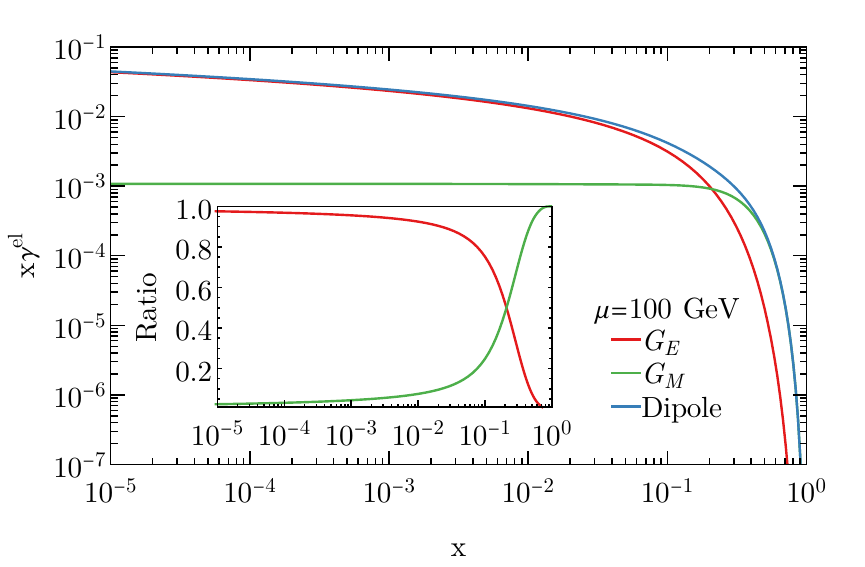}
\includegraphics[width=0.49\textwidth]{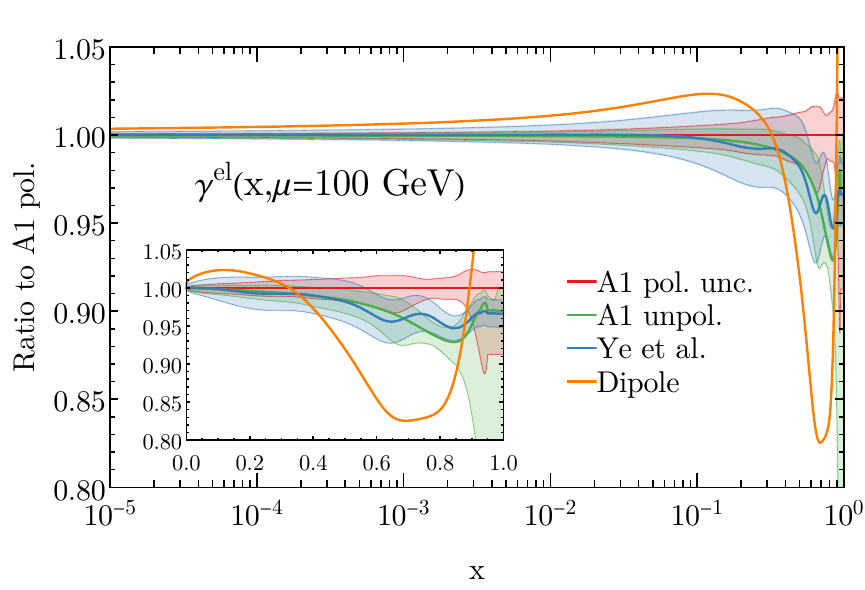}
\caption{The electric and magnetic components of dipole elastic photon at $\mu=100$ GeV (left) and the comparison with other Sach's forms, normalized to the one obtained with the A1 world fit including the polarized data (right). As with Fig.~\ref{fig:GEGM}, we emphasize that the dipole calculation shown here is {\it for comparison only}.
}
\label{fig:elphoton}
\end{figure}

LUXqed(17) \cite{Manohar:2016nzj,Manohar:2017eqh} has explored variations based on the A1 fit \cite{Bernauer:2013tpr} with all the world data involving electron scattering up to the year 2013. We show the A1 fits to the world polarized data and the unpolarized data in the upper pannel of Fig.~\ref{fig:GEGM}. We notice that the polarized fit include two-photon-exchange (TPE) corrections, while the unpolarized one does not. Corresponding ratios to the commonly-used single-parameter dipole prediction are shown in Fig.~\ref{fig:GEGM}, middle and lower panels; these ratios highlight the nontrivial $Q^2$ dependence of the realistic form factors, which can markedly differ from that of the simpler dipole ansatz.
In addition, a more recent global fit of the unpolarized world cross-section data from Ye \emph{et al.}~in Ref.~\cite{Ye:2017gyb} similarly incorporated two-photon exchange corrections. We see that $G_E$ obtained in this fit agrees better with the A1 fit of polarized data, whereas $G_M$ agrees with the A1 unpolarized one. 
The elastic photon PDF as obtained with these various prescriptions is shown in the panels of Fig.~\ref{fig:elphoton}.
In particular, the corresponding electric and magnetic components of the elastic photon PDF are shown at $\mu\!=\!100$ GeV in the left panel of Fig.~\ref{fig:elphoton}. For reference, the full calculation of $x\gamma^\mathrm{el}(x)$ using the dipole parametrization of Eq.~(\ref{eq:dipole}) is also shown. We see that the electric contribution dominates the elastic photon in the low-$x$ region, while the magnetic one takes over at large $x$.
This can be understood in terms of the asymptotic behavior of the elastic photon in Eq.~(\ref{eq:elastic}). In the small-$x$ region, the $Q^2$ integration over the electric form factor behaves as $\sim\!\!\ln(1/x)$ while the analogous integration over the $G_M$-dependent term approaches a constant. On the other hand, at large $x$, $G_E(Q^2)$ is suppressed either by $(1-x)$ or $m_p^2/Q^2$ compared with the $G_M(Q^2)$ contribution.
Meanwhile, in the right panel of Fig.~\ref{fig:elphoton} we plot the ratio at $\mu\! =\! 100$ GeV of the
elastic photon PDF computed with each form factor parametrization explored in this analysis with respect
to that calculated using the A1 polarized fit.
We see that the fit of Ye \emph{et al.}~\cite{Ye:2017gyb} gives a larger uncertainty band than that of A1, due to its broader uncertainty in the low-$Q^2$ region. At the same time, the central elastic photon associated with the Ye \emph{et al.}~form factor fit agrees better with the A1 unpolarized calculation at large $x$.
The dipole form factor result extends beyond the error bands for most $x$ regions, corresponding to an overestimate of the elastic photon at small $x$ and extremely large $x$, and an underestimate at moderately large $x$.
Following LUXqed(17) \cite{Manohar:2016nzj,Manohar:2017eqh}, we retain the A1 polarized fit as our default choice, while the corresponding uncertainty is estimated by the larger deviation obtained either from the error band of the A1 polarized fit or from the central of A1 unpolarized fit. We also release an extra PDF set based on the Ye {\it et al.}'s form factors, but it is not included in the PDF uncertainty estimation.

\begin{figure}\centering
	\includegraphics[width=0.65\textwidth]{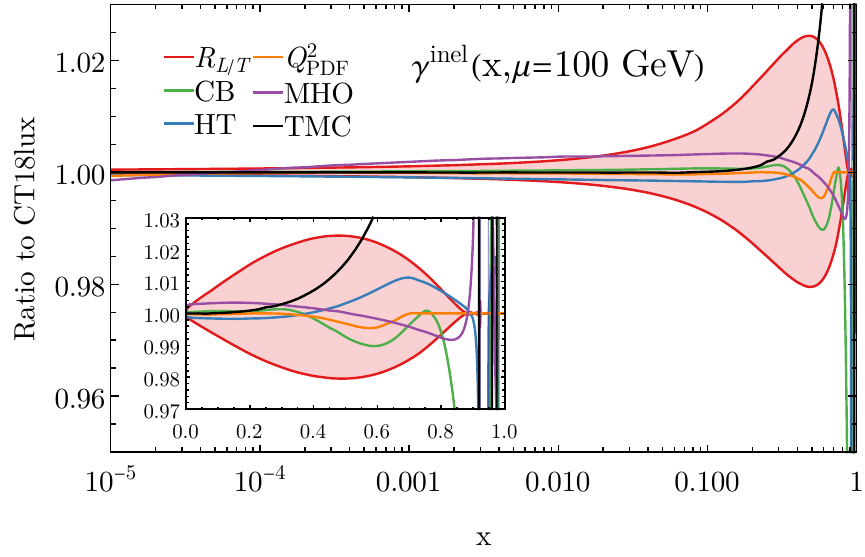}
	\caption{The uncertainty of inelastic photon induced by various inputs.}
	\label{fig:inelastic}
\end{figure}

{\bf Higher-twist effects}.
The structure functions are formally determined as Fourier transforms of
hadronic matrix elements of quark-level electroweak current operators,
\begin{equation}
W_{\mu\nu}
= {1 \over 4 \pi} \int d^4 z\,e^{i q \cdot z}
\langle p \left|
\left[ J_\mu^\dagger(z), J_\nu(0) \right]
\right| p \rangle
= \mathcal{L}^i_{\mu\nu}(p,q)\, F_i (x,Q^2)\,
\label{eq:Wmunu_def}
\end{equation}
in which the index $i$ on the RHS can be expanded over a complete tensor basis and
$\mathcal{L}^i_{\mu\nu}$ represents the set of unique Lorentz structures constructed from the momenta 
$p$ and $q$ in that basis.
At lower energies, various soft quark-parton correlations within the target can be
included in the matrix element of Eq.~(\ref{eq:Wmunu_def}),
which have the effect of increasing the {\it twist} (difference between operator
spin and dimension) of the matrix element. For unpolarized matrix elements,
the next nonzero term in the twist expansion beyond leading twist appears at twist-4, which
is power-suppressed by $\sim\!1/Q^2$. Computing or fitting these separately may be
an involved task, but one quick alternative would be to use an existing determination
such as that from the CJ15 NLO fit~\cite{Accardi:2016qay}, which fitted an $x$-dependent higher-twist (HT) correction,
\begin{equation}
F^{\HT}_2(x,Q^2) = F^{\rm LT}_2(x,Q^2) \left(1 + {C_\mathrm{HT}(x) \over Q^2} \right)\ ,
\end{equation}
where $C_\mathrm{HT}(x) = h_0 x^{h_1} (1 + h_2 x)$. While there is an obvious model
uncertainty associated with this choice of parametric form, there is already a
parametric uncertainty that can be used to generate extreme scenarios for the
HT correction to $F_{2}$. The absolute higher-twist corrections to $F_2(x,Q)$, based on the CJ15 NLO fit \cite{Accardi:2016qay}, are shown as solid lines in left plot of Fig.~\ref{fig:HTTMC}. We see clearly that, as expected, the size of the HT correction is maximized at lower values of $Q^2$, particularly for
large $x$, but relatively suppressed as the $Q^2$ scale increases. The sharp dip around $x\sim0.48$ is due to a localized sign change in $C_{\rm HT}(x)$ in the CJ15
NLO fit.  Similarly, we expect a higher-twist contribution to $F_L$, which is parameterized with one parameter as
\begin{equation}
F_L^{\HT}(x,Q^2)=F_L^{\rm LT}\left(1+\frac{A_{\HT}}{Q^2}\right).
\end{equation}
Recent studies on the DIS data suggest $A_{\HT}=5.5\pm0.6$~\cite{Abt:2016vjh}.
The MMHT group obtained a smaller value as $A_{\HT}=4.3~\GeV^2$ \cite{Harland-Lang:2016yfn}. Here we take the larger value from Ref. \cite{Cooper-Sarkar:2016foi}, with the impact shown as the dashed lines in Fig.~\ref{fig:HTTMC}, left plot. The corresponding corrections to the inelastic photon PDF arising from these HT corrections are represented by the blue curves in Fig.~\ref{fig:inelastic}. The overall size of the uncertainty is generally small, a fact which can be understood as originating in the $\sim\!1/Q^2$ suppression of HT corrections, which here only enter the
calculation in the high-$Q^2$ continuum region. Nevertheless, the HT uncertainty peaks at $\gtrsim\!1\%$ for $x\!\sim\!0.65$, and remains a relevant consideration
for precision in the very high-$x$ regime.\\

{\bf Target-mass corrections}.
At leading twist, additional operators can be inserted into the matrix element associated with
the hadronic tensor, $W_{\mu\nu}$ of Eq.~(\ref{eq:Wmunu_def}), which, in turn, affect the low-$Q$
behavior of the unfolded structure functions. These additional operators are insertions of
covariant derivatives:

\begin{figure}
	\includegraphics[width=0.49\textwidth]{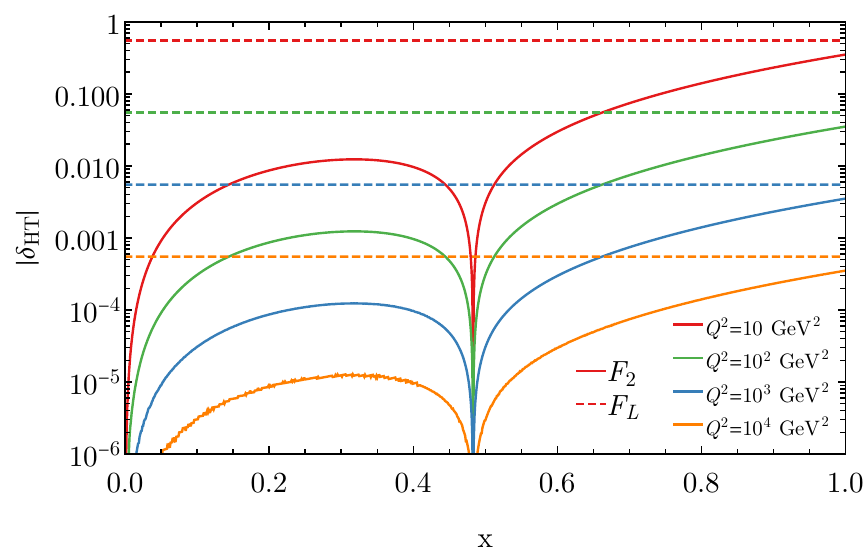}
	\includegraphics[width=0.49\textwidth]{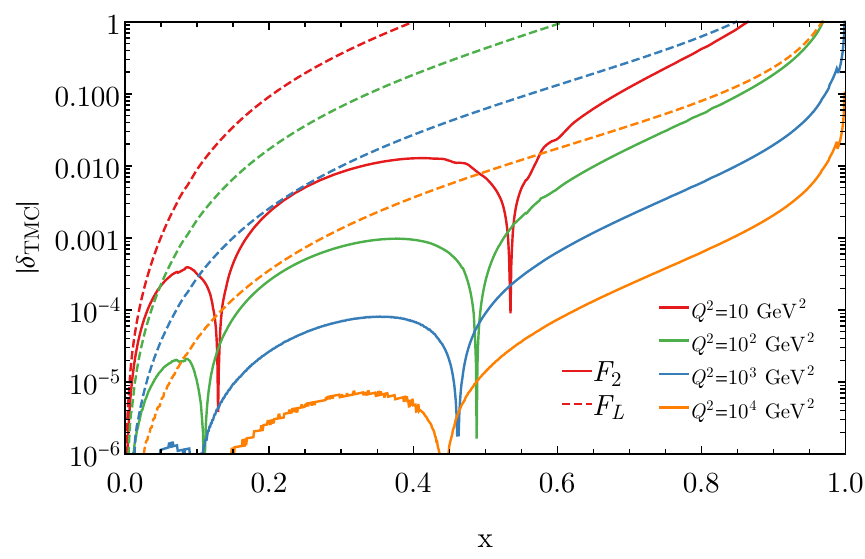}
	\caption{The higher-twist corrections (left) and the target mass corrections (right) to the structure functions $F_{2,L}$.}
	\label{fig:HTTMC}
\end{figure}

\begin{equation}
\bar{\psi}\gamma^{\mu} D^{\mu_1} \cdots D^{\mu_n} \psi\ ,
\end{equation}
which increase both the {\it spin} and {\it dimension} of the matrix element,
thus leaving the {\it twist} unchanged.  These corrections are referred to
as ``kinematical higher twist'' corrections, and can be unfolded in the operator
product expansion (OPE) of Georgi-Politzer \cite{Georgi:1976ve}, as done here. The net effect is target-mass dependent
corrections (TMCs) that go as $\sim m_p^2/Q^2$ and dominate mass-corrected
structure functions at high $x$.  There, in principle, can be substantial
prescription dependence in the implementation of these corrections, leading
to another potential uncertainty. A number of these structure-function level prescriptions have been computed
and reviewed in Refs.~\cite{Schienbein:2007gr,Brady:2011uy}.
For example, there can be substantial variation in $F_L$, depending upon the
specific prescription. The target mass corrections to $F_{2,L}$, defined as
\begin{equation}
\delta_{\rm TMC}=\frac{F_{2,L}^{\rm TMC}}{F_{2,L}}-1,
\end{equation}
are shown in Fig.~\ref{fig:HTTMC}, right plot, calculated according to the standard OPE formalism implemented in APFEL \cite{Bertone:2013vaa}. We remind the reader that the TMCs can be either positive or negative, but we only show the relative absolute deviations here. Similar to the HT case, the several sharp dips are due to sign changes on
the log-scale over which we plot. Owing to the kinematical suppression of target-mass effects, which result in a na\"ive rescaling of
computed structure functions in the Georgi-Politzer formalism \cite{Georgi:1976ve},
\begin{equation}
x \to \xi(x,Q^2) = {2 x \over 1 + \sqrt{1 + x^2m_p^2/Q^2}}\ ,
\end{equation}
TMCs introduce a kinematical dependence on the ratio $x^2m_p^2/Q^2$. In particular, at larger $x$ and smaller $Q^2$, we expect the most significant impact, especially for quantities that are steeply-falling functions at $x\to 1$; this is true, for instance, of the longitudinal structure functions, $F_L$, which can receive a sizable correction induced by rescaling, $x \to \xi$. The TMCs to the inelastic photon PDF are shown as black lines in Fig.~\ref{fig:inelastic}. We see the impact can dominate the uncertainty at very high $x$, {\it e.g.}, for $x>0.6$.\\

\begin{figure}
	\includegraphics[width=0.49\textwidth]{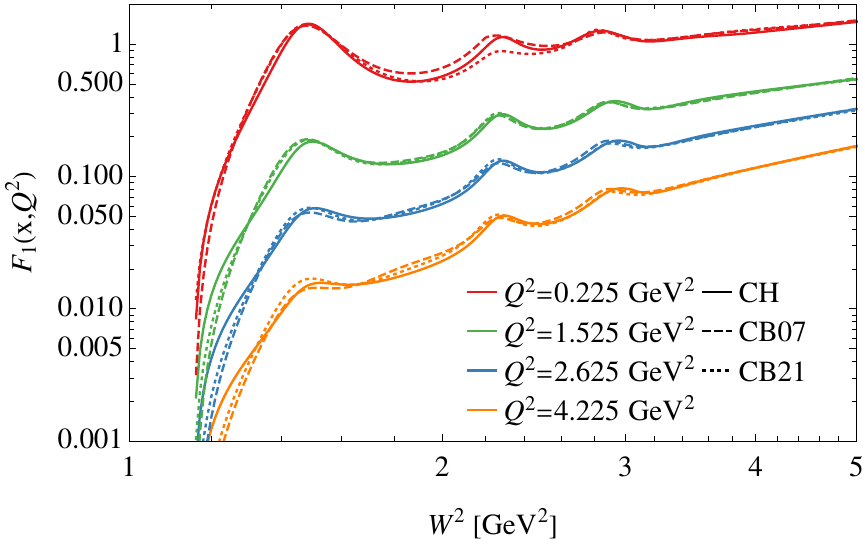}
	\includegraphics[width=0.49\textwidth]{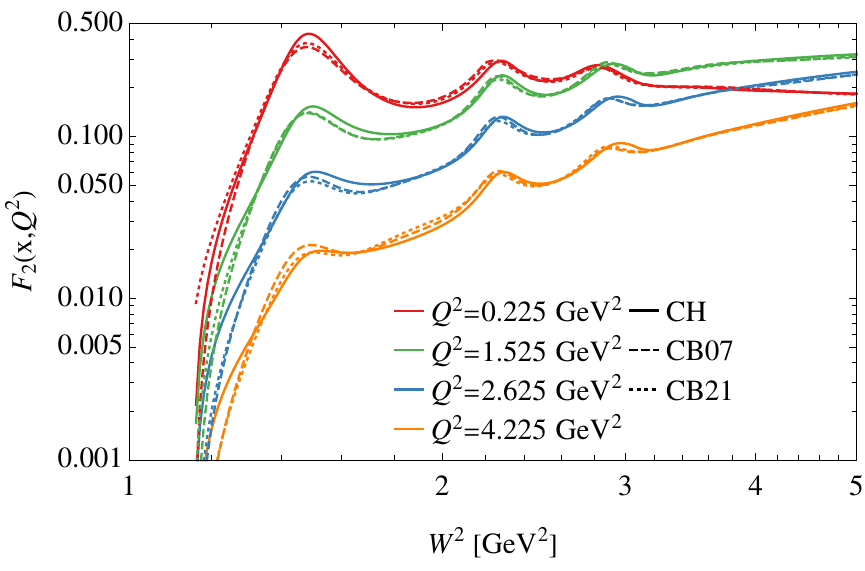}
	\caption{The structure functions of $F_{1,2}$ in the resonance region. 
	}
	\label{fig:SFs}
\end{figure}

{\bf Scattering from nucleon resonances}.
At lower energies, $Q^2<Q_{\rm PDF}^2$ or $W^2\! <W_{\rm high}^2$, scattering from nucleon resonances ({\it e.g.}, the $\Delta$, the Roper)
dominates the $\gamma p$ cross section and therefore contribute to $F_{2}$ and $F_L$ (or $F_1$ equivalently).\footnote{The relation among these three structure functions follows
	\begin{equation}
	F_L=F_2(1+4x^2m_p^2/Q^2)-2xF_1.
	\end{equation}
}  These are typically
described with a combination of Breit-Wigner parametric forms on a smooth background
(for instance, informed by Regge Theory).  Even more so than the elastic contributions, uncertainties
from resonances are likely to be truly data-driven, with little variation from underlying theory
or models. The structure functions $F_{1,2}$ in the resonance regions at several representative scales are shown in Fig.~\ref{fig:SFs}. Here we show the invariant squared mass of the final-state hadronic system, $W^2$, which is directly used in the experimental measurements, such as CLAS data \cite{Osipenko:2003bu} or the Christy-Bosted phenomenological fit \cite{Christy:2007ve}. The conversion from $W^2$ to the Bjorken-$x$ can be easily performed as
\begin{equation}
W^2=m_p^2+Q^2\frac{1-x}{x}.
\end{equation}
The ``CH" curves in Fig.~\ref{fig:SFs} bridge the CLAS resonance to the HERMES continuum with a smooth transition:
\bea
F_{a}=\begin{cases}
	F_{a}^\mathrm{res} & W^2<W^2_\mathrm{low},\\
	(1-\rho)F_{a}^\mathrm{res}+\rho F_{a}^\mathrm{cont} \, , & W^2_\mathrm{low}<W^2<W^2_\mathrm{high},\\
	F_{a}^\mathrm{cont} & W^2>W^2_\mathrm{high},
\end{cases}
\eea    
where $a=1,2$ (or $L$) and $\rho$ is 
\begin{equation}
\rho=2\omega^2-\omega^4, ~ {\rm with} ~ \omega=\frac{W^2-W^2_\mathrm{low}}{W^2_\mathrm{high}-W^2_\mathrm{low}}.
\end{equation}
The transition points are $W^2_\mathrm{low}=3~\GeV^2$ and $W^2_\mathrm{high}=4~\GeV^2$, respectively. 
The original Christy-Bosted fit was released in 2007 \cite{Christy:2007ve}, which is denoted as ``CB07" in Fig.~\ref{fig:SFs}. A recent update includes more data from proton and mostly nuclei cross sections \cite{Christy:2021abc}, denoted as ``CB21" in the figure. We show the corresponding inelastic photon of CB21 as the green curves in Fig.~\ref{fig:inelastic}. We see the variation from our default choice, CLAS, is very mild, and only 1\% around $x\sim 0.5$. It only deviates significantly when $x>0.85$, where the structure functions become unreliable, and so does inelastic photon.\\

{\bf The $R_{L/T}$}. In the resonance and low-$Q^2$ continuum region, the longitudinal structure function $F_L$ is modeled, similarly to LUXqed(17) \cite{Manohar:2016nzj,Manohar:2017eqh}, as
\begin{equation}
F_L(x,Q^2)=F_2(x,Q^2)\left(1+\frac{4x^2m_p^2}{Q^2}\right)\frac{R_{L/T}(x,Q^2)}{1+R_{L/T}(x,Q^2)}\ ,
\end{equation}
where $R_{L/T}=\sigma_L/\sigma_T$. 
The specific $R_{L/T}$ is provided by LUX group, who adopted the HERMES convention \cite{Airapetian:2011nu} and the $R_{1998}$ fit provided by E143 Collaboration \cite{Abe:1998ym}. The uncertainty is assigned conservatively to be $\pm50\%$, with the corresponding inelastic photon shown as the red bands in Fig.~\ref{fig:inelastic}. Around $x\sim0.45$, the uncertainty induced by $R_{L/T}$ can be as large as 2\%, which dominates around this region.\\

{\bf Matching scale, $Q_{\rm PDF}^2$, or continuum choice}. The default matching scale from the low-$Q^2$ HERMES to high-$Q^2$ pQCD continuum region is set as $Q_{\rm PDF}^2=9~\GeV^2$. LUX varied this scale down to $Q_{\rm PDF}^2=5~\GeV^2$ in order to estimate the corresponding uncertainty due to this parametric choice. Different from the LUXqed(17) \cite{Manohar:2016nzj,Manohar:2017eqh} and NNPDF3.1luxQED \cite{Bertone:2017bme} calculations, the MMHT2015qed photon PDF was initialized at $\mu_0=1~\GeV$ is therefore fully determined by the low-energy SFs, taken from the fits of HERMES \cite{Airapetian:2011nu} and CLAS \cite{Osipenko:2003bu} (or CB \cite{Christy:2007ve}). The corresponding uncertainty there is quantified by taking the uncertainty bands of the GDP-11 fit provided by the HERMES Collaboration \cite{Airapetian:2011nu}. 
At higher scales, $\mu$, $\gamma(x,\mu)$ in MMHT2015qed is entirely determined by DGLAP evolution.
For CT18lux, we choose to follow the LUX approach to quantify the matching-scale uncertainty, which is shown as the orange curve appearing in Fig.~\ref{fig:inelastic}. We find the variation from $Q_{\PDF}^2=9~\GeV^2$ down to $5~\GeV^2$ produces a very mild, subpercent effect which is maximally peaked at high $x\!\sim\!0.6$.\\

{\bf Missing higher-order (MHO) correction}. As explained in App.~\ref{app:sep}, the missing higher-order uncertainty is quantified by varying the separation scale $M^2(z)$ between the default choice $\mu^2/(1-z)$ and the alternative one $\mu^2$.
The variation of the inelastic photon PDF from the CT18lux central set is shown as the purple line in Fig.~\ref{fig:inelastic}. We find that, with the expected cancellation of the divergent $\log[1/(1-z)]$ contribution captured by the modified $\msbar$-conversion term in Eq.~(\ref{eq:con}), we obtain only very mild variations in the resulting photon PDF at nominal $x$ values. The only exception occurs at $x\!>\!0.85$, where the absolute photon PDF becomes extremely small, as indicated in Fig.~\ref{fig:CT18lux}, such that the calculation becomes unreliable.\\

\begin{figure}
	\centering
	\includegraphics[width=0.65\textwidth]{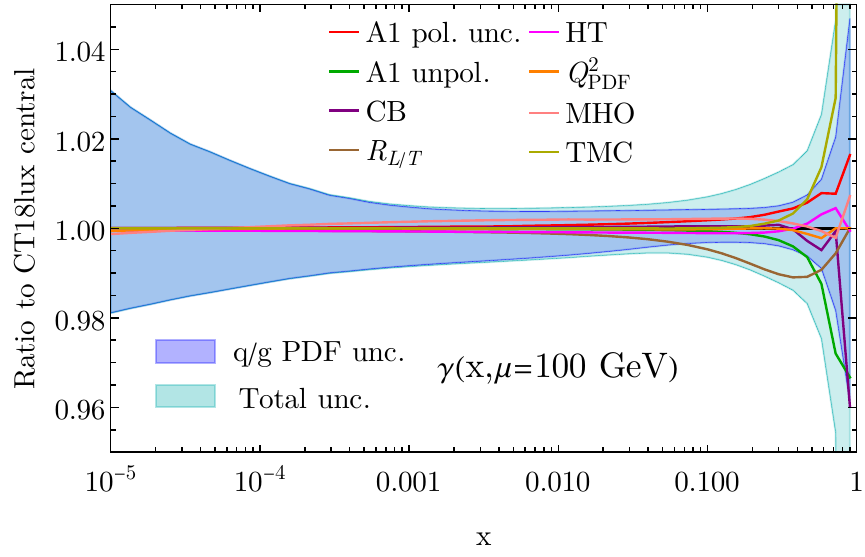}
	\caption{The individual contributions to the full $\gamma$-PDF uncertainty in our CT18lux calculation. The various
		sources of uncertainty are discussed in the accompanying text of Sect.~\ref{sec:uncSource}, and are added on top
		of the uncertainties associated with variations of the quark and gluon PDFs in the LUX master expression in
		Eq.~(\ref{eq:LUX}).}
	\label{fig:CT18luxUnc}
\end{figure}

To summarize this discussion, in Fig.~\ref{fig:CT18luxUnc} we present the complete set of contributions to the CT18lux photon-PDF
uncertainty, contributed from each of the various sources discussed above. Here, we have summed the elastic and inelastic components and show the total one.

% - - - - - - - - - - - - - - - - - - - - - - - - - - - - - - - - - - - - - - - - - - - - - - - - - - -
% - - - - - - - - - - - - - - - - - - - - - - - - - - - - - - - - - - - - - - - - - - - - - - - - - - -
\section{The Photon PDF from a DGLAP-driven approach: CT18qed}
\label{sec:ct18qed}
In this section, we present the photon PDF based on a DGLAP evolution approach, which consistently includes QED effects.
We call the resulting photon PDF sets ``CT18qed.''  After first reviewing our numerical procedure for evaluating PDFs in
CT18qed (Sec.~\ref{sec:ct18qedNum}) and discussing the CT18qed PDFs themselves and their
uncertainties (Sec.~\ref{sec:ct18qedPDFs}), we turn to several issues related to the determination of PDFs in the DGLAP-driven approach, including
the enforcement of momentum conservation (Sec.~\ref{sec:ct18qedMOM}), subtleties comparing the CT18lux and CT18qed calculations at different
perturbative orders in the combined QCD and QED expansion, and the impacts of QED evolution within the CT global analysis. 

\subsection{Numerical procedure}
\label{sec:ct18qedNum}

The CT18qed PDFs are constructed in such a way as to first separate the photon PDF into elastic and inelastic components, in analogy with the CT14QEDinc and CT14QED PDFs~\cite{Schmidt:2015zda}. The elastic photon distribution is directly calculated by applying the LUX formula, cf.~Eq.~(\ref{eq:elastic}), while the inelastic photon distribution, together with the (anti)quark and gluon partons of the proton, are predicted by applying the DGLAP equations to evolve PDFs from an initial scale, $\mu_0\! \sim\! [\mathrm{few\, GeV}]$, to an arbitrary scale at higher energies, $\mu$.

\begin{itemize}

	\item After separating the photon PDF into elastic and inelastic components as noted above, the elastic photon distribution at any scale, $\mu$, is directly evaluated by applying the appropriate elastic LUX formula, Eq.~(\ref{eq:elastic}). At $\mu_0=1.3$ GeV, the elastic photon contributes to about $\langle x\gamma^{\rm el}\rangle(\mu^2_0)=0.15\%$ momentum fraction of the proton. 
	
	\item At $\mu_0=1.3$ GeV, we determine the inelastic photon distribution according to the LUX master formula, Eq.~(\ref{eq:LUX}), excluding the elastic photon component in this stage of the calculation; here, the structure functions $F_{2,L}$ in the DIS region are calculated directly from the CT18 NNLO PDFs \cite{Hou:2019efy}. At our
	starting scale, the inelastic photon contributes $\langle x\gamma^{\inel}\rangle(\mu^2_0)=0.066\%$ to the total momentum of the proton. We also note that, since CT18 has one central set and 58 error sets, the resulting CT18qed PDFs also contain 1+58 PDF sets corresponding to the central prediction and additional Hessian error sets associated with the underlying uncertainty arising from the quark and gluon PDFs.
	
	\item At $\mu_0=1.3$ GeV, the shape and normalization of the valence quark and gluon PDF are fixed to those obtained in CT18, except that we require the total momentum fraction carried by the quark sea to be $1-\langle x(g + q^\mathrm{val} + \gamma^{\el+\inel})\rangle(\mu^2_0)$, where $q^{\rm val}=(u-\bar{u})+(d-\bar{d})$. This ensures consistency with the momentum sum rule at  $\mu_0=1.3$ GeV. This procedure is implemented by adjusting the normalizations of sea-quark PDFs in the fashion typically used in CT global fits so as to respect total momentum conservation.

	\item We then evolve the quark, gluon and {\it inelastic} photon PDFs from the starting scale ($\mu_0$) to an arbitrary scale $\mu$ ($\! >\! \mu_0$) by applying NNLO QCD plus
	NLO QED DGLAP evolution equations implemented inside the APFEL package~\cite{Bertone:2013vaa}. Consequently, a new set of quark and gluon PDFs, slightly different from those obtained in CT18, are generated with the photon PDF. 
	Crucially, owing to the properties of the combined QCD+QED splitting kernel, the total momentum carried by quark and gluon degrees-of-freedom and the {\it inelastic} photon will remain invariant, having been fixed to our initial choice of $\langle x(g+\Sigma+\gamma^{\inel})\rangle(\mu^2)=1-\langle x\gamma^{\el}\rangle(\mu^2_0)$. Since the momentum fraction carried by the {\it elastic} photon will change only very slightly with increasing $\mu$, decreasing by about $\delta\langle x\gamma^{\el}\rangle(\mu^2)=0.015\%$ at $\mu=10$ TeV, the CT18qed PDFs are guaranteed to satisfy the momentum sum rule to greater accuracy than in CT18lux. 
	
	\item We also compare this scenario outlined above with a different choice for the starting scale of $\mu_0=3$ GeV. In this case, the input quark and gluon PDFs at the scale $\mu_0=3$ GeV are those predicted by the standard CT18 PDFs (with $\mu_0=1.3$ GeV) after having been evolved from 1.3 GeV to 3 GeV. We note that in this case, the charm PDF is non-zero at the starting scale of $\mu_0=3$ GeV. For reasons explained in the next subsection, we will from here on take this PDF set with $\mu_0=3$ GeV as the default choice of CT18qed, and use the name CT18qed to refer to this set in particular, if no other specifications are given. We will use the name CT18qed1.3GeV, instead, to refer specifically to the PDF set with $\mu_0=1.3$ GeV as the starting scale, when a distinction is required. 
	
	\item We stress that the CT18qed PDFs described above are based on a boostrap logic, with an underlying assumption that the effects of QED evolution can be regarded as a minor perturbation about the CT18 global minimum. We validate this assumption in the context of an actual QED refit. Various impacts of the inclusion of QED in a fit, including those upon the $\chi^2$ and fitted PDFs, as well as the redistribution of momentum fractions due to the photonic radiation, are explored. Ultimately, we find these effects to be negligible, such
	that the CT18qed PDF sets above remain effectively unchanged.
\end{itemize}

\begin{figure}[ht]
	\centering
	\includegraphics[width=0.65\textwidth]{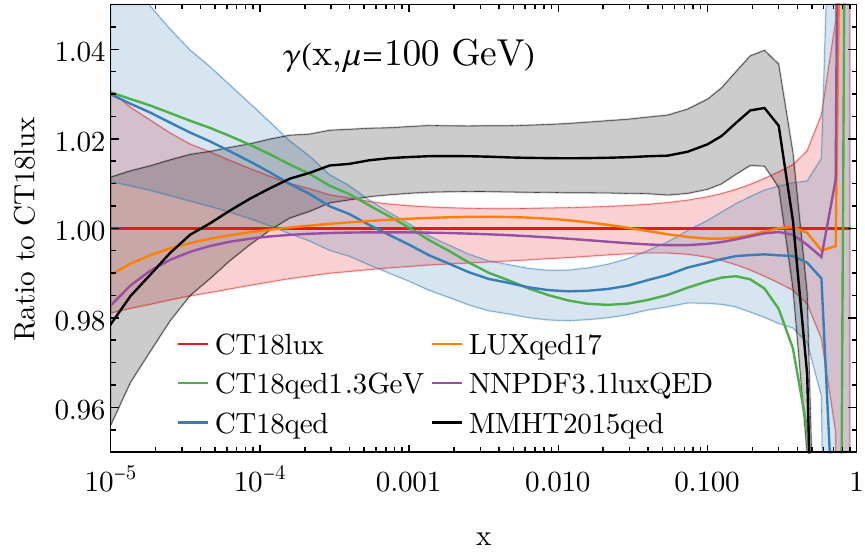}
	\caption{A comparison of the photon PDFs among the CT18lux, CT18qed(1.3GeV), LUXqed17 \cite{Manohar:2017eqh}, NNPDF3.1luxQED \cite{Bertone:2017bme}, and MMHT15qed \cite{Harland-Lang:2019pla}.
	}
	\label{fig:photon}
\end{figure}

\subsection{The proton's photonic content in CT18qed}
\label{sec:ct18qedPDFs}
\begin{figure}\centering
	\includegraphics[width=0.49\textwidth]{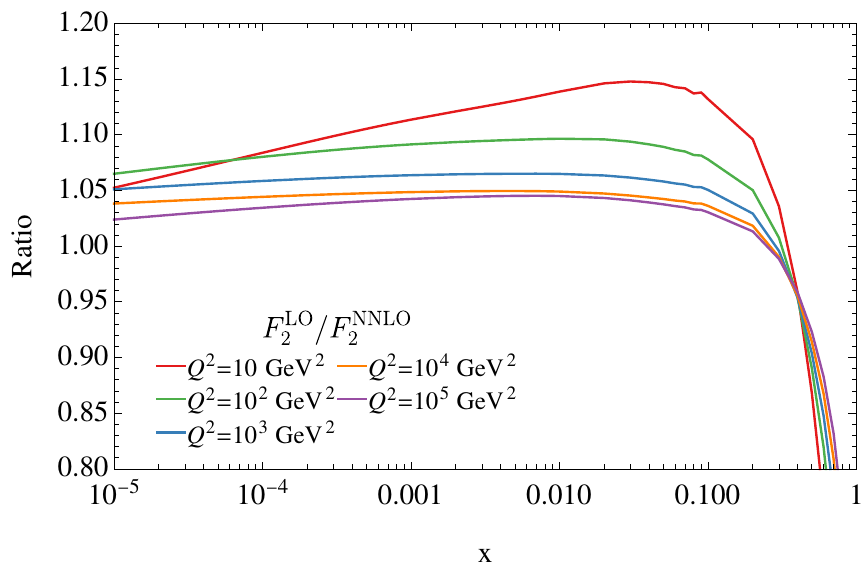}
	\includegraphics[width=0.49\textwidth]{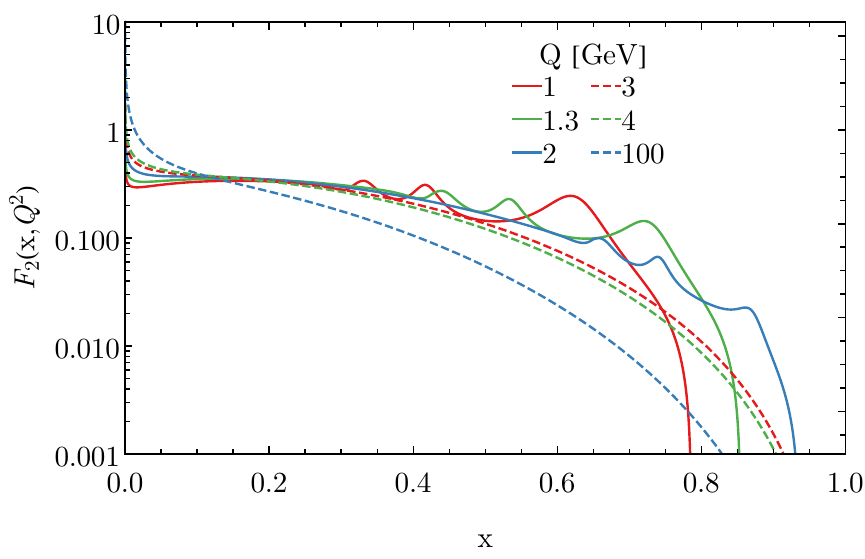}
	\caption{The ratios of structure function $F_2$ at LO to the NNLO in the high-$Q^2$ pQCD region (left).
		The pQCD NNLO $F_2(x,Q^2)$ compared with the one in the low-$Q^2$ non-perturbative region (right).}
	\label{fig:F2}
\end{figure}

\begin{figure}\centering
	\includegraphics[width=0.65\textwidth]{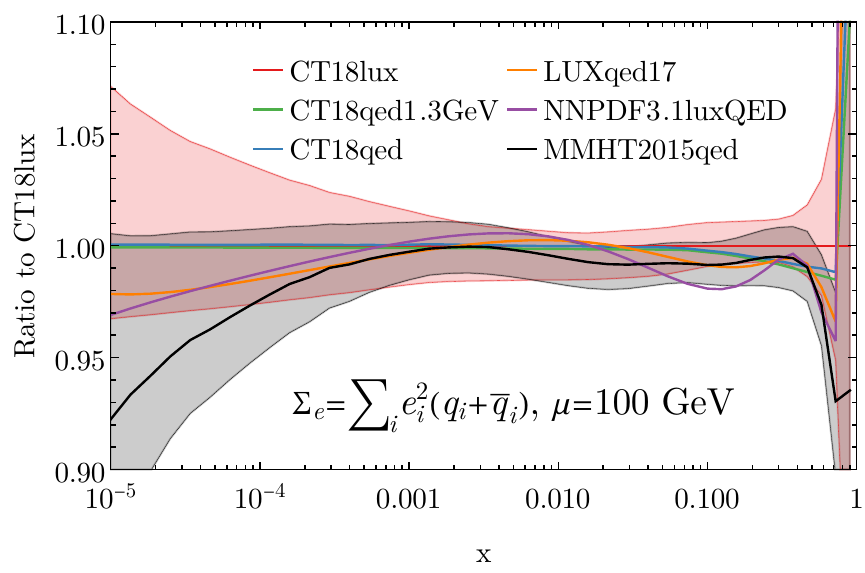}
	\caption{As a companion to Fig.~\ref{fig:photon}, we plot a comparison of the charge-weighted singlet PDFs of several photon PDF sets. 
	}
	\label{fig:ChargeSinglet}
\end{figure}

{\bf CT18qed PDFs}.
As outlined above, it is possible to determine the photon PDF at any perturbative scale through combined QCD+QED
evolution from a given initial boundary condition at the starting scale $\mu_0$. 
Various published photon PDF sets were obtained by choosing different $\mu_0$ and their associated boundary conditions. 
In Fig.~\ref{fig:photon}, we compare the CT18qed (with $\mu_0=1.3$ GeV and 3 GeV)
and CT18lux photon PDFs with other existing photon PDFs at $\mu=100$ GeV. It shows that in the intermediate-$x$ region, the CT18lux photon is between LUXqed17 (similar to NNPDF3.1luxQED) and MMHT2015qed, while CT18qed gives a smaller photon distribution than all other sets. In contrast, at extremely large $x$, the MMHT15qed and CT18qed calculations, both of which are based on a low-scale evolution approach, give relatively smaller photon PDFs than the others. 
Meanwhile, at small $x$, CT18qed gives a larger photon PDF than CT18lux. This can be qualitatively understood in terms of the approximation of solution to Eq. (\ref{eq:DGLAP}) as
\begin{equation}
\gamma\sim\int^{\mu^2}_{\mu_0^2}\frac{\dd Q^2}{Q^2}\frac{\alpha}{2\pi}
\sum_i e_i^2p_{\gamma q}\otimes (q_i+\bar{q}_i).
\end{equation}
Recall that structure functions at the leading order, \emph{i.e.}, $\calO(\alpha_s^0)$, are
\begin{equation}
F_2(x,Q^2)=\sum_i e_i^2(q_i+\bar{q}_i), ~ F_L(x,Q^2)=0.
\end{equation}
Hence, in this approximation,  
the DGLAP solution agrees well with the LUX prediction, when the LO structure functions are used and the $\msbar$ conversion term is ignored in Eq. (\ref{eq:LUX}). 
In Fig.~\ref{fig:F2}, left plot, 
we directly compare the LO with the full NNLO calculation of $F_2$ employed in the LUX approach. It shows that at high scales $\mu$, 
the ratio $F_2^{\rm LO}/F_2^{\rm NLO}$ is larger than one in the small $x$ region, and becomes much smaller than one in the large $x$ region. 
This explains why the CT18qed photon PDF is larger than CT18lux at small $x$ values and becomes smaller when $x$ increases.   
Another noticeable feature of  
Fig.~\ref{fig:photon} is that 
CT18qed photon PDF drops very fast as $x$ approaches to 1.
In order to understand the reduction of CT18qed at large $x$ values relative to CT18lux, we need to keep in mind that both the elastic
and inelastic photon PDFs drop rapidly when $x\to1$.
Given that the elastic components are the same for CT18lux and CT18qed
and essentially scale-invariant (see Fig.~\ref{fig:CT18lux}), the main difference between the two comes from the
inelastic contribution, especially from the $\msbar$-conversion term.
In CT18qed with $\mu_0=1.3~\GeV$,\footnote{It is also true for the $\mu_0=3$ GeV in the extremely large $x$ region of $W^2>W^{2}_{\rm high}$ in Fig. \ref{fig:xQ2plane}.} the starting scale of $\mu_0$ falls within the low-$Q^2$ continuum and resonance regions and therefore receives substantial non-perturbative contributions from the effects reviewed in Sec.~\ref{sec:uncSource}. 
In the right plot of Fig.~\ref{fig:F2}, we compare  a few non-perturbative $F_2$ (solid lines) with perturbative ones (dashed lines). We see that the non-perturbative $F_2$ is significantly larger than the perturbative one.\footnote{We remind the reader that the CLAS fit is bounded by a threshold condition for nucleon resonance production, $W^2>(m_p+m_\pi)^2$,
	which gives a boundary of $x$ at low $Q^2$. The $F_2$ will drop to zero beyond this point, which explains the sharp drop of the solid non-perturbative $F_2$ in Fig.~\ref{fig:F2}.} Therefore, we expect the corresponding absolute value of low-$Q^2$ $\msbar$-conversion terms to be significantly larger than the one in the high-$Q^2$ pQCD region. Recalling the negative sign of the $\msbar$ conversion term, this explains the significant reduction of CT18qed at large $x$ values compared with CT18lux. The same scenario occurs with MMHT2015qed at large $x$, as shown in Fig.~\ref{fig:photon}.
Finally, MMHT2015qed gives a smaller photon PDF in the small $x$ region as compared to CT18qed, which is due to the comparatively lesser value of its charge-weighted singlet distribution, $\Sigma_e$, which is depicted in Fig.~\ref{fig:ChargeSinglet}.

\begin{figure}\centering
	\includegraphics[width=0.65\textwidth]{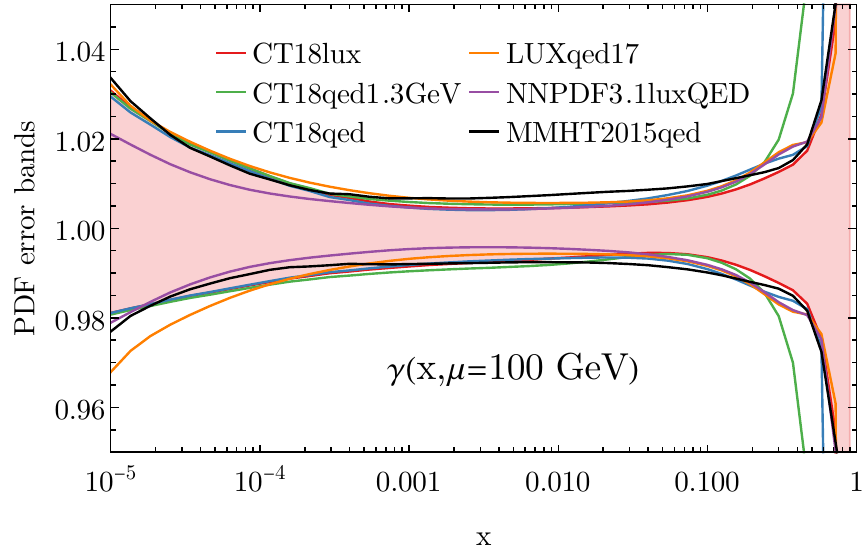}
	\caption{As a companion to Fig.~\ref{fig:photon}, we plot the self-normalized uncertainty bands for each of the photon PDFs examined in this analysis.
	}
	\label{fig:norm_unc}
\end{figure}

\begin{figure}
	\centering
	\includegraphics[width=0.49\textwidth]{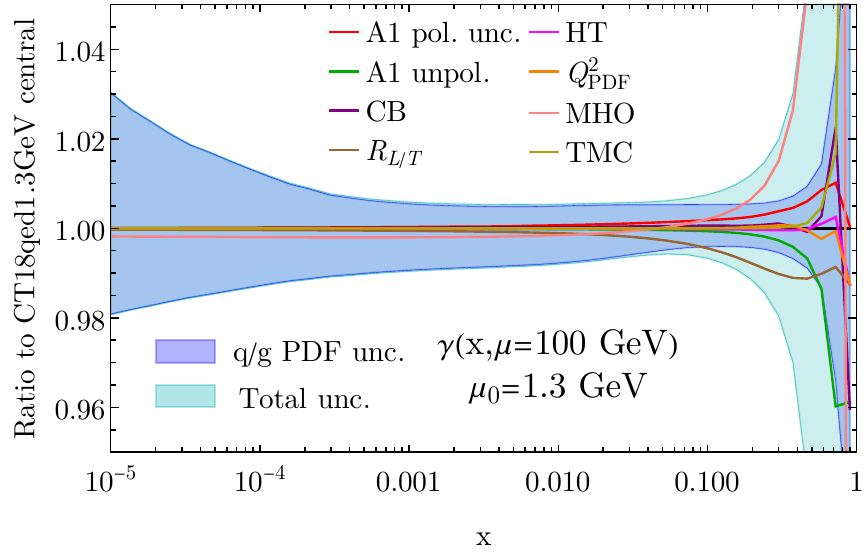}
	\includegraphics[width=0.49\textwidth]{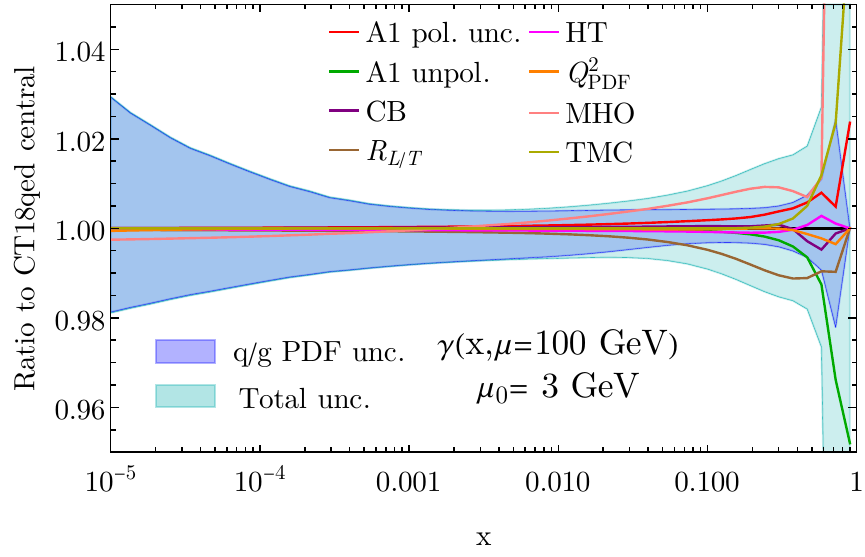}
	\includegraphics[width=0.49\textwidth]{photon_PDFunc_CT18lux_100GeV}
	\caption{Separated contributions to the full uncertainty of the photon PDF as obtained in CT18qed(1.3GeV) (upper)
		vs.~the analogous result in CT18lux (lower), which we again display here for comparison. We note that the contributions discussed here
		arise from various low-energy sources as enumerated in Sect.~\ref{sec:uncSource}.
	}
	\label{fig:CT18qed}
\end{figure}

{\bf The CT18qed PDF total uncertainty}.
The next question to ask is ``What is the photon PDF total uncertainty in CT18qed?'' 
Before detailing various potential sources of the photon PDF uncertainty, we first compare the photon PDF uncertainties as predicted by various global analysis groups. In Fig.~\ref{fig:norm_unc}, we show the self-normalized uncertainty bands to the corresponding central for each of the photon PDFs examined in this analysis. 

The photon PDF uncertainty induced by quark and gluon degrees-of-freedom can be calculated, as described above, in the same way as the PDF uncertainty for the quark and gluon PDFs --- by applying the master formula presented in Ref.~\cite{Nadolsky:2001yg} to the complete set of quark and gluon PDF error eigensets.
In fact, the uncertainty induced by the DGLAP-evolved quark and gluon PDFs has already been shown in Fig.~\ref{fig:phUncqg}, which roughly gives the same size of error bands
as compared with CT18lux and other groups, like LUXqed17 and MMHT2015qed. Similarly to the investigation in Sec.~\ref{sec:CT18lux}
for CT18lux, we investigate all other low-energy sources of uncertainty, which we summarize and compare side-by-side with CT18lux in Fig.~\ref{fig:CT18qed}. Much as we
observed for CT18lux, in the small-$x$ region the CT18qed photon uncertainty is dominated by the uncertainties in the quark and gluon PDFs. In the large-$x$ region, however, most of the
low-energy error sources are also significant, inducing uncertainties that are comparable to those seen in CT18lux, with the exception of the $M^2(z)$ variation ascribed to missing higher-order (MHO) effects.

For CT18qed, with $\mu_0=1.3$ GeV,  the MHO contribution becomes dominant at $x\!\gtrsim\!0.2$, giving larger error bands than CT18lux in this region. However, due to the much smaller MHO contribution, the total uncertainty band of photon PDF in CT18qed with $\mu_0=3~\GeV$ only becomes large at much larger $x$ values ($x\!\gtrsim\!0.6$).
As discussed in App.~\ref{app:sep}, the MHO uncertainty is estimated by shifting from the separation scale $M^2(z)=\mu^2/(1-z)$ to $\mu^2$. The corresponding $\msbar$ conversion term should be changed as Eq.~(\ref{eq:con}), which is semi-analytically integrated over the interval $Q^2\in[\mu^2,\mu^2/(1-z)]$, using the stationary approximation to replace the $Q$-dependent $F_2(x/z,Q^2)$ with the  fixed-$\mu^2$ contribution $F_{2}(x/z,\mu^2)$. 
%\carl{Keping, check that this is still correct.} 
In the pQCD DIS region, $F_2$ respects the stationary condition with a small logarithmic violation resulted from higher-order corrections. However, this condition is not respected that well in the non-perturbative region, when $Q^2<Q^2_{\rm PDF}$, show in Fig.~\ref{fig:F2}. 
Hence, the MHO contribution in  CT18qed with $\mu_0=1.3$ GeV is much larger than that with $\mu_0=3$ GeV, particularly in the large $x$ region, as shown in Fig.~\ref{fig:CT18qed}. For this reason, we have chosen to take the PDF set with $\mu_0=3$ GeV as the default CT18qed, and the one with $\mu_0=1.3$ GeV as an alternative set, which is dubbed as CT18qed1.3GeV for distinction. 

% - - - - - - - - - - - - - - - - - - - - - - - - - - - - - - - - - - - - - - - - - - - - - - - - - - - - - - - - - - - - - - - - - - - - - - - - - - - - - - - - - -

\subsection{Momentum conservation and QCD+QED evolution in CT18qed}
\label{sec:ct18qedMOM}
\begin{figure}\centering
	\includegraphics[width=0.49\textwidth]{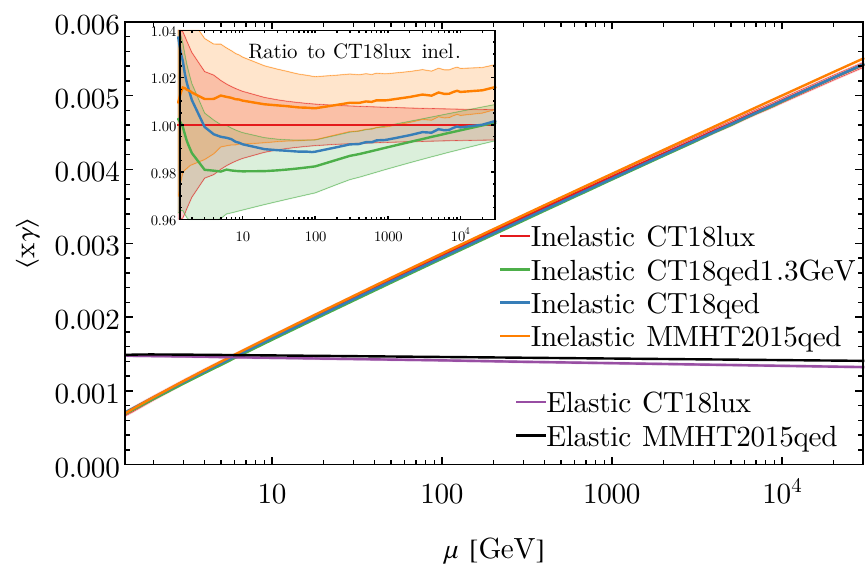}
	\includegraphics[width=0.49\textwidth]{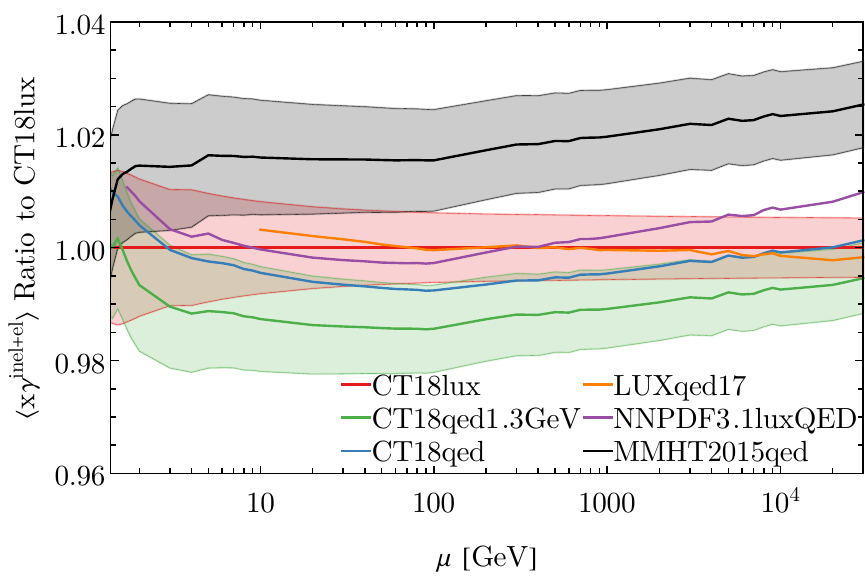}		
	\caption{The momentum fractions of the elastic, inelastic, and total photons. The turning point around $\mu\sim2$ GeV is due to a small discontinuity at the threshold of $m_\tau$. %\kp{I am not sure what causes the turning around $\mu\sim2$GeV, possibly because of $m_\tau$ or because of the $\mu$ node in the LHAPDF table.}
		%\cpy{Remove CT18qed4GeV curves.}
	}
	\label{fig:mom}
\end{figure}

{\bf Photon PDF moments and the momentum sum rule.} The proton's partonic constituents are expected to satisfy the momentum sum rule given by Eq.~(\ref{eq:momSR}).
CT18qed, like MMHT2015qed, determines an initial photon distribution at a low scale before consistently evolving all parton flavors under a combined
QCD+QED kernel to higher scales in a fashion that closely preserves the total proton momentum. In Sec.~\ref{sec:ct18qedNum} we outlined this procedure before showcasing
the resulting PDFs in Sec.~\ref{sec:ct18qedPDFs}. 
Here, we compare 
the first moments
of the photon and other PDFs, and relate this to aspects of the QCD+QED evolution framework. In addition, we compare properties of evolution in CT18qed with MMHT2015qed
as well as the other fitting groups.
As noted before, we enforce the momentum sum rule in CT18qed at $\mu_0$ as
\begin{equation}
%\langle x(q+g+\gamma^{\rm inel})\rangle(\mu_0)=1~\textrm{and}~
\langle x(g+\Sigma+\gamma^{\rm inel+el})\rangle(\mu^2_0)=1\ ,
\label{eq:mom-inel}
\end{equation}
making use of the very mild scale dependence of the first moment of the {\it elastic} photon PDF, $\langle x \gamma^{\rm el} \rangle(\mu^2)$.
In turn, the DGLAP evolution of the remaining components of Eq.~(\ref{eq:mom-inel}) is such that the total {\it inelastic} momentum, $\langle x(\Sigma+g+\gamma^{\rm inel})\rangle(\mu^2)$,
remains fixed due to the momentum-conserving properties of the splitting functions:
\begin{equation}
\int_0^1\dd x x\left(P_{ii}+\sum_{j}P_{ji}\right)=0\ .
\end{equation}
The scale dependence of the separate (in)elastic contributions of the photon momentum are plotted in the plot of Fig.~\ref{fig:mom}.
We remind the reader that the combination $\langle x(\Sigma+g+\gamma^{\rm inel+el})\rangle(\mu^2)$ does not exactly conserve momentum due to the very minor
scale variations in $\gamma^{\rm el}(\mu^2)$ depicted in the lower set of curves shown in Fig.~\ref{fig:mom}, left plot. 
\begin{table}[]
	%\centering
	\hspace{-0.65cm}\begin{tabular}{c|c|c|c|c|c}
		\hline
		$\mu_{\min}$ [GeV]  & 1.3               & 1.3               & 1                 & 10                & 1.65               \\
		$\mu$   [GeV]  & CT18qed           & CT18lux           & MMHT2015qed       & LUXqed17          & NNPDF3.1luxQED     \\ \hline
		1              & --                & --                & $0.196\pm0.003$   & --                & --                 \\
		1.3            & $0.215\pm0.003$   & $0.215\pm0.003$   & $0.215\pm0.003$   & --                & --                 \\ 
		1.65           & $0.227\pm0.003$   & $0.227\pm0.003$   & $0.230\pm0.003$   & --                & $0.229\pm0.003$    \\
		10             & $0.314\pm0.003$   & $0.317\pm0.003$   & $0.323\pm0.003$   & $0.319\pm0.003$   & $0.317\pm0.003$    \\
		100            & $0.419\pm0.003$   & $0.424\pm0.003$   & $0.432\pm0.004$   & $0.425\pm0.003$   & $0.424\pm0.003$    \\
		1000           & $0.522\pm0.004$   & $0.527\pm0.003$   & $0.538\pm0.004$   & $0.528\pm0.004$   & $0.529\pm0.003$    \\ \hline
	\end{tabular}
	\caption{The averaged photon momenta, $\langle x\gamma\rangle(\mu^2)~[\%]$, at a number of scales as obtained in CT18qed and CT18lux (leftmost
		columns) as well as in several other recent analyses. We remind the reader that $\mu_{\min}$ here is the lowest energy scale of the corresponding LHAPDF grids, which is same as the DGLAP initialization scales, $\mu_0$, of CT18lux, CT18qed1.3GeV and MMHT2015qed, but different from the $\mu_0$ of CT18qed, LUXqed(17) and NNPDF3.1luxQED.
	}
	\label{tab:momfrac}
\end{table}
As a companion to these plots, in Table~\ref{tab:momfrac}, we display the first moments of the photon PDF as obtained by CT18qed and CT18lux at a number
of relevant scales, and compare with the corresponding results published by MMHT2015qed, LUXqed17, and NNPDF3.1luxQED.

For completeness, we also show the comparison of the total photon momentum fractions among different PDFs in the right plot of Fig.~\ref{fig:mom}, with the specific numbers
at a few typical scales listed in Table~\ref{tab:momfrac}. We see the overall size agrees very well among different photon PDF sets, both for the absolute values
as well as for the uncertainties. MMHT2015qed gives 2$\sim$3\% larger than other groups, due to its larger elastic photon. (See  Fig.~\ref{fig:elastic} in Appendix B.)
LUXqed17 gives
very much the same as CT18lux, while CT18qed is slightly smaller. One important feature appears for the inelastic photon that the low-$\mu_0$ DGLAP approach in CT18qed yields a smaller photon at a low scale than LUX one in CT18lux, but gradually exceeds when energy
increases up to certain scales. 

Here, we note that CT18lux violates the momentum sum rule of Eq.~(\ref{eq:momSR}) very weakly, as the photon is added on top of the existing quark and gluon distributions without making
compensating adjustments to the latter. This small violation can be quantified by the averaged momentum fraction carried by the photon,
\begin{equation}
\langle x \gamma\rangle(\mu^2)=\int_0^1\dd x\, x\gamma(x,\mu^2)\ ,
\end{equation}
for which the scale dependence of the separated inelastic and elastic components are shown in Fig.~\ref{fig:mom}. At the CT18lux starting scale, $\mu_0=1.3~\GeV$,
these separate contributions are
\begin{equation}
\langle x\gamma^{\rm el}\rangle(\mu^2_0)=0.15\%\ , \ \ \ \ ~\langle x\gamma^{\rm inel}\rangle(\mu^2_0)=0.066\%\ \, 
\end{equation}
which are 0.15\% and 0.11\%, respectively, when 
$\mu_0=3~\GeV$.
We note that $\langle x\gamma^{\rm inel}\rangle (\mu^2)$ receives a contribution from the $q\to q\gamma$ splitting, which grows logarithmically with $\mu^2$. 
In contrast, $\langle x\gamma^{\rm el}\rangle (\mu^2)$ remains nearly constant, with a small decrease that varies as $\alpha(\mu_0^2)/\alpha(\mu^2)$. 

\begin{figure}\centering
	\includegraphics[width=0.49\textwidth]{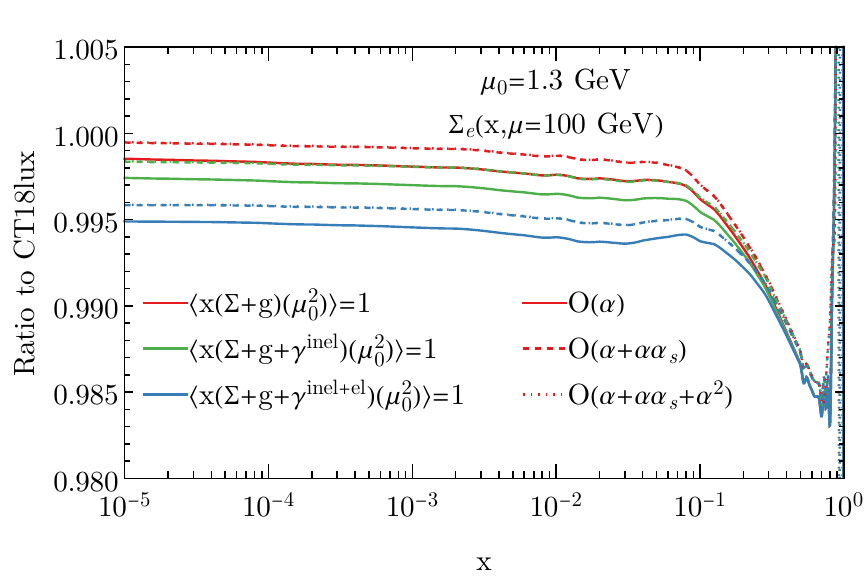}
	\includegraphics[width=0.49\textwidth]{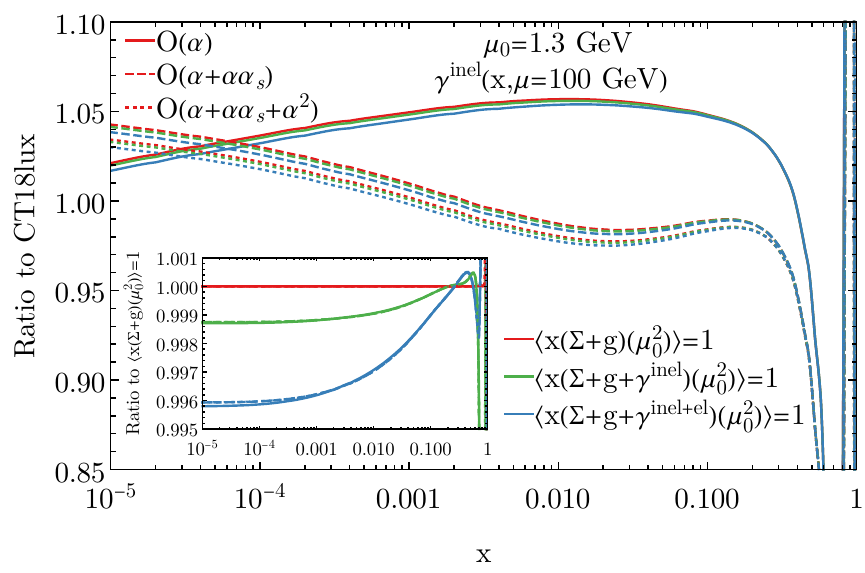}	
	\caption{The ratios of the ${\rm QCD}\otimes\textrm{QED}$ PDFs compared with the CT18lux ones.}
	\label{fig:renorm}
\end{figure}

In order to examine the size of the impact resulting from enforcing the momentum sum rule, we compare our default treatment, Eq.~(\ref{eq:mom-inel}), 
with an intermediate one, 
$\langle x(g+\Sigma+\gamma^{\rm inel})\rangle(\mu^2_0)=1$, and a CT18lux-like constraint, $\langle x(g+\Sigma)\rangle(\mu^2_0)=1$, in Fig.~\ref{fig:renorm}. Here we show
the charge-weighted singlet, $\Sigma_e$, on the left and the inelastic photon, $\gamma^{\rm inel}$, on the right, normalized to the CT18lux PDFs. Due to the reduction of the momentum fraction of the quark sea by the photon, the overall relative decrease in the charge-weighted singlet PDF is roughly $2.5\langle x\gamma\rangle(\mu_0^2) \sim 0.5\%$, especially in the small-$x$ region, as shown in the left plot of Fig.~\ref{fig:renorm}.
Reciprocally, the inelastic photon is reduced by roughly the same amount as we explicitly show in the right-hand plot of Fig.~\ref{fig:renorm}.

{\bf Comparison of CT18lux and CT18qed at a different order of QED evolution.}
Critical aspects of the scale dependence and agreement among photon PDF calculations follow from the implementation of perturbative
QCD+QED evolution in CT18qed vs.~the framework in other fits.
For example, in comparing different orders of QED evolution in generating CT18qed with the CT18lux inelastic photon in the left plot
Fig.~\ref{fig:renorm}, we find that the NLO DGLAP kernel gives better agreement than the LO one. It is understood
that the LUX formula for the photon PDF includes up to one perturbative order higher than the traditional DGLAP approach \cite{Manohar:2017eqh}. 
We also explained already that the CT18qed gives a larger inelastic photon at small $x$ but significantly smaller at large $x$.

The LO and NLO [$\calO(\alpha\alpha_s)$ and $\calO(\alpha^2)$] QED inelastic photon of CT18qed, together with the charge-weighted singlet distribution, $\Sigma_e=\sum_i e_i^2(q_i+\bar{q}_i)$, are compared against CT18lux in Fig.~\ref{fig:renorm}. We see the redistribution of the proton momentum to the inelastic photon only impacts the inelastic photon and charge-weighted singlet by the corresponding overall factor, which is negligible once compared to the impact of the QED splitting, $q\to q\gamma$. When turning on the NLO QED evolution, the inelastic photon receives negative corrections, while the remaining $\Sigma_e$ becomes larger as less photon is radiated off quarks. More specifically, the $\calO(\alpha\alpha_s)$ corrections reduce $\gamma^{\rm inel}$ by approximately 6\% around $x\!\sim\!0.02$, and the $\calO(\alpha^2)$ corrections reduces this by another $\sim$1\%. The $\calO(\alpha^2)$ corrections to $\Sigma_e$ are effectively invisible, as they coincide with the $\calO(\alpha\alpha_s)$ effect in Fig.~\ref{fig:renorm}, right plot. 
This comparison clearly shows that the impact of including higher-order QED effects into the DGLAP evolution equations is much larger than fixing the amount of momentum violation due to the
incorporation of the photon PDF.

\subsection{QED evolution in the global fit}
%\section{Theoretical description of data in CT18lux and CT18qed}
\label{sec:fit}

\begin{table}[htbp]
	\hspace{-10pt}\begin{tabular}{c|lr|c|c|c|c}
		\hline
		\textbf{ID}  & \textbf{Experimental data set} & Ref. & $N_{\rm pt}$  & CT18lux  & CT18qed & QED fit\tabularnewline
		\hline
		\hline
		160 & HERAI+II 1 fb$^{-1}$, H1 and ZEUS comb.                                   & \cite{Abramowicz:2015mha}   & 1120  &   1406 & 1405 & 1405 \tabularnewline\hline
		101 & BCDMS $F_{2}^{p}$                                                         & \cite{Benvenuti:1989rh}     &  337  &    375 &  381 &  377 \tabularnewline\hline
		102 & BCDMS $F_{2}^{d}$                                                         & \cite{Benvenuti:1989fm}     &  250  &    281 &  283 &  281 \tabularnewline\hline
		104 & NMC $F_{2}^{d}/F_{2}^{p}$                                                 & \cite{Arneodo:1996qe}       &  123  &    126 &  126 &  126 \tabularnewline\hline
		108 & CDHSW $F_{2}^{p}$                                                         & \cite{Berge:1989hr}         &   85  &   85.6 & 86.6 & 86.6 \tabularnewline\hline         
		109 & CDHSW $x_B F_{3}^{p}$                                                     & \cite{Berge:1989hr}         &   96  &   86.4 & 87.1 & 86.0 \tabularnewline\hline
		110 & CCFR $F_{2}^{p}$                                                          & \cite{Yang:2000ju}          &   69  &   78.4 & 77.6 & 77.7 \tabularnewline\hline
		111 & CCFR $x_B F_{3}^{p}$                                                      & \cite{Seligman:1997mc}      &   86  &   33.4 & 32.3 & 33.9 \tabularnewline\hline
		124 & NuTeV $\nu\mu\mu$ SIDIS                                                   & \cite{Mason:2006qa}         &   38  &   18.6 & 18.8 & 18.4 \tabularnewline\hline
		125 & NuTeV $\bar\nu \mu\mu$ SIDIS                                              & \cite{Mason:2006qa}         &   33  &   38.4 & 38.5 & 37.8 \tabularnewline\hline
		126 & CCFR $\nu\mu\mu$ SIDIS                                                    & \cite{Goncharov:2001qe}     &   40  &   29.8 & 29.7 & 29.8 \tabularnewline\hline
		127 & CCFR  $\bar\nu \mu\mu$ SIDIS                                              & \cite{Goncharov:2001qe}     &   38  &   19.8 & 19.7 & 19.8 \tabularnewline\hline
		145 & H1 $\sigma_{r}^{b}$                                                       & \cite{Aktas:2004az}         &   10  &   6.81 & 6.81 & 6.91 \tabularnewline\hline
		147 & Combined HERA charm production                                            & \cite{Abramowicz:1900rp}    &   47  &   58.7 & 58.7 & 57.7 \tabularnewline\hline
		169 & H1 $F_{L}$                                                                & \cite{Collaboration:2010ry} &    9  &   17.0 & 17.0 & 16.9 \tabularnewline\hline
		201 & E605 Drell-Yan  $s\dd^2\sigma/(\dd\sqrt{\tau}\dd y)$                      & \cite{Moreno:1990sf}        &  119  &    103 &  104 &  103 \tabularnewline\hline
		203 & E866 Drell-Yan  $\sigma_{pd}/(2\sigma_{pp})$                              & \cite{Towell:2001nh}        &   15  &   16.2 & 16.4 & 16.6 \tabularnewline\hline
		204 & E866 Drell-Yan  $Q^3\dd^2\sigma_{pp}/(\dd Q\dd x_F)$                      & \cite{Webb:2003ps}          &  184  &    244 &  245 &  246 \tabularnewline\hline
		225 & CDF Run-1 lepton $A_{\rm ch}$, $p_{T\ell}>25$ GeV                         & \cite{Abe:1998rv}           &   11  &   9.04 & 9.30 & 9.17 \tabularnewline\hline
		227 & CDF Run-2 electron $A_{\rm ch}$, $p_{T\ell}>25$ GeV                       & \cite{Acosta:2005ud}        &   11  &   13.5 & 12.8 & 13.4 \tabularnewline\hline
		234 & D\O~ Run-2 muon $A_{\rm ch}$, $p_{T\ell}>20$ GeV                          & \cite{Abazov:2007pm}        &    9  &   8.91 & 10.2 & 9.36 \tabularnewline\hline
		260 & D\O~ Run-2 $Z$ rapidity                                                   & \cite{Abazov:2007jy}        &   28  &   16.8 & 16.8 & 16.8 \tabularnewline\hline
		261 & CDF Run-2 $Z$ rapidity                                                    & \cite{Aaltonen:2010zza}     &   29  &   49.1 & 50.5 & 49.1 \tabularnewline\hline
		266 & CMS 7 TeV $4.7\mbox{ fb}^{-1}$, muon $A_{\rm ch}$, $p_{T\ell}>35$ GeV     & \cite{Chatrchyan:2013mza}   &   11  &   7.72 & 8.23 & 7.92 \tabularnewline\hline
		267 & CMS 7 TeV $840\mbox{ pb}^{-1}$, electron $A_{\rm ch}$, $p_{T\ell}>35$ GeV & \cite{Chatrchyan:2012xt}    &   11  &   11.0 & 12.4 & 12.0 \tabularnewline\hline
		268 & ATLAS 7 TeV $35\mbox{ pb}^{-1}$, $W/Z$ cross sec., $A_{\rm ch}$           & \cite{Aad:2011dm}           &   41  &   44.8 & 44.1 & 44.0 \tabularnewline\hline
		281 & D\O~ Run-2 $9.7\mbox{ fb}^{-1}$, electron $A_{\rm ch}$, $p_{T\ell}>25$ GeV& \cite{D0:2014kma}           &   13  &   22.9 & 23.6 & 22.4 \tabularnewline\hline
		504 & CDF Run-2 inclusive jet production                                        & \cite{Aaltonen:2008eq}      &   72  &    125 & 126  &  124 \tabularnewline\hline
		514 & D\O~ Run-2 inclusive jet production                                       & \cite{Abazov:2008ae}        &  110  &    114 & 113  &  114 \tabularnewline\hline
	\end{tabular}
	\caption{The $\chi^2$ of CT18lux and CT18qed for the data sets include in the CT18 NNLO global analyses \cite{Hou:2019efy}. The CT18lux shares the same $\chi^2$ as the CT18 PDF, as the quark and gluon PDFs remain unchanged. 
		\label{tab:EXP_1}
		}
\end{table}

\begin{table}[tb]
	\hspace{-10pt}\begin{tabular}{c|lr|c|c|c|c}
		\hline
		\textbf{ID }  & \textbf{Experimental data set} & Ref.  & $N_{\rm pt}$  & CT18lux & CT18qed & QED fit  \tabularnewline
		\hline
		\hline
		245 & LHCb 7 TeV 1.0 fb$^{-1}$, forward $W/Z$                                  & \cite{Aaij:2015gna}        &  33  &   53.4 & 49.9 & 53.9 \tabularnewline\hline
		246 & LHCb 8 TeV  2.0 fb$^{-1}$, forward $Z\rightarrow e^{-} e^{+}$            & \cite{Aaij:2015vua}        &  17  &   25.5 & 23.7 & 25.5 \tabularnewline\hline
		% 248 & ATLAS 7 TeV 4.6 fb$^{-1}$, $W/Z$ combined                  & \cite{Aaboud:2016btc}      &  34  &   8.60 & 8.07  &\tabularnewline\hline
		249 & CMS 8 TeV 18.8 fb$^{-1}$, muon $A_{\rm ch}$                              & \cite{Khachatryan:2016pev} &  11  &   12.4 & 15.5 & 11.7 \tabularnewline\hline
		250 & LHCb 8 TeV 2.0 fb$^{-1}$, forward $W/Z$                                  & \cite{Aaij:2015zlq}        &  34  &   73.2 & 69.2 & 72.6 \tabularnewline\hline 
		253 & ATLAS 8 TeV 20.3 fb$^{-1}$, $Z$ $p_T$                                    & \cite{Aad:2015auj}         &  27  &   30.0 & 29.4 & 31.1 \tabularnewline\hline
		542 & CMS 7 TeV 5 fb$^{-1}$, single incl. jet $R=0.7$                          & \cite{Chatrchyan:2014gia}  & 158  &   195  &  193 &  195 \tabularnewline\hline
		544 & ATLAS 7 TeV  4.5 fb$^{-1}$, single incl. jet  $R=0.6$                    & \cite{Aad:2014vwa}         & 140  &   202  &  200 &  204 \tabularnewline\hline
		545 & CMS 8 TeV 19.7 fb$^{-1}$, single incl. jet $R=0.7$                       & \cite{Khachatryan:2016mlc} & 185  &   213  &  220 &  210 \tabularnewline\hline
		573 & CMS 8 TeV 19.7 fb$^{-1}$, $t\bar{t}$ $(1/\sigma)\dd^2\sigma/(\dd p_T^{t}\dd y^t)$ & \cite{Sirunyan:2017azo} &16& 18.9 & 18.8 & 18.9 \tabularnewline\hline
		580 & ATLAS 8 TeV 20.3 fb$^{-1}$, $t\bar{t}$ $\dd\sigma/\dd p_{T}^{t}$ and $\dd\sigma/\dd m_{t\bar{t}}$   & \cite{Aad:2015mbv}&15&9.51 & 9.49 & 9.70 \tabularnewline\hline
		& Total $\chi^2$ for all 39 data sets & &3681   & 4293 & 4302 & 4296 \\
		\hline
	\end{tabular}
	\caption{Like Tab.~\ref{tab:EXP_1}, the $\chi^2$ of the CT18lux and CT18qed for newly-included LHC measurements in the CT18 NNLO global analyses. %\kp{Some numbers of CT18lux are different from CT18 paper. The total $\chi^2$ are 4293 vs 4304.}
		\label{tab:EXP_2} }
\end{table}

\begin{figure}
    \centering
    \includegraphics[width=0.49\textwidth]{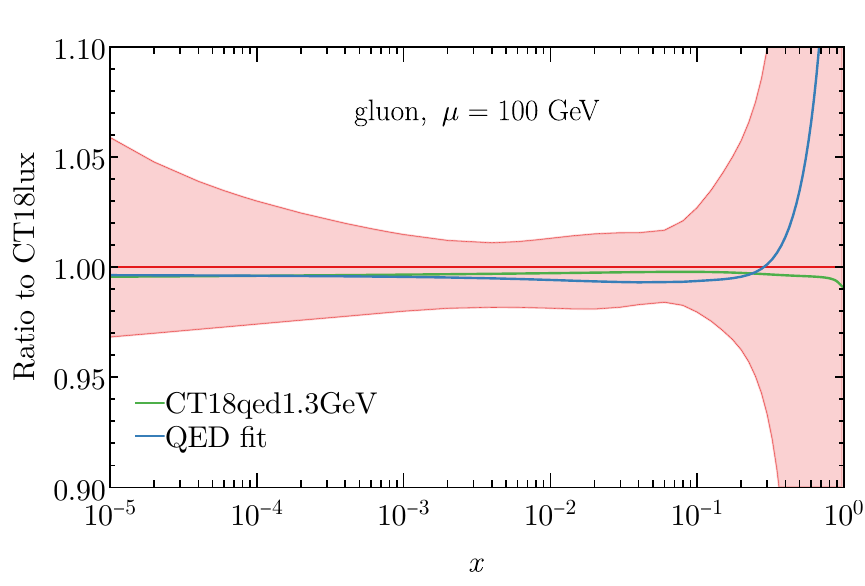}
    \includegraphics[width=0.49\textwidth]{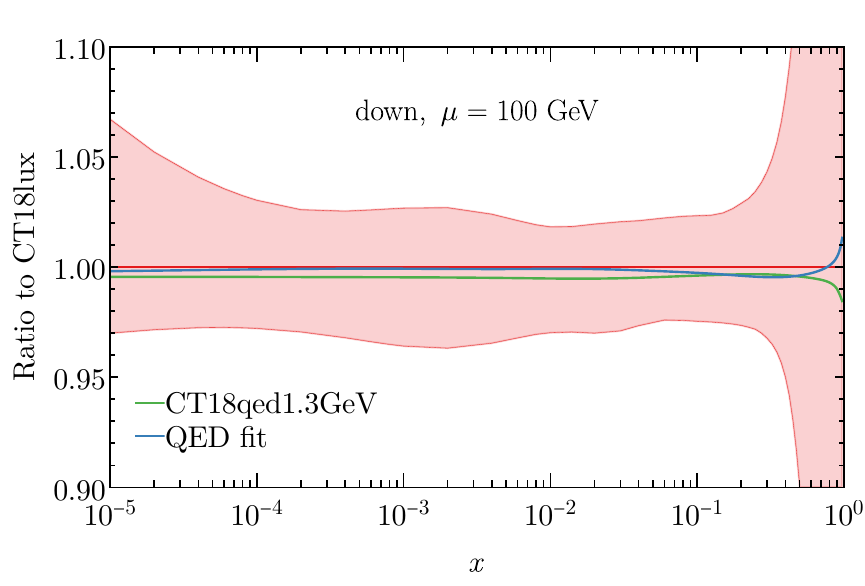}
    \includegraphics[width=0.49\textwidth]{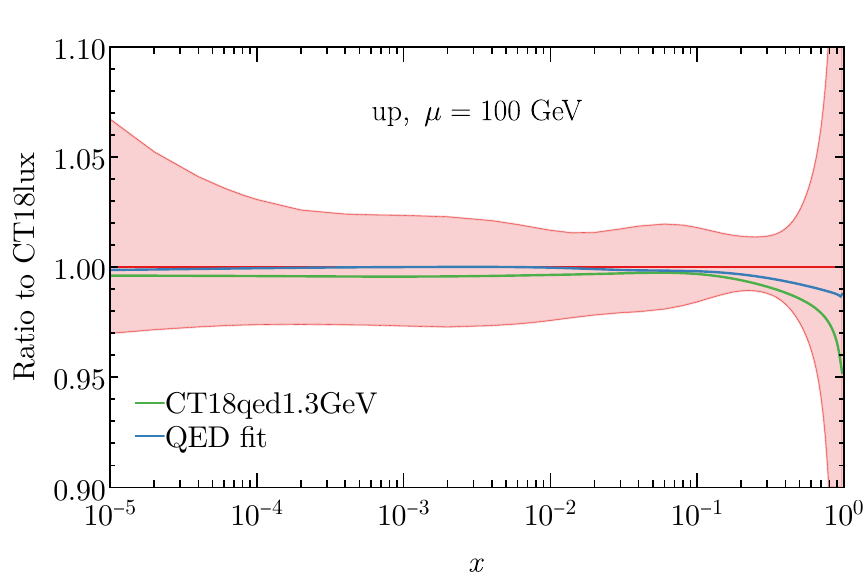}
    \includegraphics[width=0.49\textwidth]{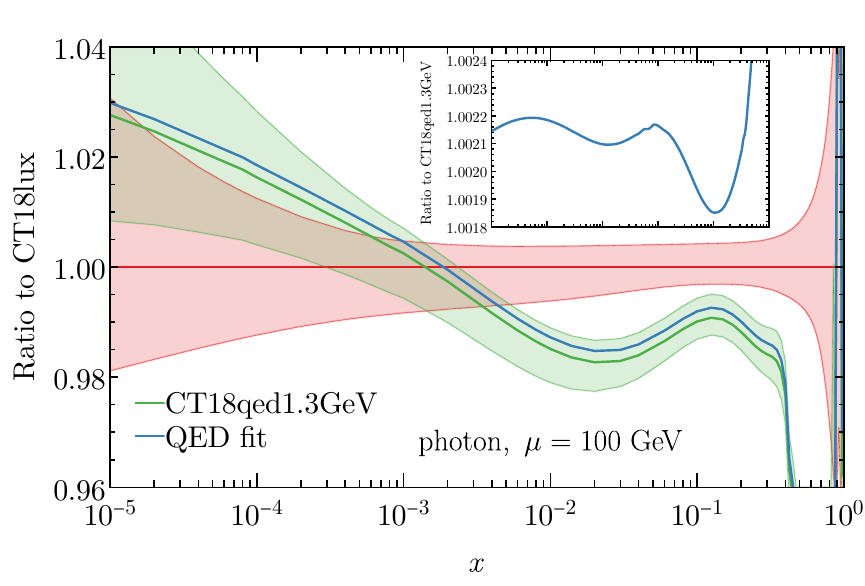}    
    \caption{The PDFs with QED fits compared with the CT18lux and CT18qed1.3GeV.}
    \label{fig:CT18QEDfit}
\end{figure}

In principle, the QED-corrected DGLAP evolution discussed above can play a role in the CT global analysis. In order to examine this possibility, we perform a new global fit, in which the QED corrections to the splitting kernels are included alongside the NNLO QCD contributions, and the QED corrections to DIS structure functions are also applied, similarly to the treatments of NNPDF3.1luxQED~\cite{Bertone:2017bme}, MMHT2015qed~\cite{Harland-Lang:2019pla}, and MSHT20qed~\cite{Cridge:2021pxm}. 
The resulting $\chi^2$ values for fitting to 
the CT18 default data sets are presented in the rightmost column of Tabs.~\ref{tab:EXP_1}--\ref{tab:EXP_2}, and the corresponding PDFs in Fig.~\ref{fig:CT18QEDfit}.
In order to compare with the CT18 NNLO global fit on the same footing, the initialization scale is chosen to be the same, {\it i.e.}, we set $\mu_0=1.3~\GeV$. 
The results of the unfitted CT18lux (having the same $\chi^2$ as CT18 NNLO) and CT18qed calculations are also shown for comparison. In particular,
the $\chi^2$ values for CT18lux and CT18qed are given explicitly in Tabs.~\ref{tab:EXP_1}--\ref{tab:EXP_2} in the fourth and fifth columns from left, respectively. As we demonstrated with the example of the charge-weighted singlet PDF in Fig.~\ref{fig:ChargeSinglet} and right panel of Fig.~\ref{fig:renorm} already, introducing the photon as a new component within the QED evolution takes away only a very small fractional momentum from the quark PDFs, especially at large $x$ due to the $(q\to q\gamma)$ splitting. As a result, we observe a small increase in $\chi^2$ under the unfitted CT18qed calculation, amounting to $\Delta \chi^2\! \approx\! 10$ units out of $\approx\!4293$ for the full data set consisting of $\sum N_\mathrm{pt}\!=\!3681$ points. In terms of specific data sets, the theoretical description remains unchanged for most experiments, while the $\chi^2$ values of those data most sensitive to the $d/u$ PDF ratio ({\it e.g.}, the CMS 8 TeV muon charge asymmetry~\cite{Khachatryan:2016pev}) or large-$x$ PDF behavior ({\it e.g.}, LHCb 7 TeV forward $W/Z$ production~\cite{Aaij:2015gna} and BCDMS $F_2^p$~\cite{Benvenuti:1989rh}), are shifted by a few units.

In the new refit carried out in the presence of QED corrections (corresponding to the `QED fit' column  of Tabs.~\ref{tab:EXP_1}--\ref{tab:EXP_2}), the total $\chi^2$ was reduced marginally, but, in the
end, the description of the global data set remains substantially unchanged. In terms of the refitted PDFs of specific parton flavor, we see in Fig.~\ref{fig:CT18QEDfit} that the reduction to the $u$- and $d$-quark PDFs that occurred under CT18qed1.3GeV is mostly restored to match the behavior seen with CT18lux, as preferred by the fitted data. In this case, the reduction to the total partonic momentum carried by the quarks and gluon due to the presence of the photon PDF derives mainly from the gluon and sea quarks, roughly by an overall factor. This momentum is taken by the photon PDF, which receives a slight enhancement at the level of $\langle x\gamma\rangle(\mu_0^2)\sim0.21\%$, compared with CT18qed, in which the momentum fraction is only taken from the sea quarks manually. In this sense, the role of the momentum sum rule makes the photon momentum fraction in the QCD+QED fit lie in between those obtained by the redistribution of the gluon momentum adopted by LUXqed(17)~\cite{Manohar:2016nzj,Manohar:2017eqh} and of the sea quark momentum adopted in our default CT18qed.

The analysis above serves as validation of the boostrap logic leading to CT18qed, in which the QED evolution can be viewed as a minor perturbation giving only a very small variation in the quark and gluon PDFs from the CT18 global minimum. We therefore regard the publicly-released CT18qed set(s) as our primary QED calculation in light of this small change.

\section{Implications for photon-initiated processes at the LHC}
\label{sec:pheno}
\begin{figure}
	\centering
	\includegraphics[width=0.49\textwidth]{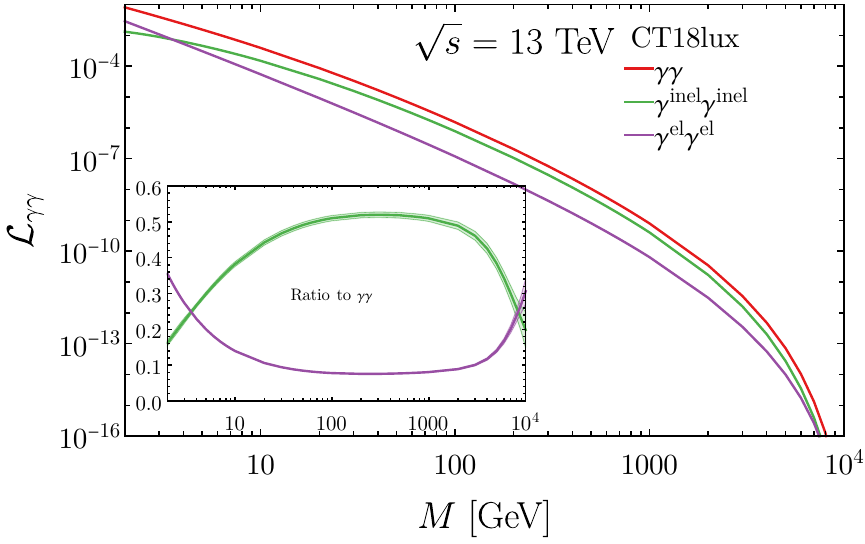}	
	\includegraphics[width=0.49\textwidth]{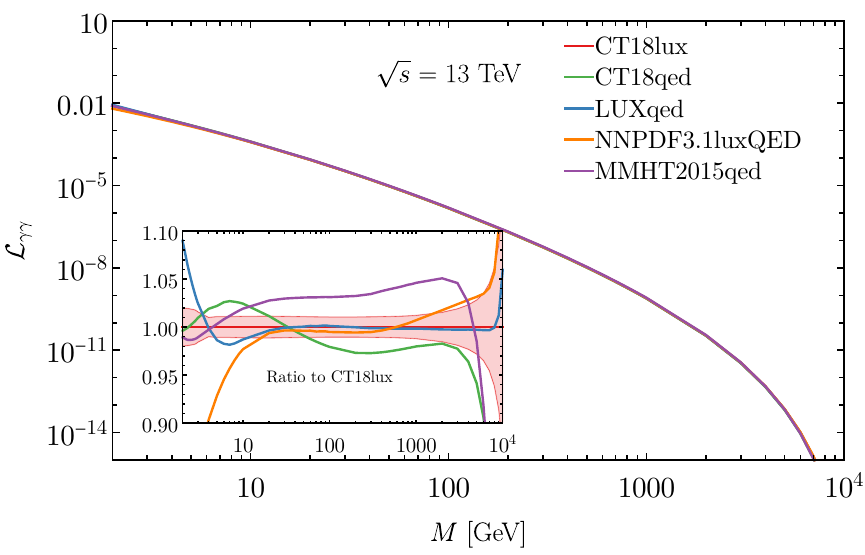}
	\includegraphics[width=0.49\textwidth]{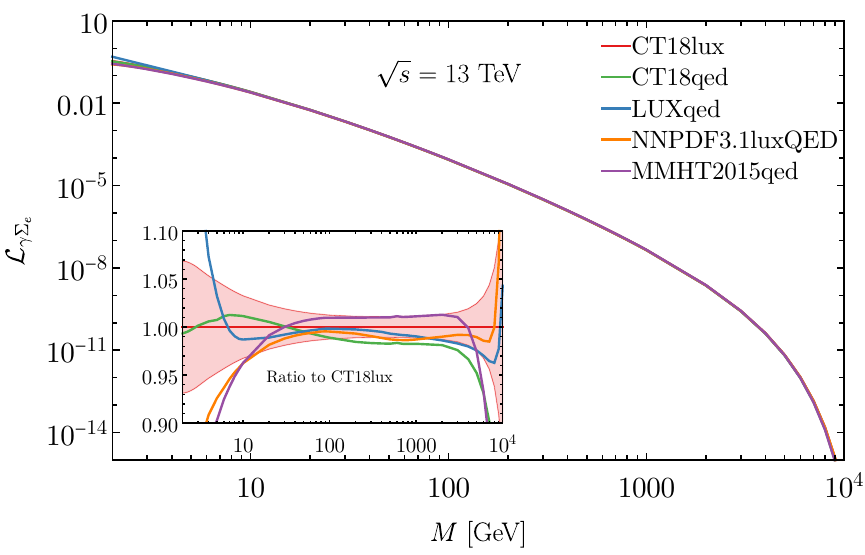}
	\includegraphics[width=0.49\textwidth]{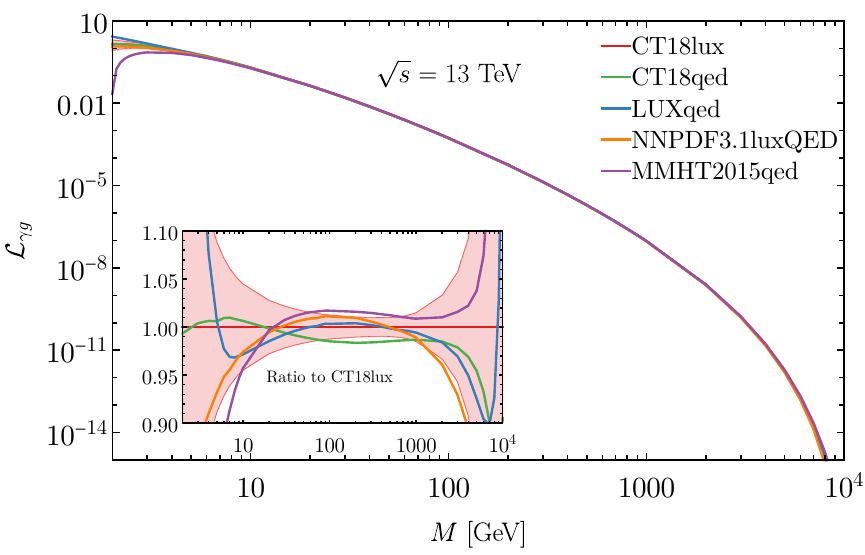}		
	\caption{The parton luminosities of $\mathcal{L}_{\gamma\gamma}$, $\mathcal{L}_{\gamma\Sigma_e}$, and $\mathcal{L}_{\gamma g}$ at a $\sqrt{s}=13$ TeV $pp$ machine.}
	\label{fig:lumi}
\end{figure}

Having developed and applied a combined QCD+QED formalism to obtain the photon PDF in the preceding sections,
we now examine the sensitivity of observables at the LHC to this photon PDF and explore the phenomenological consequences. 
Before going to specific SM processes, we first compute the parton-parton luminosities involving the photon. The definition
of the parton-parton luminosities can be generically taken as \cite{Campbell:2006wx}
\begin{equation}
\mathcal{L}_{ij}(s,M^2)\equiv\frac{1}{s}\frac{1}{1+\delta_{ij}}
\int_{\tau}^1\frac{\dd x}{x}
\left[f_{i}(x,\mu^2)f_{j}(\tau/x,\mu^2)+(i\leftrightarrow j)\right]\ ,
\end{equation}
where $\tau=M^2/s$. Usually, the scale is chosen as $\mu^2=M^2$.
The parton-parton luminosities $\mathcal{L}_{\gamma\gamma},\mathcal{L}_{\gamma\Sigma_e},\mathcal{L}_{\gamma g}$ at a 13 TeV $pp$ collider are shown in Fig.~\ref{fig:lumi}.
First, we examine the CT18lux parton luminosities separately calculated from the elastic and inelastic contributions to the photon PDF compared with the total photon PDF in the upper left of Fig.~\ref{fig:lumi}. We see that the elastic photon makes a sizeable contribution in both the low- and high-$M$ limits, due to the relatively large size of the elastic photon at lower scales as well as in the large-$x$ region.
We also compare parton-parton luminosities based on the total photon PDF input but using different existing PDF frameworks in Fig.~\ref{fig:lumi}. In general, we find CT18lux agrees quite well with LUXqed17 and NNPDF3.1luxQED, whereas CT18qed gives comparatively smaller parton luminosities and MMHT2015qed is somewhat larger. At higher invariant masses, $M\!>\!1$ TeV, CT18qed and MMHT2015qed give smaller values of $\mathcal{L}_{\gamma\gamma}$, while NNPDF3.1luxQED is larger, a feature we trace to the difference between the low and high values for the initialization scales in the DGLAP vs.~LUX approaches, respectively. In comparison, MMHT2015qed yields a larger $\mathcal{L}_{\gamma g}$ while NNPDF3.1luxQED is smaller, as a result of the large-$x$ gluon behaviors.

In the following, we examine the impact of the photon PDF upon collider phenomenology as represented by a number of Standard Model processes. For these, we take the production of high-mass Drell-Yan and $W^+W^-$ pairs, Higgs-associated $W^+$ production, and $t\bar{t}$ pair production as typical examples sensitive to $\mathcal{L}_{\gamma\gamma}, \mathcal{L}_{\gamma\Sigma_e}$(or $\mathcal{L}_{\gamma\Sigma}$), and $\mathcal{L}_{\gamma g}$. 

\subsection{High-mass Drell-Yan production}

\begin{figure}
	\centering
	\includegraphics[width=0.4\textwidth]{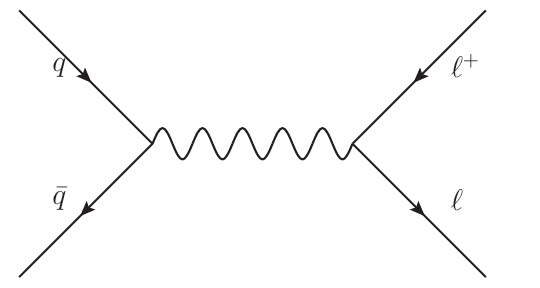}
	\includegraphics[width=0.4\textwidth]{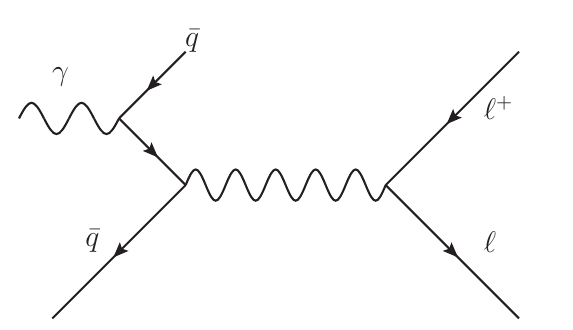}
	\includegraphics[width=0.3\textwidth]{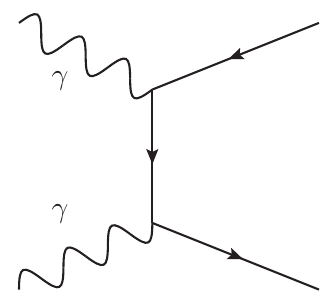}
	\includegraphics[width=0.3\textwidth]{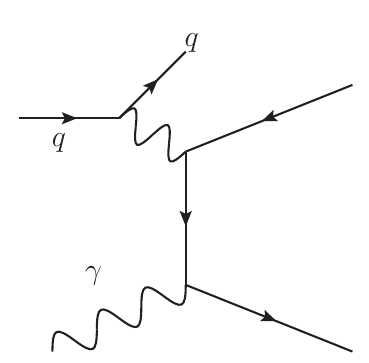}	
	\caption{The representative Feynman diagrams for the Drell-Yan lepton pair production through the leading order QCD, single and double photon initiated processes.}
	\label{feyn:DY}
\end{figure}
\begin{figure}
	\centering
	\includegraphics[width=0.49\textwidth]{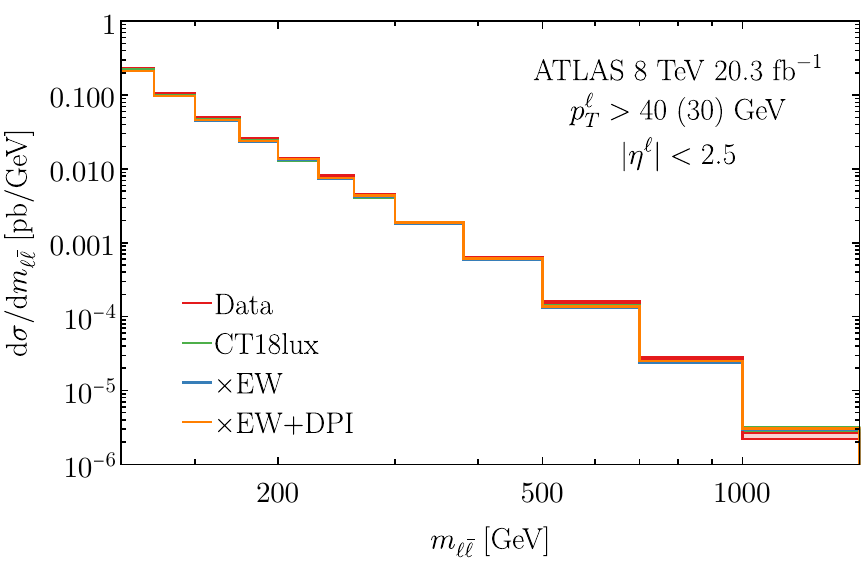}
	\includegraphics[width=0.49\textwidth]{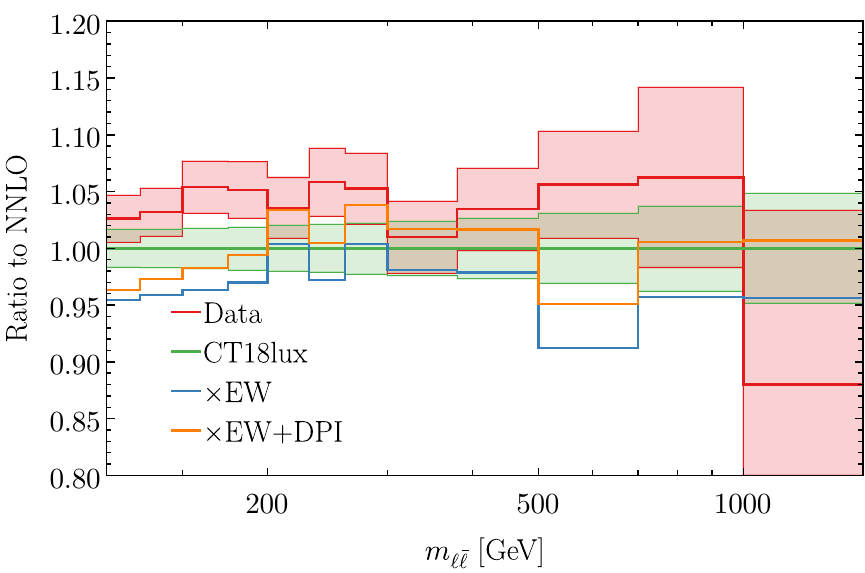}
	\includegraphics[width=0.49\textwidth]{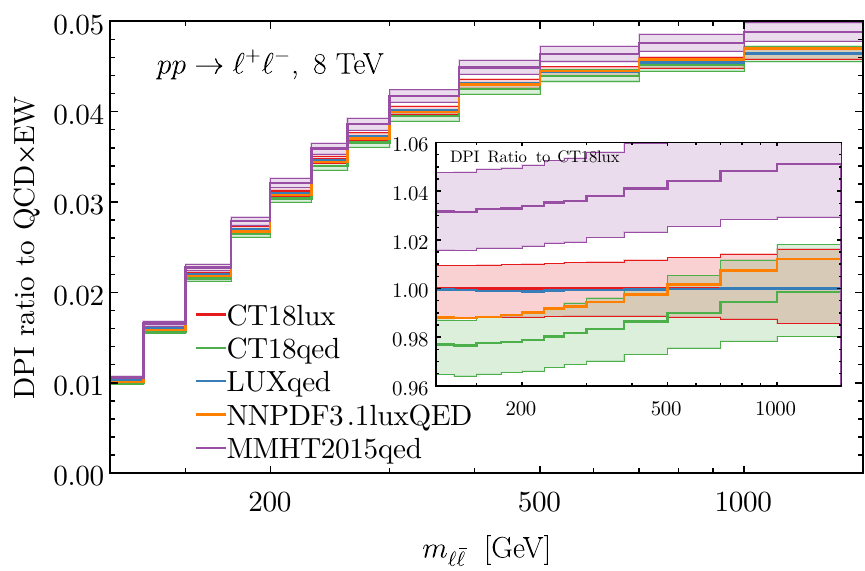}
	\includegraphics[width=0.49\textwidth]{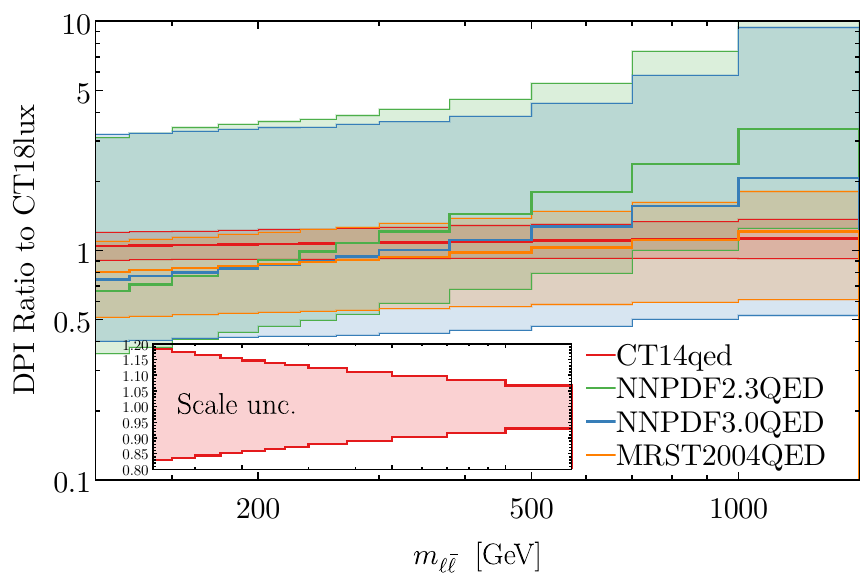}	
	\caption{The NLO EW corrections and double-photon-initiated (DPI) contribution with reference to the NNLO QCD calculation for the ATLAS 8 TeV very high-mass Drell-Yan production.}
	\label{fig:ATL8DYhiM}
\end{figure}

We start with high-mass Drell-Yan production, which has been extensively measured by both the CMS \cite{Sirunyan:2018owv} and ATLAS \cite{Aad:2016zzw} experiments at the LHC. Representative Feynman diagrams for this process are shown in Fig.~\ref{feyn:DY}.
In our theoretical predictions, we generally assume the ATLAS 8 TeV fiducial cuts~\cite{Aad:2016zzw},
\begin{equation}
\label{eq:DYfid}
p_{T}^{\ell}>40~(30)~\GeV,\ \ \  ~|\eta_\ell|<2.5,\ \ \  ~M_{\ell\bar{\ell}}>116~\GeV, 
\end{equation}
in order to directly compare with data as illustrated below. The quantity appearing inside paraentheses in Eq.~(\ref{eq:DYfid}) indicates the transverse momentum cut on sub-leading leptons. The NNLO QCD and NLO EW\footnote{The NLO EW corrections include the single-photon-initiated (SPI) processes, $\gamma q\to \ell^+\ell^- j$ in Fig. \ref{feyn:DY}, middle diagram, which are estimated with the LUXqed \cite{Manohar:2016nzj,Manohar:2017eqh} in the CT18 global analysis \cite{Hou:2019efy}.} corrections are already examined in the CT18 NNLO global analysis as documented in Ref.~\cite{Hou:2019efy}. We repeat this same computation here, but now with single and double photon initiated contribution(s) --- {\it i.e.}, processes involving ligh-by-quark and light-by-light scattering as shown in Fig.~\ref{feyn:DY}---  updated according to the various existing photon PDF calculations.
The absolute differential cross section for production of Drell-Yan pairs, $\dd\sigma/\dd m_{\ell\bar{\ell}}$, is plotted in the upper left of Fig.~\ref{fig:ATL8DYhiM}. In particular, we observe that, with increasing di-lepton invariant masses over the range $116\!<\!m_{\ell\bar{\ell}}\!<\!1500$ GeV, the absolute cross section monotonically decreases through $\sim$5 orders-of-magnitude.
To better illustrate variations in the cross section of the electroweak corrections, we therefore normalize $\dd\sigma/\dd m_{\ell\bar{\ell}}$ to the NNLO QCD calculation as shown in Fig.~\ref{fig:ATL8DYhiM}, upper right. In this context, we see that NLO EW corrections are responsibly for an approximate $\sim$5\% negative effect around $m_{\ell\bar{\ell}}\sim1$ TeV. The inclusion of double-photon-initiated (DPI) processes, on the other hand, largely counteracts this effect, increasing the absolute cross section by a similar, $\sim$5\% shift within this $m_{\ell\bar{\ell}}$ region.

Next, we compare the purely DPI cross sections obtained with different assumed photon PDFs in Fig.~\ref{fig:ATL8DYhiM}, lower plots. As mentioned before, CT18qed refers to the DGLAP-driven calculation using an initialization scale of $\mu_0=3$ GeV. In this case, we see the CT18qed result lies about 2\% below the CT18lux prediction, a fact which can be understood in terms of the behavior of the photon PDF itself: the DPI cross section goes approximately as $\sim\! 2\! \times\! \gamma(x)$, with the photon PDF being just under $\sim$1\% smaller in CT18qed relative to CT18lux, as shown in Fig.~\ref{fig:photon}. Meanwhile, CT18lux gives almost the same predictions as LUXqed17, while NNPDF3.1luxQED produces a slightly smaller DPI cross section at low $m_{\ell\bar{\ell}}$, but higher at large $m_{\ell\bar{\ell}}$; this can be attributed to the difference between the DGLAP evolution and the LUX approaches. The MMHT2015qed cross section is 3$\sim$5\% larger than that based on CT18lux.
We note also that the PDF uncertainty for CT18lux, CT18qed, and MMHT2015qed are roughly the same, lying within the range 1$\sim$2\%. 
In addition to these comparisons based on our CT18 QED results and other recent calculations, we also compare DPI cross sections based on the previous generation of photon PDFs, namely, MRST2004, NNPDF2.3, NNPDF3.0, and CT14qed, which we plot in Fig.~\ref{fig:ATL8DYhiM}, lower right, normalized to CT18lux as a reference. The central predictions of CT14qed give quite strong agreement with CT18lux, while the uncertainty is about 20\%. The MRST2004qed gives slightly larger uncertainty, which is about 30$\sim$50\%. The NNPDF2.3 and NNPDF3.0 calculations give significantly larger predictions at higher invariant mass, and the size of uncertainty bands can be as large as 100$\sim$200\% in this latter case.

We remind the reader that a major purpose of this phenomenological study is an exploration of the photon PDF uncertainty. In order to isolate the photon's impact, we focus on double-photon-initiated (DPI) dilepton production and the theoretical calculation is performed with leading-order matrix elements for which the representative diagram is the third graph of Fig.~\ref{feyn:DY}. It was pointed out in Ref.~\cite{Carrazza:2020gss} that this DPI process suffers from a large scale uncertainty, which is found to be $\sim\!\! 10\!-\!20\%$, shown as the inset in the lower-right panel of Fig.~\ref{fig:ATL8DYhiM}. Starting at next-to-leading order, we begin to have contributions from the single-photon-initiated process ({\it i.e.}, $q\gamma$ scattering, as shown in the last diagram of Fig.~\ref{feyn:DY}), which has an overlap with the NLO EW diagrams for Drell-Yan production ({\it e.g.}, the second diagram of Fig.~\ref{feyn:DY}). A systematic treatment to include these dynamics simultaneously at this order requires a proper subtraction to avoid double-counting, which is beyond the scope of this work. Instead of taking a partonic factorization picture, it is also demonstrated in Refs.~\cite{Harland-Lang:2019eai,Harland-Lang:2021zvr} that the photon-initiated processes can be formulated using the structure function approach, which avoids these artificial scale uncertainties and provides a precise definition of the proton's photon content.

\subsection{$WH$ production}
\begin{figure}
	\centering
	\includegraphics[width=0.6\textwidth]{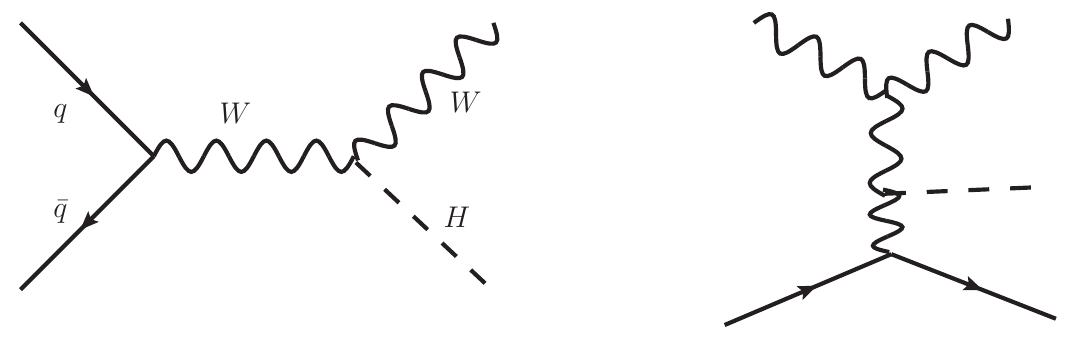}
	\caption{Representative Feynman diagrams for $W$ boson associated with Higgs production.}
	\label{feyn:WH}
\end{figure}

\begin{figure}
	\centering
	\includegraphics[width=0.49\textwidth]{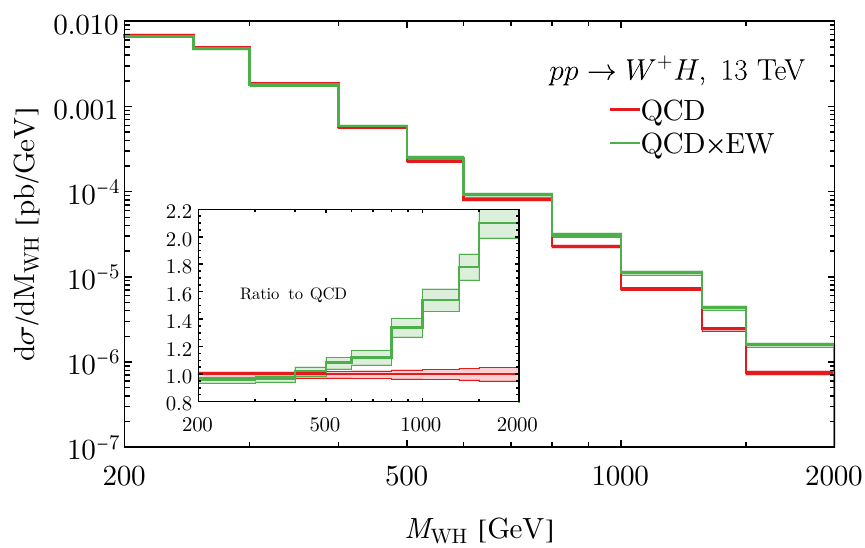}
	\includegraphics[width=0.49\textwidth]{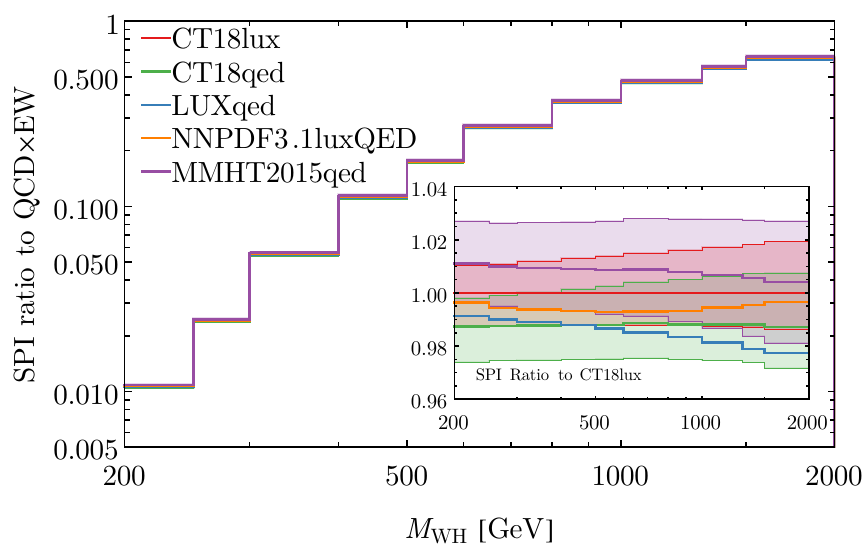}
	\includegraphics[width=0.49\textwidth]{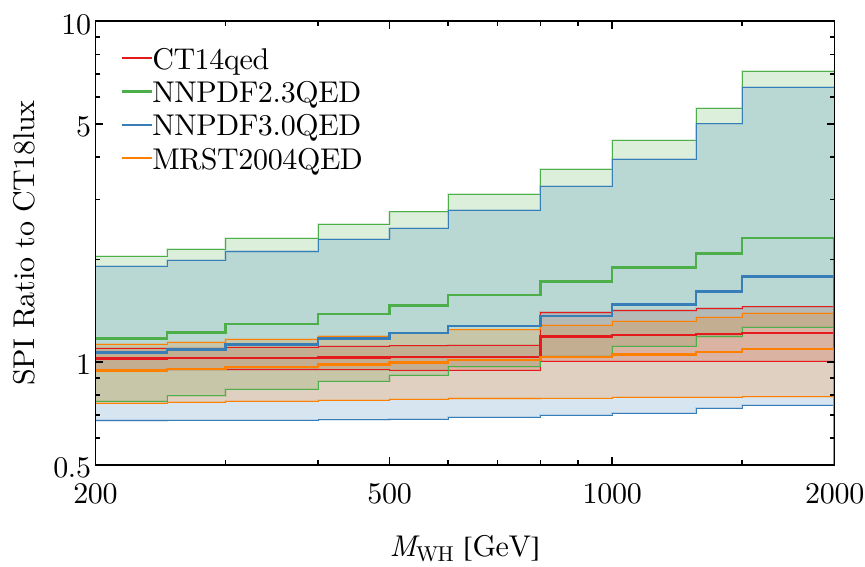}	
	\caption{The $W^+H$ production at a 13 TeV $pp$ collider.}
	\label{fig:WH}
\end{figure}

At $pp$ colliders with sufficient $\sqrt{s}$, $W$-boson-associated Higgs ({\it i.e.}, $WH$) production can proceed through a Drell-Yan-like mechanism mediated by $W$-boson exchange, as depicted in the diagram shown in Fig.~\ref{feyn:WH}, upper left. At one higher EW (or QED) order, we can also have contributions from SPI processes like that appearing in Fig.~\ref{feyn:WH}, right diagram.
The QCD calculation for $WH$ production can be achieved at NNLO with MCFM \cite{Campbell:2019dru} or \textsc{vh@nnlo}~\cite{Harlander:2018yio}, while the EW corrections considered here can be performed by means of the HAWK package \cite{Denner:2011id} or the general-purpose generator MadGraph\_aMC@NLO \cite{Frederix:2018nkq}.
Here, we consider the total inclusive\footnote{By ``total inclusive,'' we mean the full phase space without any fiducial cuts; the $W$ and $H$ bosons are assumed to be on-shell final states without subsequent decay.} cross section for $W^+H$ production at a 13 TeV $pp$ collider as a demonstration.  Following these considerations, we plot the absolute differential cross section, $\dd\sigma/\dd M_{WH}$, in Fig.~\ref{fig:WH}, upper left, performed here at NNLO in pQCD and with NLO EW effects based on the CT18lux PDF. We observe that the absolute differential cross section drops as drastically as by four magnitudes when the $WH$ invariant mass increases up to 2 TeV. At large invariant mass, the size of NLO EW correction can be as significant as $\calO(1)$ compared with the pure NNLO QCD prediction. 

In order to examine the importance of the photon-initiated contribution, we show the ratio of the SPI cross section to the total one, \emph{i.e.}, NNLO QCD and NLO EW, in Fig.~\ref{fig:WH}, upper right. We see that right above the threshold around $M_{WH}\sim$200 GeV, the SPI processes only contribute about 1\% to the total cross section. In contrast, when the $M_{WH}$ increases up to 2 TeV, the SPI contribution becomes 60\% --- exceeding even the pure QCD cross section --- which highlights the importance of the photon contribution. In addition, we also show the PDF uncertainty, which is about 2\%, significantly reduced when compared with the first generation ones shown in Fig.~\ref{fig:WH}, lower plot.

\subsection{$W^+W^-$ pair production}
\begin{figure}
	\centering
	\includegraphics[width=0.6\textwidth]{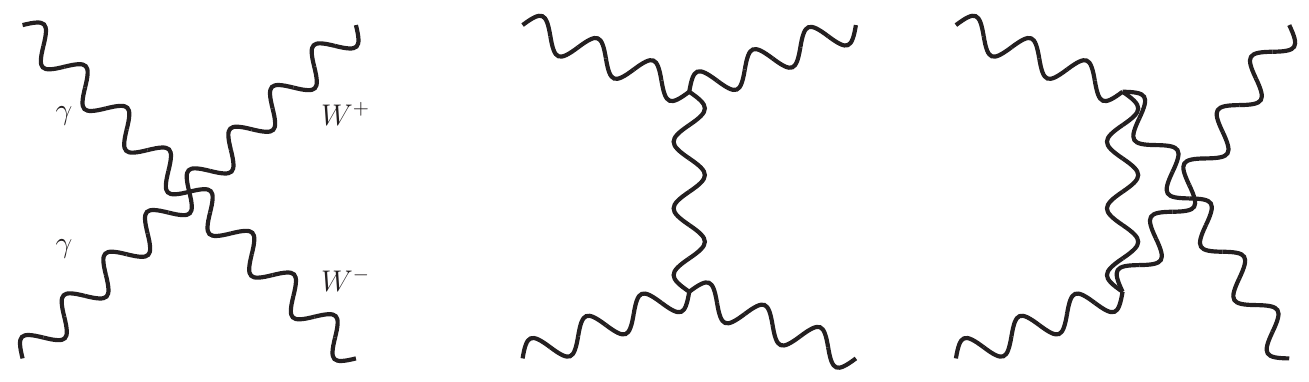}
	\caption{The Feynman diagrams for exclusive $W^+W^-$ boson pair production.}
	\label{feyn:WW}
\end{figure}

The total cross sections of exclusive or quasi-exclusive $W^+W^-$ production at the LHC via $pp\to p^{(*)}W^+W^-p^{(*)}\to p^{(*)}\mu^\pm e^{\mp}p^{(*)}$ has been measured by both the CMS~\cite{Chatrchyan:2013foa,Khachatryan:2016mud} and ATLAS~\cite{Aaboud:2016dkv} experiments.
The experimental cross sections are corrected to the full phase space, with results summarized in Tab.~\ref{tab:WW}.
The theoretical predictions are evaluated as 
\begin{eqnarray}
\sigma(pp\to p^{(*)}\mu^{\pm}e^{\mp}p^{(*)})=F\sigma(pp\to pW^+W^-p) \times {\rm BR}(W^+W^-\to \mu^{\pm}e^{\mp}X)\ \, ,
\end{eqnarray}
where $pp\to pW^+W^-p$ represents the elastic scattering process in which the proton remains intact. The Monte-Carlo simulated cross sections are taken as $\gamma\gamma\to W^+W^-\to \mu^{\pm}e^{\mp}X$ with photon ($\gamma$) distribution function being calculated from the equivalent photon approximation (EPA) \cite{Budnev:1974de}.\footnote{CMS takes CalcHEP and MadGraph, respectively, for the predictions of 7 TeV and 8TeV, while ATLAS adopts Herwig++, resulting in a smaller cross section due to different implementations of EPA. Herwig++ uses the integration of the dipole form factors \cite{Gieseke:2011na}, while both MadGraph and CalcHEP directly implement the approximated form \cite{Budnev:1974de}.}  The measured values and theoretical predictions are listed in the second and third column, respectively, of Tab. \ref{tab:WW}. The branching ratio of $W^+W^-$ pairs decaying into $\mu^{\pm}e^{\mp}X$, including $\tau$ leptonic decays was taken as BR=3.23\% \cite{Agashe:2014kda}.\footnote{In Ref. \cite{Beringer:1900zz}, the branch ratio of $W^+W^-\to\mu^{\pm}e^{\mp}X$ was taken as a slightly different value, BR=3.1\%.} The dissociation factor, $F$, reflects the effect of including also the ``quasi-exclusive" or ``proton dissociation" production contribution, was extracted from the CMS and ATLAS data of high-mass lepton-pair production $\gamma\gamma\to\ell^+\ell^-$. 
We note that, as found in Ref.~\cite{Harland-Lang:2016apc}, underlying event contributions from subsequent rescatterings of
the intact proton are expected to be small in the exclusive, non-dissociative channel and are not
explicitly calculated here.
With the $F$ and BR, we can correct the Monte-Carlo simulated EPA prediction to the exclusive cross section at the $W^+W^-$ undecayed level, listed as the EPA (fifth) column in Tab. \ref{tab:WW}. 

\begin{table}
	\hspace{-1cm}\begin{tabular}{c|c|c|c|c|c|c}
		\hline
		\multirow{2}{*}{Experiment} & \multirow{2}{*}{Data} & EPA theory & Dissociation &\multicolumn{3}{c}{Exclusive $pp\to pW^+W^-p$} \\
		& & prediction  &  factor $F$ & EPA & CT18lux & MMHT2015qed \\
		\hline
		CMS 7 TeV~\cite{Chatrchyan:2013foa} & $2.2^{+3.3}_{-2.0}$ &  $4.0\pm0.7$ & $3.23\pm0.53$ & $38\pm9$  
		& 33.7($\pm0.96\%$) & 36.8($\pm0.52\%$) \\
		CMS 8 TeV~\cite{Khachatryan:2016mud} & $10.8^{+5.1}_{-4.1}$ & $6.2\pm0.5$ & $4.10\pm0.43$ & $47\pm6$ & \multirow{2}{*}{40.9($\pm0.93\%$)} 
		& \multirow{2}{*}{44.8($\pm0.49\%$)} \\
		ATLAS 8 TeV~\cite{Aaboud:2016dkv} & $6.9\pm2.6$ & $4.4\pm0.3$ & $3.30\pm0.23$ & $41\pm4$ &\\
		\hline
	\end{tabular}
	\caption{The exclusive/quasi-exclusive $pp\to p^{(*)}W^+W^-p^{(*)}\to p^{(*)}\mu^\pm e^{\mp}p^{(*)}$ production cross sections $\sigma$ [fb] measured by CMS \cite{Chatrchyan:2013foa,Khachatryan:2016mud} and ATLAS \cite{Aaboud:2016dkv} groups.
	}
	\label{tab:WW}
\end{table}

\begin{figure}
	\centering
	\includegraphics[width=0.65\textwidth]{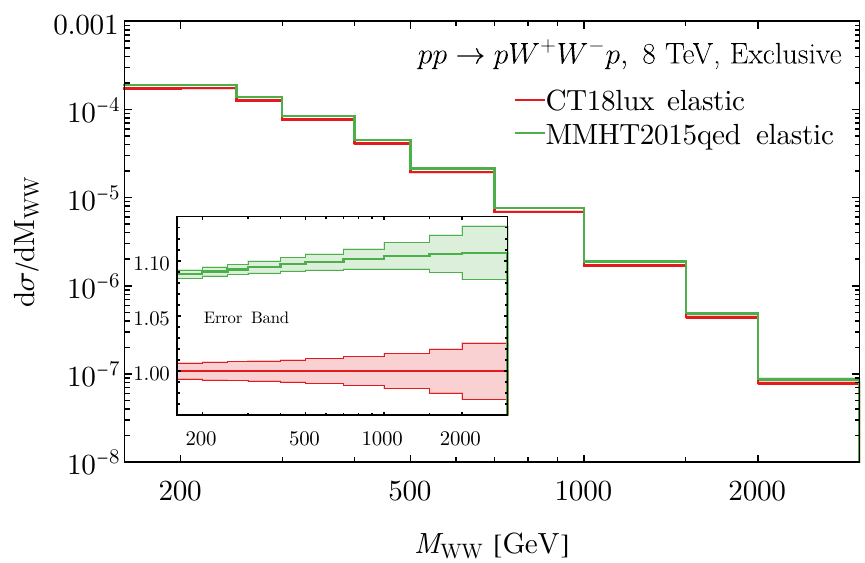}
	\caption{The exclusive $pp\to pW^+W^-p$ production at a 8 TeV $pp$ collider.}
	\label{fig:WW}
\end{figure}

The implications of the CMS (quasi-) exclusive $W^+W^-$ production data were investigated in Ref.~\cite{Ababekri:2016kkj}, in which the CMS data was found to be in good agreement with a theory prediction including both elastic and single-dissociative contributions predicted by CT14QED and CT14QEDinc PDFs~\cite{Schmidt:2015zda}.   
Here, we compare in Tab.~\ref{tab:WW} the predictions of elastic photons in CT18lux (same as CT18qed) and MMHT2015qed to CMS and ATLAS data on the exclusive $\gamma\gamma\to W^+W^-$ production cross section to the EPA predictions, as shown in the last two columns of the table.   
We find that MMHT2015qed yields an enlarged (relative to CT18lux) cross section due to its larger elastic photon, as shown in Fig.~\ref{fig:elastic}. We also quantify the uncertainty due to potential variations in the elastic photon contribution as discussed in Sec.~\ref{sec:nonDIS}, with MMHT2015qed giving a slightly smaller uncertainty.
%because of its smaller elastic photon variation shown in Fig.~\ref{fig:ElasticVar}.
In addition, we also show the invariant mass distribution, $\dd\sigma/\dd M_{WW}$, for exclusive $W^+W^-$ production in Fig.~\ref{fig:WW}. At larger invariant masses, the elastic photon of MMHT2015qed results in a similar uncertainty as CT18lux, although the central value is about 10\% larger.

Recently, the ATLAS group released a new measurement of this process based on 139 fb$^{-1}$ of 13 TeV LHC data~\cite{ATLAS:2020iwi}, which rejects the background-only hypothesis with a significance of 8.4 standard deviations. Unlike the 7 and 8 TeV measurements, however, this analysis was performed in the fiducial volume. An
investigation of the implications of these data is deferred to a future study.

\subsection{Top-quark pair production}
\begin{figure}
	\centering
	\includegraphics[width=0.5\textwidth]{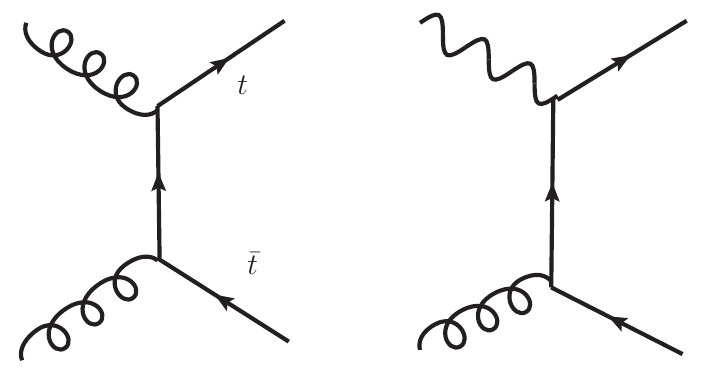}
	\caption{Representative Feynman diagrams for leading order QCD and single photon initiated production of $t\bar{t}$ pair.}
	\label{feyn:ttx}
\end{figure}

\begin{figure}
	\centering
	\includegraphics[width=0.49\textwidth]{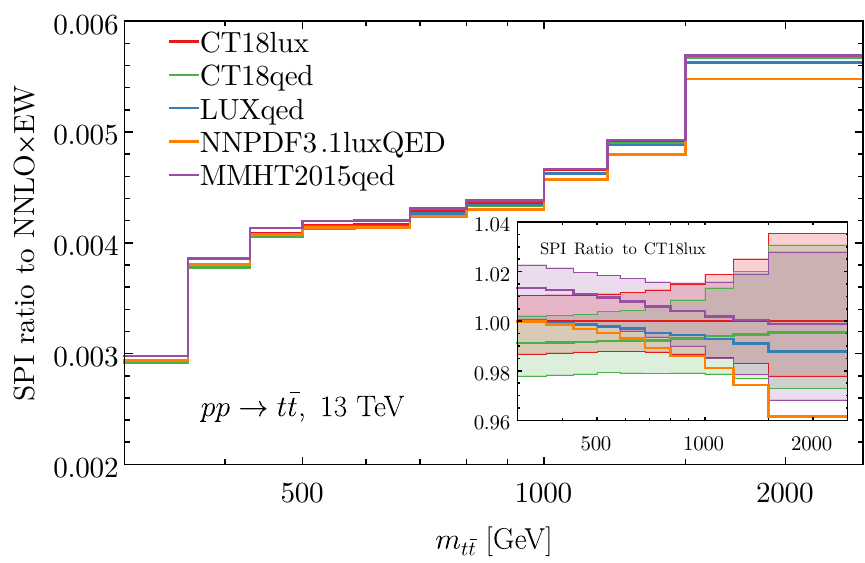}
	\includegraphics[width=0.49\textwidth]{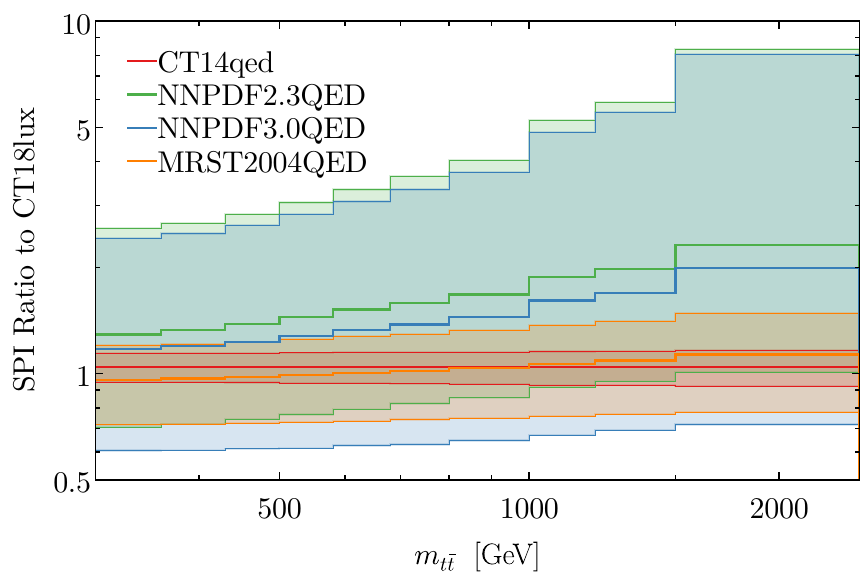}	
	\caption{Exclusive $t\bar{t}$ production at a 13 TeV $pp$ collider.}
	\label{fig:ttx}
\end{figure}

Finally, we explore the impact of the photon PDF on $t\bar{t}$ production. Representative diagrams for $t\bar{t}$ production proceeding through the leading order QCD channels (\emph{e.g.}, gluon fusion) and single-photon-initiated processes (\emph{e.g.}, photon-gluon fusion) are shown in the left and right panels of Fig.~\ref{feyn:ttx}, respectively. As before, we consider $t\bar{t}$ production at a 13 TeV $pp$ machine for specificity, with the inclusion of NNLO QCD and NLO EW (QCD$\times$EW) corrections as calculated in Ref.~\cite{Czakon:2017wor}. In this work, we examine the SPI cross section, whose ratio to the QCD$\times$EW cross section is shown in Fig.~\ref{fig:ttx}. We see by its relative size that the SPI process only contributes less than 0.6\% of the full QCD$\times$EW cross section. In this sense, the SPI contribution is safely ignorable in $t\bar{t}$ production from the perspective of contemporary precision, unlike what we observed for high-mass Drell-Yan and $W$-boson-associated Higgs production. The SPI uncertainties of CT18lux, CT18qed, and MMHT2015qed are roughly similar and about $\sim$3\% in this range, constituting a significant reduction when compared with the first generation of photon PDFs shown in Fig.~\ref{fig:ttx}, right plot. 
Needless to say that the contribution of photon-photon fusion to the production of top-quark pairs at the 13 TeV LHC is 
negligible. 

% -  -  -  -  -  -  -  -  -  -  -  -  -  -  -  -  -  -  -  -  -  -  -  -  -  -  -  -  -  -  -  -  -  -  -  -  -  -  -  -  -  -  -

\section{Conclusion}
\label{sec:conc}
In the present analysis, we have carried out a first implementation of the recent LUX QED formalism
into the larger CT PDF analysis framework.
%
%
% [observation on the essential features of LUX; motivation for the analysis]
%
The LUX approach represented a significant advance in the consistent determination of the photon
content of the proton with minimal underlying model assumptions. The LUX QED methodology was subsequently
interfaced with several QCD global analysis frameworks, each making unique choices regarding
implementation strategy and inclusion of physics ingredients in the ultimate calculation.
For this reason, as well as the recent development of QED splitting kernels at $\mathcal{O}(\alpha \alpha_s)$
and $\mathcal{O}(\alpha^2)$, it is time to update the CT analysis family with the dedicated QED study
presented here.
%
% [resulted in two primary photon PDF determinations]
%
In particular, in this study, we examined possible systematic differences that arise between
following each of the two main approaches carried out in recent LUX-based photon PDF calculations. Broadly,
these consist either of computing $\gamma(x,\mu^2)$ according to the LUX master expression in
Eq.~(\ref{eq:LUX}) at an arbitrary scale often chosen to be $\mu\!\sim\!100$ GeV for iteration
in a global fit, \emph{e.g.}, NNPDF3.1luxQED \cite{Bertone:2017bme}, 
%\carl{(I'm confused by the description of the first approach: Doesn't the Lux method use the formula to calculate the photon PDF at all scales?)};
or of instead evaluating the photon PDF at an initial scale, $\gamma(x,\mu_0^2)$, by
applying a slight modification of the LUX master formula and relying entirely on QED+QCD evolution
to determined the photon PDF at $\mu\!>\mu_0$, \emph{e.g.}, MMHT2015qed \cite{Harland-Lang:2019pla}.
%
% [CT18qed --- first among these: consistent use of QCD+QED evolution]
%
We adapted the latter of these two approaches to our CT methodology, leading to the CT18qed
photon PDF set, which we regard as the primary result of this analysis.
%
% [CT18lux]
%
For the sake of comparison, we also implement the former approach, in which the photon PDF is
everywhere computed according to the unmodified LUX master formula, designating the result CT18lux.
We release this alternative calculation alongside CT18qed, and we have compared the two against other recent
determinations in the present work.

%
% [unique features of this work; comprehensive treatment of uncertainties; low-energy sources of error]
%
Together, the CT18qed and CT18lux photon PDFs we have produced involve several novel and unique features,
in addition to the fact that we have considered both approaches comprehensively within
a single framework. An important aspect of the CT18 photon PDFs is the updated final uncertainty
we report, which follows from a critical appraisal of possible error sources following a number
of physics updates that have occurred since the original LUX publication. These include a reassessment
of the elastic form factor uncertainty, possible higher-twist and target-mass effects, and updates
to the description of the proton structure function inside the resonance region as canvassed in
Sec.~\ref{sec:uncSource}.
%
% [Qualitatative recap of the resulting PDFs.]
% [Other aspects: description of data, refitting; uncertainty; comparison to other groups.]
%
While we compared our CT18qed and CT18lux PDFs extensively in Sec.~\ref{sec:CT18lux} and \ref{sec:ct18qed}, a number of qualitative features are
worth highlighting. In particular, we obtain a strong general agreement among the calculations most closely aligned
with the original LUX approach, as can be seen in Fig.~\ref{fig:photon}, which illustrates the close similarity of CT18lux
relative to the LUXqed17 and NNPDF3.1luxQED results. The comparative photon PDF uncertainties among these calculations show some mild differences in Fig.~\ref{fig:norm_unc}, especially at low $x$, for which our CT18lux calculation is essentially intermediate between the smaller NNPDF3.1luxQED band and the larger low-$x$
uncertainty reported in LUXqed17 \cite{Manohar:2017eqh}. In the CT incarnation of the DGLAP approach, we observe some
intriguing differences in the shape and magnitude of our final PDFs, with our final pair of CT18qed PDFs, with initialization scale $\mu_0=1.3$ GeV and 3 GeV, somewhat underhanging CT18lux, for example, especially for $x\!>\!10^{-3}$. While
we again refer interested readers to Sec.~\ref{sec:CT18lux} and \ref{sec:ct18qed} for an in-depth dissection, we point
out that this behavior goes in much the opposite direction compared with that of MMHT2015qed.
%

% phenomenological implications
%
In Sec.~\ref{sec:pheno}, we traced a number of phenomenological consequences of the CT18 photon
PDFs, with a particular emphasis on $pp$ collisions at the LHC. We presented TeV-scale parton-parton luminosities,
distributions for high-mass Drell-Yan production, as well as $WH$, $W^+W^-$, and $t\bar{t}$ cross sections.
For the parton-parton luminosities shown in Fig.~\ref{fig:lumi}, we found generally robust concordance, up to uncertainties, among the various
calculations explored in this analysis --- albeit with some evidence of deviation in $\mathcal{L}_{\gamma\gamma}$
at $M\!>\! 10$ GeV, especially for MMHT2015qed, and, to a lesser extent, CT18qed. 
Similarly, we report reasonable agreement among the phenomenological calculations of the invariant-mass distributions
computed for high-mass Drell-Yan, $WH$, $W^+W^-$, and $t\bar{t}$, production, again finding a frequent excess, especially
for MMHT2015qed, as follows from its comparatively larger photon PDF. Of phenomenological importance, the PDF uncertainties in these processes can still be significant and are linked to the many nonperturbative error sources explored in this
study. Achieving the elevated precision of the perturbative calculation (to percent-level accuracy) needed for Beyond Standard Model (BSM) searches in, {\it e.g.}, the tails of invariant-mass distributions will therefore require improvements 
%\kp{up to the percent-level in order to single out} 
to these lower-energy inputs to the photon PDF calculation.
%
% [closeout --- availability of PDF sets on LHAPDF]
%

As a companion to this article, we publicly release LHAPDF6 \cite{Buckley:2014ana} grids corresponding
to the two main calculations presented above: CT18lux and CT18qed (including CT18qed1.3GeV as well) in the HEPForge repository, \url{https://ct.hepforge.org}.
Again, we stress
that these grids include both our newly-calculated photon PDF within each approach
as well as the accompanying (anti-)quark and gluon distributions, with uncertainties quantified
according to the Hessian approach combined with low-$Q^2$ resources as Eq.~(\ref{eq:unc}). These uncertainties represent those associated with
the underlying quark and gluon degrees-of-freedom, as well as the collection of
uncertainty sources reviewed in Sec.~\ref{sec:uncSource}.
%
% [recommendations for use]
%
As pointed out above, we identify our CT18qed (with $\mu_0=3$ GeV) calculation as a `first among
equals' result and advocate its primary use in phenomenological calculations
like those shown in Sec.~\ref{sec:pheno}. The DGLAP evolution of photon simultaneously with quark and gluon PDFs provides a consistent description of the photon-initiated processes together with the contribution of quark (and possibly gluon as well) partons in the perturbative expansion. As an alternative, the CT18qed1.3GeV PDFs, in which the photon PDF is initialized at the CT18 starting scale of $\mu_0=1.3$ GeV, are more appropriate for describing the photon in the low-energy range $1.3<\mu<3$ GeV but give a larger uncertainty at large $x$ values.
The remaining set that we have released simultaneously, CT18lux, uses a method that provides a particularly useful determination of the inclusive photon.

%
% [limitations; possible future directions]
%
Finally, we remind the reader that a number of aspects of this study, among
them, more detailed investigation of parton-level charge-symmetry breaking
\cite{Hobbs:2011vy}
and simultaneous determinations of the neutron's photon content, leave
room for further improvement. Rather than undertaking these in the analysis
above, we reserve these issues for future work(s), for which such considerations
maybe relevant to achieving still higher electroweak precision in
next-generation phenomenology.

% -  -  -  -  -  -  -  -  -  -  -  -  -  -  -  -  -  -  -  -  -  -  -  -  -  -  -  -  -  -  -  -  -  -  -  -  -  -  -  -  -  -  -

\acknowledgments
We thank Eric Christy for providing updated code to compute the low-$W^2$ resonance-region structure functions, as well as for valuable conversations. We also thank our the CTEQ-TEA colleagues for helpful discussions and Sergei Kulagin for useful exchanges. 
The work at MSU is partially supported by the U.S.~National Science Foundation
under Grant No.~PHY-2013791.
The work of T.~J.~Hobbs was supported at SMU by the U.S.~Department of
Energy under Grant No.~DE-SC0010129 as well as by a JLab EIC Center Fellowship; support
was also provided by
the Fermi National Accelerator Laboratory, managed and operated by Fermi Research Alliance, LLC under Contract No.~DE-AC02-07CH11359 with the U.S.~Department of Energy.
The work of K.~Xie was supported in part 
by the U.S.~Department of Energy under
grant No.~DE-FG02-95ER40896, U.S.~National Science Foundation under Grant No.~PHY-1820760, and in part by the PITT PACC.
The work of M.~Yan is supported by the National Science Foundation of China under Grant Nos. 11725520, 11675002, and 11635001.
C.-P.~Yuan is also grateful for the support from
the Wu-Ki Tung endowed chair in particle physics.

\appendix
\section*{Appendix}
% -  -  -  -  -  -  -  -  -  -  -  -  -  -  -  -  -  -  -  -  -  -  -  -  -  -  -  -  -  -  -  -  -  -  -  -  -  -  -  -  -  -  -
\section{Separation of elastic and inelastic photon PDF components}
\label{app:sep_inel}
\begin{figure}\centering
	\includegraphics[width=0.50\textwidth]{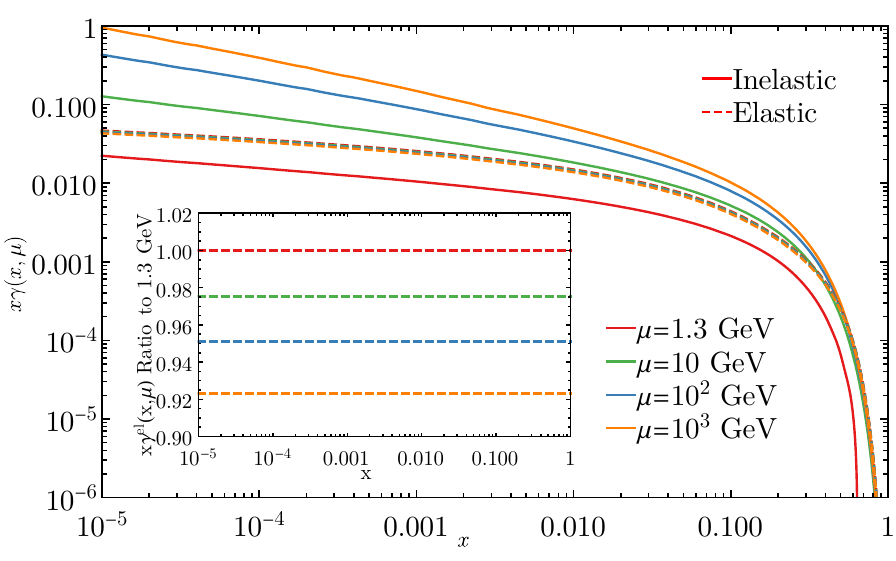}
	\includegraphics[width=0.48\textwidth]{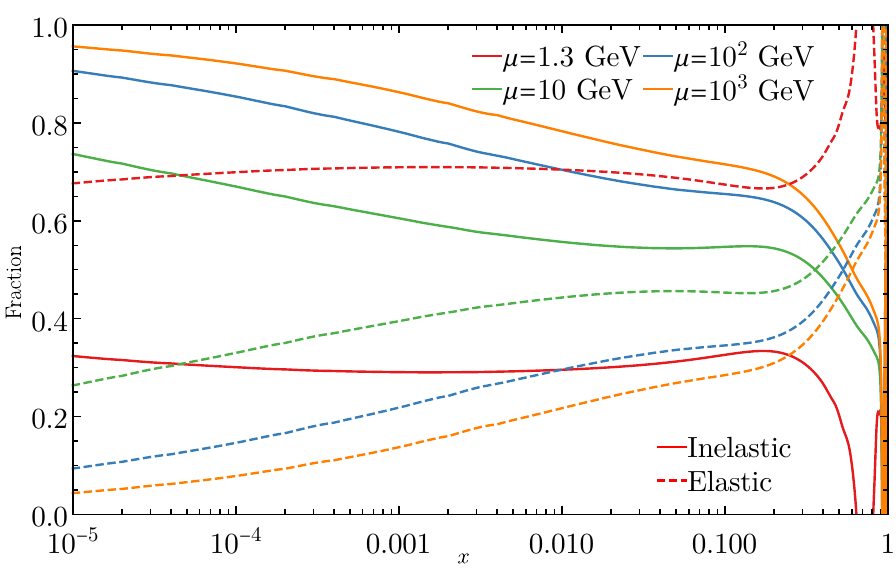}
	\caption{The CT18lux elastic and inelastic photon PDFs at $Q=1.3,10,10^2$, and $10^{3}$ GeV (left) and the corresponding fractional contribution of the (in)elastic component to the total photon PDF (right).
	}
	\label{fig:CT18lux}
\end{figure}

As discussed in Sec.~\ref{sec:intro} and~\ref{sec:LUX2DGLAP}, the photon PDF consists of elastic and inelastic components. To illustrate, we show the absolute values and corresponding PDF fractions of these two components for CT18lux at a number of scale choices\footnote{We note that the default starting scale for the CT18 NNLO PDFs is the first of these, $\mu_0=1.3$ GeV.}, namely, $\mu=1.3$, $10$, $10^{2}$, $10^{3}$ GeV in left and right plot of Fig.~\ref{fig:CT18lux}, respectively. At low energy, \emph{e.g.}, $\mu_0=1.3~\GeV$, the inelastic photon is mainly determined by the low-$Q^2$ structure functions $F_{2,L}$ directly measured in low-energy experiments, such as HERMES \cite{Airapetian:2011nu} in the continuum region and CLAS \cite{Osipenko:2003bu} (or Christy-Bosted \cite{Christy:2007ve}) in the resonance region, as illustrated in Fig.~\ref{fig:xQ2plane}. As the scale increases, the inelastic photon receives a significant contribution from the quark splitting, $q\to q\gamma$, with a $\log(\mu^2)$ enhancement. As a result, the absolute value of the inelastic photon PDF increases drastically, and, the corresponding fractional share of the inelastic photon PDF relative to the total comes to dominate at higher $\mu^2$, as shown in Fig.~\ref{fig:CT18lux}.

\begin{figure}[h]
	\centering
	\includegraphics[width=0.65\textwidth]{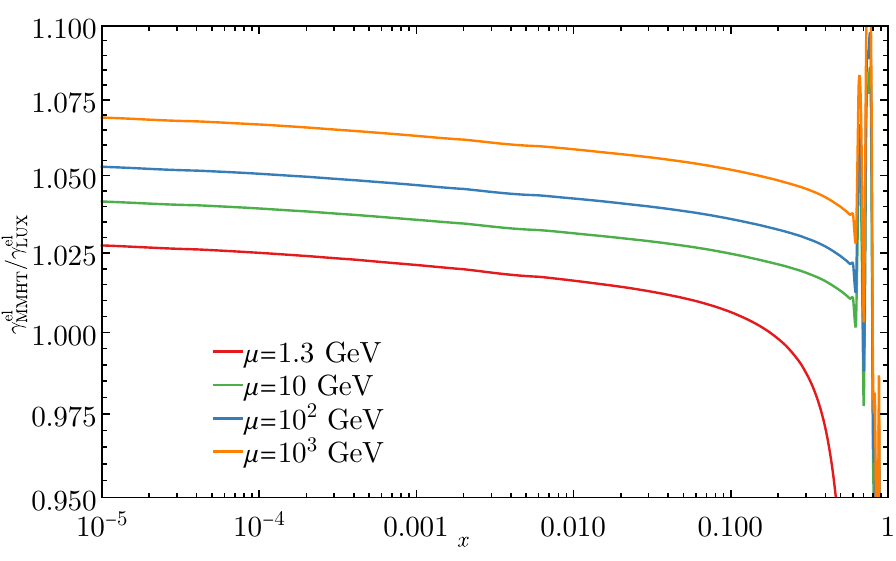}
	\caption{The ratio of elastic photon PDF of MMHT15qed and LUX (adopted in both CT18lux and CT18qed).}
	\label{fig:elastic}
\end{figure}

Similar to LUXqed(17), the elastic photon in CT18lux and CT18qed is fully determined by the elastic form factor, through
\bea
x\gamma^{\el}(x,\mu^2)=\frac{1}{2\pi\alpha(\mu^2)}\int_{\frac{x^2m_p^2}{1-x}}^{\infty}\frac{\dd Q^2}{Q^2}\alpha_{\rm ph}^2(-Q^2)
\Bigg[&\left(1-\frac{x^2m_p^2}{Q^2(1-x)}\right)\frac{2(1-x)G_E^2(Q^2)}{1+\tau}\\
&+\left(2-2x+x^2 +\frac{2x^2m_p^2}{Q^2}\right)\frac{G_M^2(Q^2)\tau}{1+\tau}\Bigg]\ ,
\label{eq:elastic}
\eea
where $\tau=Q^2/(4m_p^2)$. We notice that this elastic photon is different from the one in MMHT2015qed \cite{Harland-Lang:2019pla}. Rather than directly calculating $\gamma^{\el}$ through the LUX formalism, MMHT2015qed instead runs the elastic photon through the evolution equation,
\begin{equation}
\frac{\dd\gamma^{\el}}{\dd\log \mu^2}=p_{\gamma\gamma}\otimes \gamma^{\el}+\delta x\gamma^{\el}\ ,
\label{eq:EvEl}
\end{equation}
in which $\delta x\gamma^{\el}$ corresponds to the integrand in Eq.~(\ref{eq:elastic}) divided by $x$. The numerical difference between these two prescriptions is given in Fig.~\ref{fig:elastic}. Overall, MMHT2015qed gives a larger elastic photon than LUX (and therefore, CT18lux and CT18qed). Furthermore, the ratio increases along with the scale $\mu$. This is mainly a consequence of the fact that MMHT2015qed only includes quark contributions in $p_{\gamma\gamma}$, whereas LUX includes both quarks {\it and} leptons in the $\alpha(\mu^2)$ running. We also notice that, at large $x$ and small $\mu^2$, MMHT2015qed gives a smaller $\gamma^{\el}$. This is due to the equivalent upper integration limit, $\mu^2$, in the MMHT2015qed method, which is different from $\infty$ in Eq.~(\ref{eq:elastic}). Especially at large $x$, the lower limit, $x^2m_p^2/(1-x)$, will approach $\mu^2$, which leaves the integration over $[x^2m_p^2/(1-x),\mu^2]$ significantly smaller than the one over $[x^2m_p^2/(1-x),\infty]$. However, the effect of this relative difference in integration intervals becomes smaller with increasing $\mu^2$.

% -  -  -  -  -  -  -  -  -  -  -  -  -  -  -  -  -  -  -  -  -  -  -  -  -  -  -  -  -  -  -  -  -  -  -  -  -  -  -  -  -  -  -
\section{Physical factorization and the $\msbar$ conversion terms}
\label{app:sep}
Starting from the PDF operator definition \cite{Collins:1989gx}, the LUX group has obtained the photon PDF as \cite{Manohar:2016nzj,Manohar:2017eqh}
\begin{equation}
\begin{aligned}\label{eq:gmPDF}
&\gamma\left(x, \mu^{2}\right) =\frac{8 \pi}{x \alpha\left(\mu^{2}\right)(\mathcal{S} \mu)^{2 \epsilon}} \frac{(4 \pi)^{-\frac{D}{2}}}{\Gamma\left(\frac{D}{2}-1\right)}
\int_{x}^{1} \frac{\mathrm{d} z}{z} \int_{\frac{m_{p}^{2} x^{2}}{1-z}}^{\infty} \frac{\mathrm{d} Q^{2}}{Q^{2}} \alpha_{\mathrm{ph}, \mathrm{D}}^{2}\left(q^{2}\right)\left(Q^{2}(1-z)-x^{2} m_{p}^{2}\right)^{\frac{D}{2}-2}\times\\
&\left\{-z^{2} F_{L, D}\left(x / z, Q^{2}\right)+\left[2-2 z+z^{2}+\frac{2 m_{p}^{2} x^{2}}{Q^{2}}\right] F_{2, D}\left(x / z, Q^{2}\right)-2 \epsilon z x F_{1, D}\left(x / z, Q^{2}\right)\right\},
\end{aligned}
\end{equation}
where $q^2=-Q^2$ corresponds to the spacelike region. The calculation is performed using dimensional regularization in $D=4-2\epsilon$ dimensions. The regularization scale, $\mu$, is replaced by $\mathcal{S}\mu$ where 
$\mathcal{S}^2=e^{\gamma_E}/(4\pi)$, according to the $\msbar$ prescription.
Above, the $F_{i,D}(x,Q^2)(i=1,2,L)$ represent structure functions in the $D^\mathit{th}$ dimension. We note that this photon PDF is exact, including QED radiative
corrections. The integration in Eq.~(\ref{eq:gmPDF}) is divergent, however, when $Q^2$ is integrated up to infinity. Following the standard renormalization procedure, we can
split the $Q^2$ integral into two parts, $m_p^2x^2/(1-z)<Q^2<\mu^2/(1-z)$ and $\mu^2/(1-z)<Q^2<\infty$, which correspond to the physical factorization (PF) and
$\msbar$ conversion (con), respectively, leading to two contributions to the photon PDF,
\begin{equation}
\gamma(x,\mu^2)=\gamma^{\rm PF}(x,\mu^2)+\gamma^{\rm con}(x,\mu^2)\ .
\end{equation}
As a result, the PF term becomes finite in $D=4$ dimensions.
\begin{equation}
\begin{aligned}
\gamma^{\rm PF}\left(x, \mu^{2}\right) 
&=\frac{1}{2 \pi \alpha\left(\mu^{2}\right)} \int_{x}^{1} \frac{\mathrm{d} z}{z}\int _ { \frac { x ^ { 2 } m _ { p } ^ { 2 } } { 1 - z } } ^ { \frac { \mu ^ { 2 } } { 1 - z } } \frac { \mathrm { d } Q ^ { 2 } } { Q ^ { 2 } } \alpha ^ { 2 } ( Q ^ { 2 } ) \\
&\left[\left(z p_{\gamma q}(z)+\frac{2 x^{2} m_{p}^{2}}{Q^{2}}\right) F_{2}\left(x / z, Q^{2}\right)
-z^{2} F_{L}\left(x / z, Q^{2}\right)\right]\ .
\end{aligned}
\end{equation}
The $\msbar$ conversion term can be integrated semi-analytically as
\begin{equation}
\gamma^{\mathrm{con}}\left(x, \mu^{2}\right)=\frac{\alpha\left(\mu^{2}\right)}{2 \pi x} \int_{x}^{1} \frac{\mathrm{d} z}{z}\left\{\frac{1}{\epsilon}\left[2-2 z+z^{2}\right] F_{2}\left(x / z, \mu^{2}\right)\right\}-z^{2} F_{2}\left(x / z, \mu^{2}\right)\ ,
\end{equation}
by assuming the stationary condition, Eq. (\ref{eq:stationary}). In other words, the structure functions are assumed not to depend on $Q^2$, which is valid to lowest order in $\alpha$ and $\alpha_s$. The $1/\epsilon$ term is absorbed by the $\msbar$ counter term. At this stage, we have obtained the complete master LUX formula as in Eq.~(\ref{eq:LUX}).

In general, we could split the integration in Eq.~(\ref{eq:gmPDF}) at a separation scale $M^2(z)$. Correspondingly, the $\msbar$ conversion term should be changed to capture the difference as
\begin{equation}
\begin{aligned}
\gamma^{\operatorname{con}}(x, \mu^2,[M])=& \gamma^{\mathrm{con}}(x,\mu^2)+\frac{1}{x 2 \pi} \int_{x}^{1} \frac{\mathrm{d} z}{z} \int_{M^{2}(z)}^{\frac{\mu^{2}}{1-z}} \frac{\mathrm{d} Q^{2}}{Q^{2}} \\
& \alpha_{\rm ph}^2(-Q^2)\left\{-z^{2} F_{L}\left(\frac{x}{z}, Q^{2}\right)+z p_{\gamma q} F_{2}\left(\frac{x}{z}, Q^{2}\right)\right\}\ .
\end{aligned}    
\end{equation}
With the same stationary assumption, the modified $\msbar$ conversion term becomes
\begin{equation}\label{eq:con}
\gamma^{\operatorname{con}}(x,\mu^2,[M])
=\gamma^{\operatorname{con}}(x,\mu^2)+\frac{\alpha(\mu^2)}{2 \pi x} \int_{x}^{1} \frac{\mathrm{d} z}{z} \log \frac{\mu^{2}}{(1-z) M^{2}(z)} z p_{\gamma q} F_{2}\left(x / z, \mu^{2}\right)\ .    
\end{equation}
The change in the photon PDF originated by varying the separation scale $M[z]$ is taken to estimate the missing high order (MHO) uncertainty, which has been employed both in CT18lux and CT18qed in Sec.~\ref{sec:CT18lux} and \ref{sec:ct18qed}, respectively.

\bibliographystyle{utphys}
%\bibliography{ref}
\bibliography{ref_Xie}
\end{document}